\documentclass[letterpaper,twocolumn,10pt]{article}
\usepackage{usenix2021}

\usepackage[english]{babel}

\usepackage{epsfig}
\usepackage{balance}
\usepackage{pifont}

\usepackage{times}
\usepackage{graphicx} 
\usepackage{subcaption}
\usepackage{tabularx}
\newtheorem{theorem}{Theorem}
\usepackage{xspace}
\usepackage{listings}
\usepackage{verbatim}
\usepackage{booktabs}

\usepackage{colortbl}
\usepackage[normalem]{ulem}

\usepackage{nicefrac}
\usepackage{siunitx}
\usepackage{stackengine}
\usepackage[small, compact]{titlesec}
\usepackage{algpseudocode,amsmath}
\usepackage[noend]{algorithm2e}
\usepackage{titling}
\titlespacing*{\section}{0pt}{4pt}{3pt}
\titlespacing*{\subsection}{0pt}{3pt}{3pt}
\titlespacing*{\subsubsection}{0pt}{3pt}{3pt}

\usepackage{enumitem}
\setlist{nolistsep}

\usepackage{amsmath}
\usepackage[subtle]{savetrees}

\usepackage{color}
\definecolor{darkred}{rgb}{0.7,0,0}
\definecolor{darkgreen}{rgb}{0,0.5,0}
\hypersetup{colorlinks=true,
        linkcolor=darkred,
        citecolor=darkgreen}

\newcommand{\eg}{{e.g.,} }

\newcommand{\ie}{{i.e.}, }

\newcommand{\Fig}[1]{Fig.~\ref{fig:#1}\xspace}

\newcommand{\Sec}[1]{$\S$\ref{s:#1}\xspace}
\newcommand{\App}[1]{App.~\ref{app:#1}\xspace}

\newcommand{\ma}[1] {{\textcolor{red}{MA: #1}}}

\newcommand{\pg}[1] {{\textcolor{red}{PG: {#1}}}}
\newcommand{\rev}[1] {{\textcolor{black}{ #1}}}

\newcommand{\cut}[1]{}

\newcommand{\bfc}{BFC}

\def\compactify{\itemsep=0pt \topsep=0pt \partopsep=0pt \parsep=0pt}
\let\latexusecounter=\usecounter
\newenvironment{CompactItemize}
  {\def\usecounter{\compactify\leftmargin=17pt\latexusecounter}
   \begin{itemize}}
  {\end{itemize}\let\usecounter=\latexusecounter}
\newenvironment{CompactEnumerate}
  {\def\usecounter{\compactify\leftmargin=13pt\latexusecounter}
   \begin{enumerate}}
  {\end{enumerate}\let\usecounter=\latexusecounter}

\titleformat{\section}
  {\normalfont\large\bfseries}{\thesection}{1em}{\MakeUppercase} 

\begin{document}


\setlength{\droptitle}{-1.2cm}
\title{\bf \LARGE Backpressure Flow Control\vspace{-4mm}}
\author{\Large Prateesh Goyal$^{1}$, Preey Shah$^{2}$, Kevin Zhao$^{3}$, Georgios Nikolaidis$^{4}$,\\ \Large Mohammad Alizadeh$^{1}$, Thomas E. Anderson$^{3}$\\\\{\vspace{6mm}\large $^{1}$MIT CSAIL, $^{2}$IIT Bombay, $^{3}$University of Washington, $^{4}$Intel, Barefoot Switch Division}}\vspace{2mm}
\date{\vspace{-12mm}}

\maketitle
\begin{abstract}
    Effective congestion control for data center networks is becoming
increasingly challenging with a growing amount of latency-sensitive
traffic, much fatter links, and extremely bursty traffic.
Widely deployed algorithms, such as DCTCP and DCQCN, are still far from optimal
in many plausible scenarios, particularly for tail latency. Many operators
compensate by running their networks at low average utilization, dramatically
increasing costs.

In this paper, we argue that we have reached the practical limits
of end-to-end congestion control. Instead, we propose, implement, and evaluate a new
congestion control architecture called \emph{Backpressure Flow Control} (BFC).
BFC provides per-hop per-flow flow control, but with bounded state, constant-time
switch operations, and careful use of buffers.
We demonstrate BFC's feasibility by implementing it on \rev{Tofino2}, a state-of-the-art P4-based programmable 
hardware switch. In simulation, we show that BFC achieves near 
optimal throughput and tail latency behavior even under challenging conditions such as 
high network load and incast cross traffic. Compared to deployed end-to-end schemes, 
BFC achieves 2.3\,-\,60$\times$ lower tail latency for short flows and 1.6\,-\,5$\times$ better
average completion time for long flows.

\end{abstract}

\newcommand{\introsec}{Introduction}
\section{\introsec}
\label{s:intro}
Single and multi-tenant data centers have become one of the largest and fastest growing segments of the computer industry.
Data centers are increasingly dominating the market for all types of high-end computing,
including enterprise services, parallel computing, large scale data analysis, fault-tolerant
middleboxes, and global distributed applications~\cite{googlecloud,azure,s3}.  These workloads
place enormous pressure on the data center network to deliver, at low cost,
ever faster throughput with low tail latency even for highly bursty traffic~\cite{dean2013tail,zats2012detail}.

Although details vary, almost all data center networks today use a combination of endpoint congestion control,
FIFO queues at switches, 
and end-to-end feedback of congestion signals like delay or explicit switch state to the endpoint control loop.\footnote{We refer to schemes that rely on feedback signals delayed by an entire round-trip-time as {\em end-to-end} schemes, to contrast them with hop-by-hop mechanisms.}
As link speeds continue to increase, 
however, the design of the control loop becomes more difficult. First, more traffic fits within a single round trip,
making it more difficult to use feedback effectively. Second, traffic becomes increasingly bursty, so that
network load is not a stable property except over very short time scales. 
And third, switch buffer capacity is not keeping up with increasing link speeds \rev{(\Fig{mot_broadcom})},  making it even more challenging to handle traffic bursts. 
Most network operators run their networks at very low average load, throttle long flows
at far below network capacity, and even then see significant congestion loss.

Instead, we propose a different approach.
The key challenge for data center networks, in our view, is to efficiently 
allocate buffer space at congested network switches. 
This becomes easier and simpler when control actions
are taken per flow and per hop, rather than end-to-end. 
Despite its advantages, per-hop per-flow flow control appears to require 
per-flow state at each switch, even for quiescent flows~\cite{anderson1993high,kung1995credit},
something that is not practical at data center scale.

Our primary contribution is to show that 
per-hop per-flow flow control can be \emph{approximated} with a 
modest and limited amount of switch state, using only simple constant-time switch operations on a modern programmable switch.
\rev{Instead of all flows, we only need state for \emph{active flows}---those flows with 
queued packets. We show that, with switch-level fair queueing or 
shortest flow scheduling, 
the number of active flows is modest for typical data center workloads, 
even in the tail of the distribution. The tradeoff is that performance can degrade 
when the number of active flows exceeds the number of queues.
In practice, we advocate combining per-hop flow control with end-to-end 
congestion control to avoid pathological behavior. However,
to better illustrate the benefits and limitations of our approach, our description and
experiments focus on comparing pure per-hop control with pure end-to-end control.}

 
We have implemented our approach, {\em Backpressure Flow Control (BFC)}, on \rev{Tofino2} a 
state-of-the-art P4-based programmable switch ASIC supporting 12.8 Tbps of switching capacity ~\cite{tofino2}. \rev{Tofino2 has 32-128 independently pausable queues at 
each output port.}
Our implementation uses less than 10\% of the dedicated stateful memory on \rev{Tofino2}. 
All per-packet operations are implemented entirely in the dataplane; 
BFC runs at full switch capacity. 

To evaluate performance, we run large-scale ns-3~\cite{ns3} simulations using synthetic traces
drawn to be consistent with measured workloads from Google and Facebook data centers~\cite{homa} on an oversubscribed multi-level Clos network topology.
\rev{We synthetically add incast to these workloads to represent a challenging
scenario for both end-to-end and per-hop approaches. We consider both
throughput and tail latency performance for short, medium, and long flows.}

\rev{For our simulated workloads, BFC improves both latency for short flows and throughput for long flows. 
Compared to a wide set of deployed end-to-end systems,
including DCTCP~\cite{dctcp}, DCQCN~\cite{dcqcn}, and HPCC~\cite{hpcc}, 
BFC achieves 2.3\,-\,60$\times$ better tail flow completion times (FCTs)
for short flows, and 1.6\,-\,5$\times$ better average performance for long flows. 
ExpressPass~\cite{expresspass} achieves 35\% better short flow tail latency,
but 17$\times$ worse average case performance for long flows.
We also show that BFC performs close to an idealized fair queueing system with unbounded 
buffers and switch queues, but with limited queues and far smaller buffers.
BFC can be combined with other switch scheduling algorithms such as priority
scheduling among traffic classes.  
Unlike other receiver-driven schemes like Homa~\cite{homa}, BFC does not assume knowledge of flow sizes
and does not rely on packet spraying (which is difficult to deploy in practice). With packet spraying, Homa outperforms BFC, but without it we show BFC outperforms Homa and can enforce shortest remaining flow first scheduling more accurately.}


Our specific contributions are:

\begin{CompactItemize}
    \item A discussion of the fundamental limits of end-to-end congestion control for high bandwidth data
    center networks.
    \item A practical protocol for per-hop per-flow flow control, called BFC, that uses a small, constant
    amount of state to achieve near-optimal tail-latency performance for typical data center workloads.
    \item An implementation and proof-of-concept evaluation of BFC on a
    \rev{commercial}
    switch. To our knowledge, this is the first implementation of a per-hop per-flow flow control scheme for a multi-Tbps switch. 
\end{CompactItemize}

\section{Motivation}
\label{s:motivation}

Over the last decade, researchers and data center operators have proposed a variety of congestion control algorithms for data centers, including DCTCP~\cite{dctcp}, Timely~\cite{timely}, Swift~\cite{swift}, DCQCN~\cite{dcqcn}, and HPCC~\cite{hpcc}. The primary goals of these protocols are to achieve high throughput, low tail packet delay, and high resilience to bursts and incast traffic patterns.  Operationally, these protocols rely on {\em end-to-end} feedback loops, with senders adjusting their rates based on congestion feedback signals echoed by the receivers.
Irrespective of the type of signal (e.g., ECN marks, multi-bit INT information~\cite{hpcc, kim2015band}, delay), the feedback delay for these schemes is a network round-trip time (RTT). This delay has an important role in the performance of end-to-end schemes. In particular, senders require at least one RTT to obtain feedback, and therefore face a hard tradeoff in deciding the starting rate of a flow. They can either start at a high rate and risk causing congestion, or start at a low rate and risk under-utilizing the network. Moreover, even after receiving feedback, senders can struggle to determine the right rate if the state of the network (e.g., link utilization and queuing delay) changes quickly compared to the RTT. 

We argue that three trends are making these problems worse over time, and will make it increasingly difficult to achieve good performance with end-to-end protocols.

\begin{figure}[t]
     \centering
    \includegraphics[width=0.8\columnwidth]{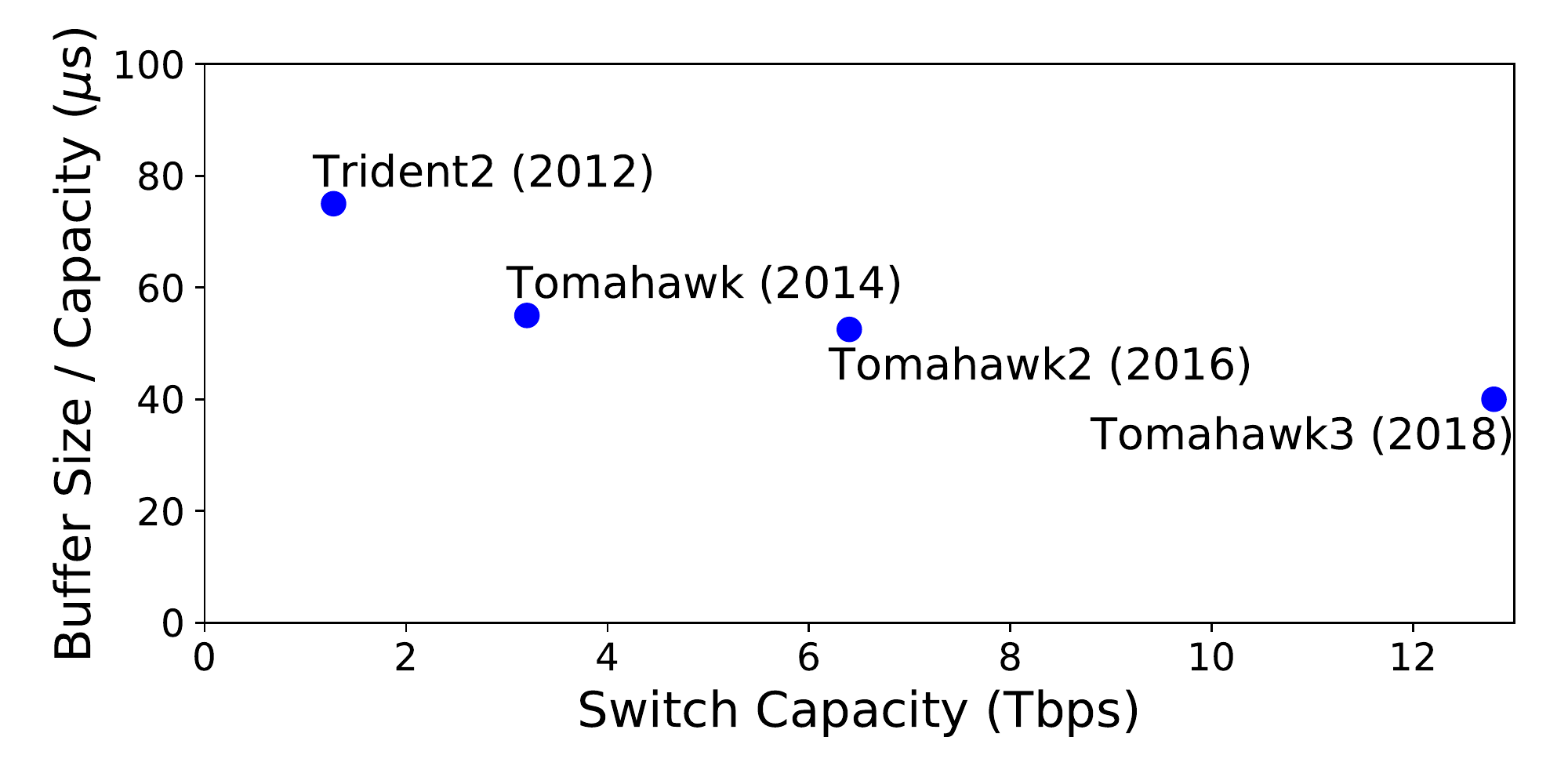}
    \vspace{-4.5mm}
    \caption{\small  Hardware trends for top-of-the-line data center switches from Broadcom. Switch capacity and link speed have been growing rapidly, but buffer size is not keeping up with increases in switch capacity.}
    \label{fig:mot_broadcom}
    \vspace{-5mm}
 \end{figure}

\noindent\textbf{Trend 1: Rapidly increasing link speed.}
\Fig{mot_broadcom} shows the switch capacity of top-of-the-line data center switches manufactured by \rev{Broadcom~\cite{broadcom,switchtrend, jimwarner}}.  
Switch capacity and link speeds have increased by a factor of 10 over the past six years with no signs
of stopping.

\begin{figure}[t]
     \centering
    \includegraphics[width=0.8\columnwidth]{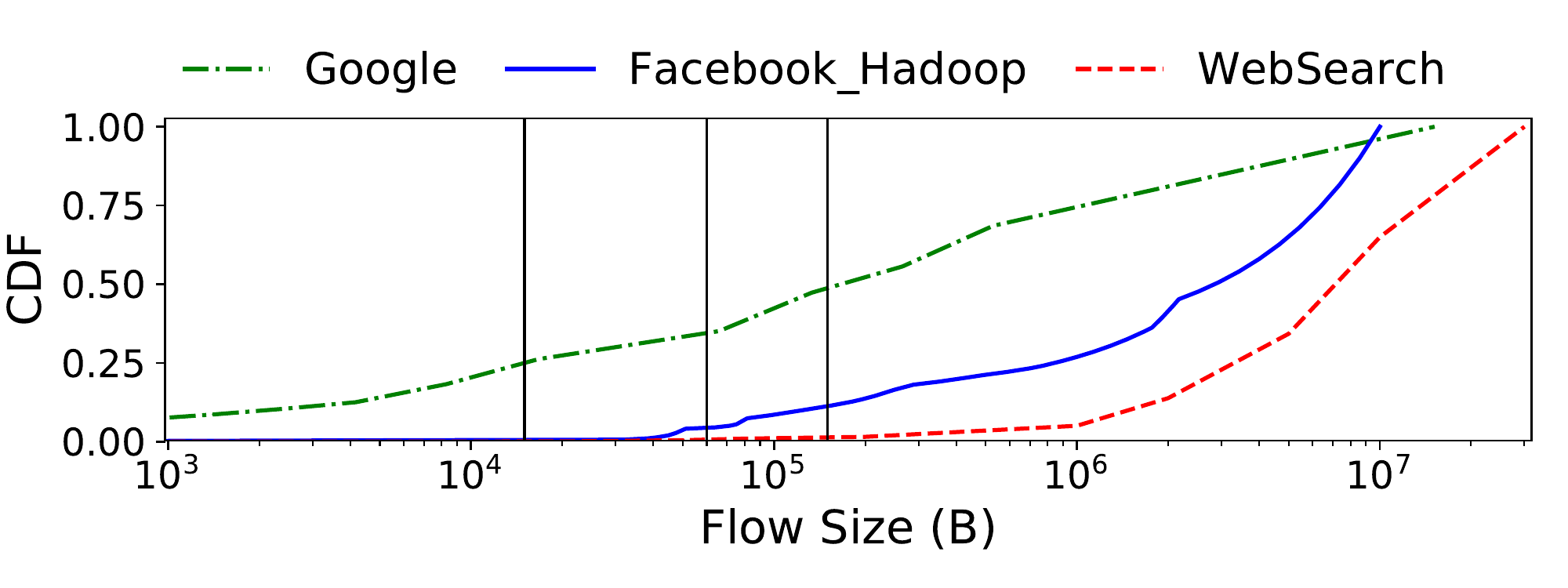}
    \vspace{-4mm}
    \caption{\small Cumulative bytes contributed by different flow sizes for three different industry workloads. The three vertical lines show the BDP for a 10~Gbps, 40~Gbps, and 100~Gbps network, assuming a 12~$\boldsymbol{\mu}$s RTT.}
    \label{fig:mot_flowsize}
    \vspace{-4mm}
 \end{figure}

 \noindent\textbf{Trend 2: Most flows are short.}
\Fig{mot_flowsize} shows the byte-weighted cumulative distribution of flow sizes for three industry data center workloads~\cite{homa}: (1) All applications in a Google data center, (2) Hadoop cluster in a Facebook center, and (3) a WebSearch workload.  Each point is the fraction of all bytes sent that belong to flows smaller than a threshold for that workload. For example, for the Google workload, flows that are shorter than 100~KB represent nearly half of all bytes. \rev{As link speed increases, a growing fraction of traffic belongs to  flows that complete quickly relative to the RTT. For example, most Facebook Hadoop traffic is likely to fit within one round trip within the next decade.} While some have argued 
that data center flows are increasing in size~\cite{atulhotos}, the trend is arguably in the opposite direction with the growing use of RDMA for fine-grained remote memory access. 

\noindent\textbf{Trend 3: Buffer size is not scaling with switch capacity.}
 \Fig{mot_broadcom} shows that the \rev{total} switch buffer size relative to its capacity has decreased by almost a factor of 2 (from 75\,$\boldsymbol{\mu}$s to 40\,$\boldsymbol{\mu}$s) over the past six years. 
 With smaller buffers relative to link speed, buffers now fill up more quickly, making it more difficult for end-to-end congestion control to manage those buffers. 
 
\subsection{Limits of End-to-End Congestion Control}
 \label{s:mot-example}
 This combination\,---\,very fast links, short flows, and inadequate buffers\,---\,creates the perfect storm for end-to-end congestion control protocols. Flows that complete within one or a few RTTs (which constitute an increasingly larger fraction of traffic) either receive no feedback, or last for so few feedback cycles that they cannot find the correct rate~\cite{sperc}. For longer flows, the rapid arrival and departure of cross-traffic creates significant fluctuations in available bandwidth at RTT timescales, making it difficult to find the correct rate. The result is loss of throughput and large queue buildup. Insufficient switch buffering further exacerbates these problems, leading to packet drops or link-level pause events  (PFC~\cite{PFC}) that spread congestion upstream. 

\begin{figure}[t]
    \centering
    \includegraphics[width=0.9\columnwidth]{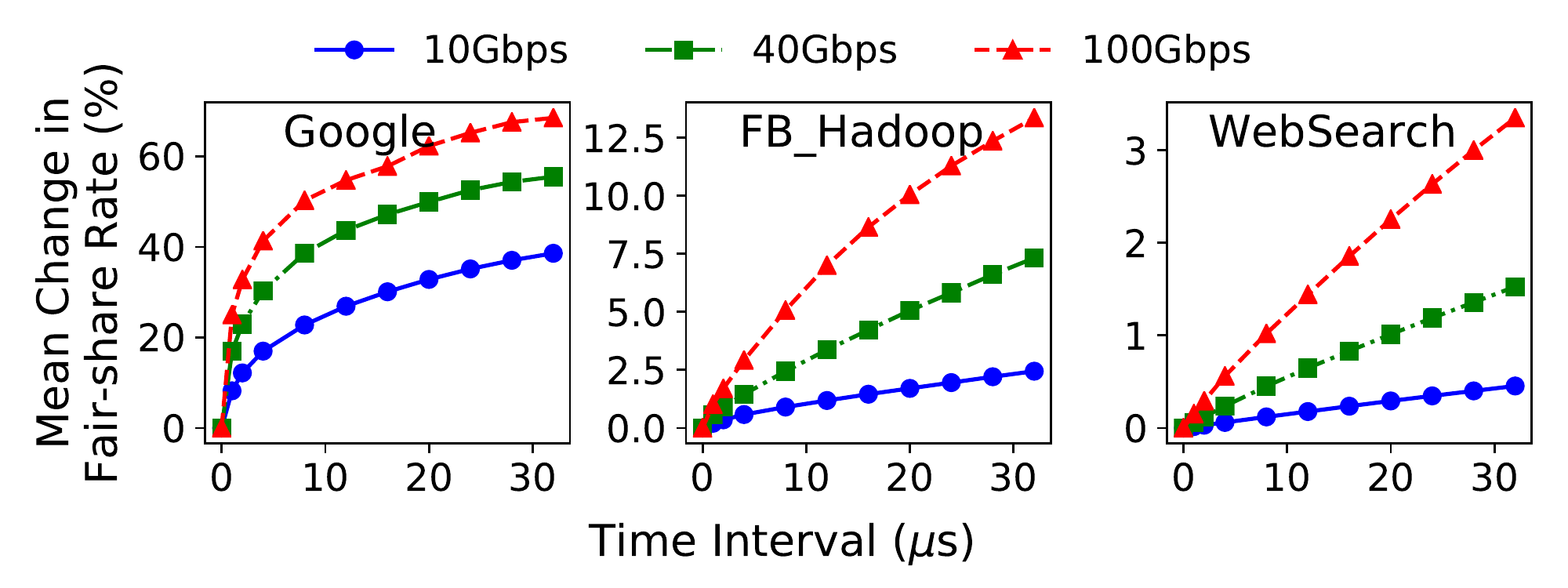}
    \vspace{-4mm}
    \caption{\small \rev{Mean percent change in fair-share rate as a function of workload, delay, and bandwidth.}}
    \label{fig:mot:ll:autocorr}
    \vspace{-4.5mm}
\end{figure}

\rev{To understand these issues, we consider an experiment with a long-lived flow competing on a single link against cross-traffic derived from the Google, Facebook, and WebSearch workloads. We repeat the experiment at 10, 40, and 100~Gbps, with the average load of the cross-traffic flows set to be 60\% of the link capacity in each case. \Fig{mot:ll:autocorr} plots the relative change in the fair-share rate 
of the long-lived flow over different time intervals.\footnote{\rev{The fair-share rate ($f(t)$) for a link of capacity $C$ shared by $N(t)$ flows is $C/N(t)$. The relative change in $f(t)$ over time interval $I$ is given by $\mid\frac{f(t+I)-f(t)}{f(t)}\mid$.}} 
Congestion control protocols  struggle to track the fair-share rate when it varies significantly over their feedback delay (typically an RTT). As link speeds increase or flows become shorter, the fair-share rate changes more rapidly (since flows arrive and finish more quickly), and hence congestion control becomes more difficult.}

\begin{table}
\small
\begin{center}
\begin{tabular}{c|c|c}
 Scheme & Throughput (\%) & 99\% Queuing Delay ($\boldsymbol{\mu}$s)\\ 
 \hline
 \hline
 BFC & 37.3 & 1.2 \\
  \hline
 HPCC & 22.9 & 23.9 \\
  \hline
 DCQCN & 10.0 & 30.4 \\
\end{tabular}
\end{center}
\vspace{-5mm}
\caption{\small \rev{For a shared 100\,Gbps link, BFC achieves close to ideal throughput (40\%) for the long flow, with low tail queuing delay.}}
\label{tab:mot_bfc_noincast}
\vspace{-4mm}
\end{table}

\rev{Table~\ref{tab:mot_bfc_noincast} considers one configuration in detail,
with a single long flow sharing a 100\,Gbps link with cross-traffic drawn from the Facebook distribution at 60\% average load. 
The minimum RTT (hence, feedback delay) is 8\,$\mu$s.
We consider both the single packet (99$^{th}$ percentile) queuing delay and throughput for the long flow, for our approach (BFC) and two end-to-end protocols (DCQCN and HPCC).
BFC is able to achieve close to the maximum possible throughput for the long-lived flow (40\%) with low tail delay, while the end-to-end protocols fall short in both respects.}

\subsection{Existing Solutions are Insufficient}

There are several existing solutions that go beyond end-to-end congestion control. We briefly discuss the most prominent of these approaches and why they are insufficient to deal with the challenges described above.

\smallskip
\noindent{\bf Priority flow control (PFC).}
One approach to handling increased buffer occupancy would be to use PFC, a hop-by-hop flow control mechanism.\footnote{For simplicity, we focus on the case where there is congestion among the traffic at a particular priority level.} With PFC, if the packets from a particular input port start building up at a congested switch (past a configurable threshold), the switch sends a ``pause'' frame upstream, stopping that input from sending more traffic until the switch has a chance to drain stored packets. This prevents switch buffers from being overrun. Unfortunately, PFC has a side effect: head-of-line (HoL) blocking~\cite{dcqcn}. For example, incast traffic to a single server can cause PFC pause frames to be sent one hop upstream towards the source of the traffic. This stops {\em all} the traffic traversing the paused link, even those flows  that are destined to other uncongested egress ports. These flows will be delayed until the packets at the congested port can be drained. Worse, as packets queue up behind a PFC, additional PFC pause frames can be triggered at upstream hops, widening the scope of HoL blocking.

\smallskip
\noindent{\bf Switch scheduling.}
Several efforts use switch scheduling to overcome the negative side-effects of elephant flows on the latency of short flows. These proposals range from approximations of fair queuing (e.g., Stochastic Fair Queuing~\cite{sfq}, Approximate Fair Queuing~\cite{afq}) to scheduling policies that prioritize short flows (e.g., pFabric~\cite{pfabric}, QJump~\cite{qjump}, Homa~\cite{homa}).
\rev{Our work is orthogonal to the choice of switch scheduling policy, and we present results with priority scheduling and shortest flow first. Scheduling by itself does nothing to reduce buffer occupancy; buffers can fill, causing packet drops or HoL blocking, regardless of scheduling.} 


\smallskip
\noindent{\bf Receiver-based congestion control.}
Because sender-based congestion control schemes generally perform poorly on incast workloads, some researchers have proposed shifting to a scheme where the receiver prevents congestion by explicitly allocating credits to senders for sending traffic.
Three examples are NDP~\cite{ndp}, pHost~\cite{phost} and Homa~\cite{homa}.
\rev{BFC makes fewer assumptions than these approaches.
Homa, for example, assumes knowledge of the flow size distribution
and flow length, so that it can assign flows to near-optimal priority queues; this is
unavailable with today's TCP socket interface and not all applications know flow lengths in advance~\cite{bai2015information, dhukic2019advance}. Homa uses packet spraying
to achieve better load balancing, so that
congestion primarily occurs at the last hop, where the receiver has complete
visibility. However, congestion-free operation of the core is difficult to engineer
for widely deployed oversubscribed and asymmetric networks~\cite{jupiter,pktcorrupt,minimalrewire}.
Packet spraying can also cause packet reordering, which is incompatible
with high-speed end host software and hardware packet handling~\cite{snap,tas}.
Other proposals suggest collecting credits generated by a flow's receiver (congestion-controlled by all switches on the flow's path) before 
sending~\cite{expresspass}; at high link speeds, the network state changes rapidly over the feedback delay, making it difficult for the receiver to determine the right rate for credits, similar to sender-based protocols.} 

\subsection{Revisiting Per-hop, Per-Flow Flow Control}
\label{s:phpfcontrol}
Our approach is inspired by work in the early 90s on hop-by-hop credit-based flow control for managing
gigabit ATM networks~\cite{anderson1993high, kung1995credit}. Credit-based flow control was also introduced by multiprocessor hardware designs of the same era~\cite{chaos,dash,cm5}. In these systems, each switch methodically tracks its buffer space, granting permission to send at an upstream switch if and only if there is room in its buffer. In ATM, packets of different flows are buffered in separate queues and are scheduled according to the flows' service requirements. The result is a network that has no congestion loss by design. 

An ideal realization of such a per-hop, per-flow flow control scheme has several desirable properties: 

\noindent{\bf (1) Fast reaction:}  When a flow starts experiencing congestion at a switch, the upstream switch can reduce its rate within a 1-Hop RTT, instead of the end-to-end RTT that it takes for standard congestion control schemes. 
Likewise, when capacity becomes available at a switch, the upstream switch can increase its rate within a 1-Hop RTT (provided the upstream switch has packets from that flow). 
Assuming a hardware implementation, the 1-hop RTT consists of the propagation latency and the switch pipeline latency\,---\,typically 1-2~$\mu$s.\footnote{For example, a 100~m cable has a propagation latency of 500~ns, and a typical data center switch has a pipeline latency around 500~ns~\cite{broadcom, tofino}.} This is substantially smaller than the typical end-to-end RTT in data centers (e.g., 10-20~$\mu$s), which in addition to multiple switch hops includes the 
latency at the endpoints.

\noindent{\bf (2) Buffer pooling:} During traffic bursts, a per-hop per-flow flow control mechanism throttles traffic upstream from the bottleneck. This enables the bottleneck switch to tap into the buffers of its upstream neighbors, thereby significantly increasing the ability of the network to absorb bursts.
 
\noindent {\bf(3) No HoL blocking:} Unlike PFC, there is no HoL blocking or congestion spreading with per-hop per-flow flow control, because switches isolate flows in different queues and perform flow control for each of them separately.

\noindent {\bf (4) Simple control actions:} Flow control decisions in a per-hop per-flow flow control system are simpler to design and reason about than end-to-end congestion control algorithms because: {\em (i)} whether to send or pause a flow at a switch depends only on feedback from the immediate next-hop switch (as opposed to multiple potential points of congestion with end-to-end schemes), {\em (ii)} concerns like fairness are dealt with trivially by scheduling flows at each switch, and therefore flow control can focus exclusively on the simpler task of managing buffer occupancy and ensuring high utilization. 

Despite these compelling properties, per-hop per-flow flow control schemes have not been widely deployed, in part because of their high implementation complexity and resource requirements. ATM schemes require per-connection state and large buffers, which are not feasible in today's data center switches. \rev{We observe, however, that per-connection switch state is not actually required. Indeed, much of the time, per-connection state is for flows that have no packets queued at the switch, and therefore don't need to be flow controlled.}

\rev{We define an \emph{active flow} to be a flow with one or more packets queued
at the switch. A result of queuing theory is that the number of 
active flows is surprisingly small for a switch using fair queuing~\cite{kleinrock1976queueing, kortebi2005evaluating}. 
In particular, for an M/G/1-PS (Processor Sharing) queue with Poisson flow arrivals operating at average load $\rho < 1$, the number of active flows has a geometric distribution with mean $\frac{\rho}{1-\rho}$,  independent of the link speed or the flow size distribution. Even at load $\rho=0.9$, the expected number of active flows is only 9. The intuition behind this fact is that a fair queued switch will tend to process short flows quickly, completing them and keeping the number of active flows small.} 

\begin{figure}[t]
     \centering
\begin{subfigure}[t]{0.45\textwidth}
    \includegraphics[trim={0 0 0 4mm},clip,width=\textwidth]{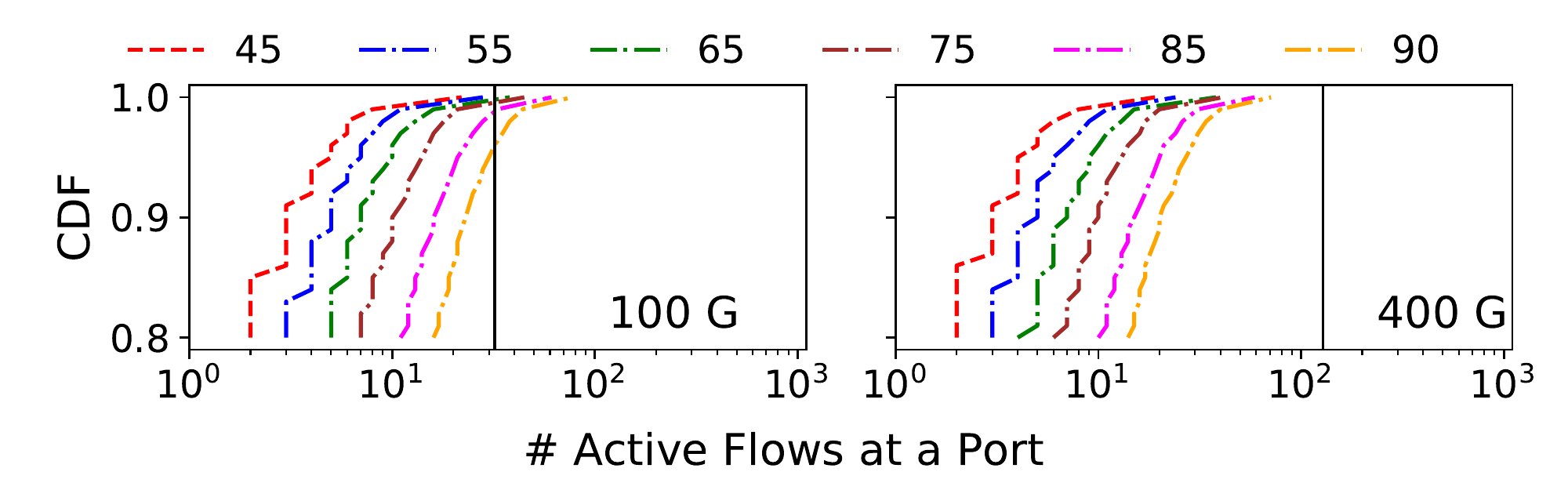}   
    \vspace{-7mm}
        \caption{\small Number of active flows vs. link speed, with fair queuing}
        \label{fig:active:linkspeed}
    \end{subfigure}
 \begin{subfigure}[t]{0.45\textwidth}   
 \includegraphics[trim={0 0 0 4mm},clip,width=\textwidth]{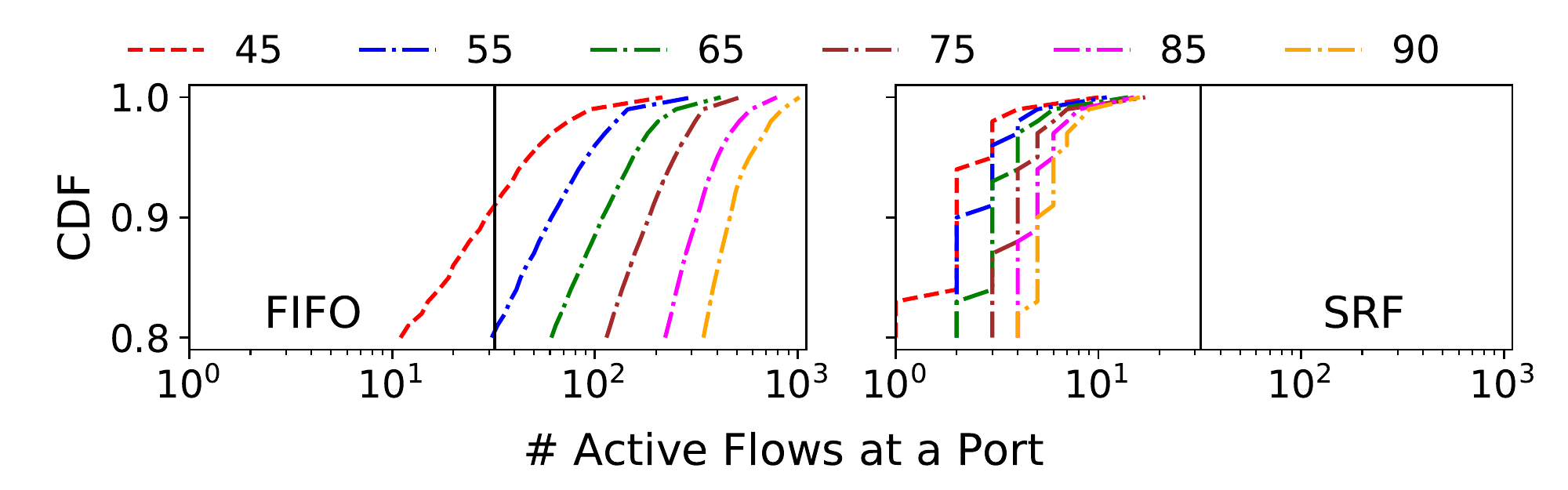}   
    \vspace{-7mm}
        \caption{\small Number of active flows vs. scheduling policy, 100G}
        \label{fig:active:policy}
    \end{subfigure}
    \vspace{-2mm}
    \caption{\small \rev{Number of active flows for different load, link speed, and scheduling policy. Lines correspond to different loads. Flow sizes are from the Google distribution with lognormal ($\sigma=2$) inter-arrival times.}}
    \label{fig:activeflows}
    \vspace{-3mm}
 \end{figure}

\rev{Data center network workloads are often more bursty than Poisson, leading to
longer queues and more active flows. However, the basic principle still holds.
\Fig{activeflows} shows the cumulative distribution of the number of active flows for a single bottleneck link operating at different loads and link
speeds, using the Google flow size distribution and (bursty) log-normal flow inter-arrival times. 
The upper graph assumes fair queuing and includes a vertical bar for the number
of queues per port on Tofino2.
At 100~Gbps, the number of active flows significantly exceeds the number of queues only for 
loads above 85\%, and then only modestly; importantly, the distribution is invariant
to link speed, and the trend is for faster links to have more queues. 
The result holds even more strongly with 
shortest remaining flow first (SRF) scheduling. By contrast, with FIFO queuing, 
even a single long 
flow can cause a large number of small flows to back up behind it, and therefore the number of active flows is much larger.} 

\cut{
\rev{Our principal contribution is a practical design called Backpressure Flow Control (BFC). BFC differs from ATM approaches in several ways: it uses on/off pause frames rather than credits; it carefully manages limited queues and buffer space; 
and, critically, it keeps
state only for active flows.} BFC can be implemented using only constant-time switch operations on a modern programmable switch. BFC is not perfectly lossless, but it makes losses extremely rare. \rev{BFC eliminates HoL blocking as long as the number of active flows is less than the number of queues at a congested port. Occasionally, this isn't possible (e.g., in a large-scale incast), but even then, BFC's performance degrades gracefully and it usually outperforms prior approaches (\S\ref{app:limits}).}
}

\section{Design}
\label{s:design}
Our goal is to design a practical system for per-hop, per-flow flow control for data center networks. We first describe the constraints on our design (\S\ref{s:design_goals}). We then sketch a plausible strawman proposal  that surprisingly turns out to not work well at all (\S\ref{s:strawman}), and we use that as motivation for our design (\S\ref{s:bfc_overview}).

\begin{figure}[t]
     \centering
    \includegraphics[trim= 0.25cm 0.5cm 0.25cm 1.5cm, clip, width=\columnwidth]{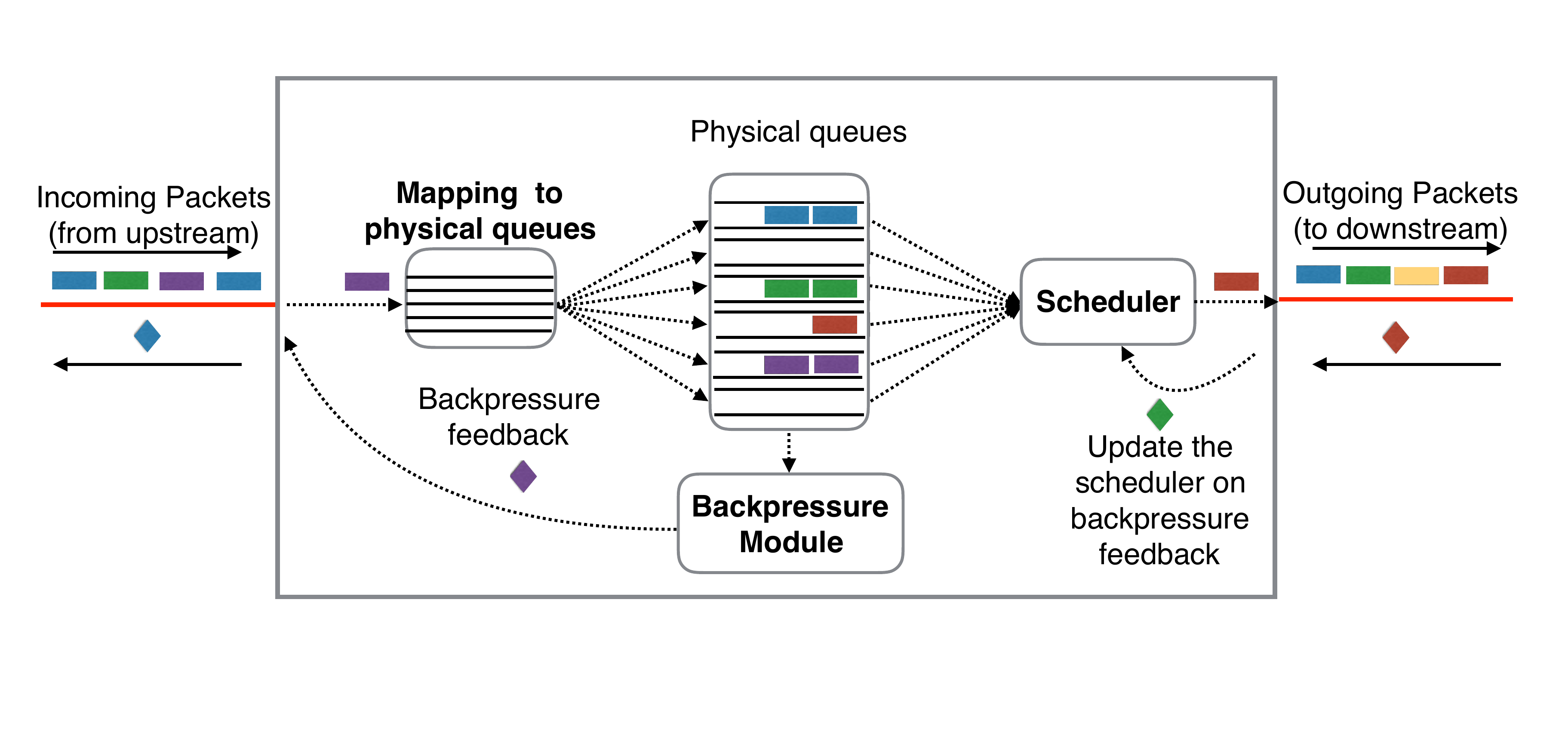}
    \vspace{-13mm}
    \caption{\small Logical switch components in per-hop, per-flow flow control.}
    \label{fig:design_diagram}
    \vspace{-6mm}
 \end{figure}
 
\subsection{Design Constraints}
\label{s:design_goals}

\Fig{design_diagram} shows the basic components of a per-hop, per-flow flow control scheme (per port). (1) {\em Mapping to physical queues:} When a packet arrives at the switch, the switch routes the packet to an egress port and maps it to a FIFO queue at that port. This assignment of flows to queues must be consistent, that is, respect packet ordering. (2) {\em Backpressure module:} Based on queue occupancy, the switch generates backpressure feedback for some flows and sends it upstream. (3) {\em Scheduler:} The scheduler at each egress port forwards packets from queues while respecting backpressure feedback from the downstream switch. 

ATM per-hop per-flow flow control systems~\cite{anderson1993high,kung1995credit} roughly followed this architecture, but they would be impractical for modern data centers. 
First, they assumed per-flow queues and state, but modern switches have a 
limited number of queues per egress port~\cite{afq,bosshart2014p4} 
and modest amounts of table memory~\cite{bosshart13, chole2017drmt}. 
In particular, it is not possible to maintain switch state for all live connections.
Second, earlier schemes did not attempt to minimize buffer occupancy. 
Instead, they sent backpressure feedback only when the switch was about to run out of buffers. 
On a buffer-constrained switch, this can result in buffer exhaustion\,---\,buffers held by straggler
flows can prevent other flows from using those buffers at a later time.

\noindent\textbf{Hardware assumptions.} Modern data center switches have made strides towards
greater flexibility~~\cite{sivaraman2016programmable,arista}, but they are not infinitely malleable and have real resource constraints. We make the following assumptions based on the capabilities of \rev{Tofino2}.
\begin{CompactEnumerate}
    \item We assume the switch is programmable and supports stateful operations.
    \rev{Tofino2} can maintain millions of register entries, and supports simple constant-time per-packet operations to update the state at line rate~\cite{domino}.
    \item The switch has a limited number of FIFO queues per egress port, meaning that flows must be multiplexed onto queues. \rev{Tofino2 has 32/128 queues per 100/400G port}.
    The assignment of flows to queues is programmable. The scheduler can use deficit round-robin
    or priorities among queues, but packets within a queue are forwarded in FIFO order.
    \item Each queue can be independently paused and resumed without slowing down forwarding from other queues. When we pause a queue, that pauses {\em all} of the flows assigned to that queue. The switch can pause/resume each queue directly within the dataplane. 
    
\end{CompactEnumerate}

\subsection{A Strawman Proposal}
\label{s:strawman}

We originally thought stochastic fair queuing~\cite{sfq} with per-queue backpressure might meet our goals: use a hash function on the flow header to consistently 
assign the packets of each flow to a randomly-chosen FIFO queue at its egress port, and pause a queue whenever its buffer
exceeds the 1-hop bandwidth-delay product (BDP). 
For simplicity, use the same hash function at each switch. 

This strawman needs only a small amount of state for generating the backpressure feedback and no state for queue assignment. 
However, 
with even a modest number of active flows, \rev{the birthday paradox implies that}
there is a significant chance that any specific flow will land in an already-occupied FIFO queue. These collisions hurt latency for two reasons: (1) The packets for the flow will be delayed behind unrelated packets from other flows; for example, a short flow may land behind a long flow. (2) Queue sharing can cause HoL blocking. If a particular flow is paused (because it is congested downstream), all flows sharing the same queue will be delayed. 
To prevent collisions from affecting tail latency performance, the strawman requires significantly more queues than active flows. For example, at an egress port with $n$ active flows, to achieve fewer than 1\% collisions, we would need roughly 100$n$ queues.

\subsection{Backpressure Flow Control (BFC)}
\label{s:bfc_overview}

Our design achieves the following properties:

\smallskip
\noindent\textbf{Minimal HoL blocking:}
We assign flows to queues dynamically. As long as the number of active flows at an egress is less than the
number of queues, (with high probability) no two flows share a queue and there is no HoL blocking. When a new flow arrives at the switch, it is assigned to an empty queue if one is available, sharing
queues only if all are in use.

\smallskip
\noindent{\bf Low buffering and high utilization:} BFC pauses a flow at the upstream when the queue occupancy exceeds a small threshold. BFC's pause threshold is set aggressively to reduce buffering. With  coarse pausing like PFC, pausing aggressively hurts utilization, but BFC only pauses those flows causing congestion (except when collisions occur). The remaining flows at the upstream can continue transmitting, avoiding under-utilization. 

\smallskip
\noindent\textbf{Hardware feasibility:}  
BFC does not require per-flow state, and instead uses an amount of memory proportional to the number of physical queues in the switch. To allow efficient lookup of the state associated with a flow, the state is stored in a flow table, an array indexed using a hash of the flow identifier. The size of this array is set in proportion to the number of physical queues. In our \rev{Tofino2} implementation, it consumes less than 10\% of the dedicated stateful memory. Critically, the mechanism for generating backpressure and reacting to it is simple and the associated operations can be implemented entirely in the dataplane at line rate.

\smallskip
\noindent\textbf{Generality:} BFC does not make assumptions about the network topology or where congestion can occur, and \rev{does not require packet spraying like NDP~\cite{ndp} or Homa~\cite{homa}. Furthermore, it does not assume knowledge of flow sizes or deadlines. Such information can be incorporated into BFC's design to improve small flow performance (see \App{homa_comp}), at a cost in deployability.}


\smallskip
\noindent\textbf{Idempotent state:} Because fiber packets can be corrupted in flight~\cite{pktcorrupt}, BFC ensures that pause and resume state is maintained idempotently, 
in a manner resilient to packet loss.

\subsubsection{Assigning flows to queues}
\label{s:qassignment}

To minimize sharing of queues and HoL blocking, 
we dynamically assign flows to empty queues. 
As long as the flow is active (has packets queued at the switch), 
subsequent packets for that flow will be placed into the same FIFO queue. 
Each flow has a unique 5-tuple of the source and destination addresses, 
port numbers, and protocol; we call this the
flow identifier (FID). BFC uses the hash of the FID to track a flow's queue assignment.
To simplify locating an empty queue, BFC maintains a bit map of empty queues. When the
last packet in a queue is scheduled, BFC resets the corresponding bit for that queue.

With dynamic queue assignment, a flow can be assigned to different queues at different switches. To pause a flow, BFC
pauses the queue the flow came from at the upstream
switch (called the upstream queue).  The pause applies to all flows sharing the same upstream queue with the paused flow. We describe the pause mechanism in detail in \Sec{backpressure}.
The packet scheduler uses deficit round robin to implement fair queuing among the  queues that
are not paused.

Since there is a limited number of queues, it is possible that all queues have
been allocated when a new flow arrives, at which point HoL blocking is unavoidable.
For hardware simplicity, we assign the flow to a random queue in this case.
Packets assigned to the same queue are scheduled in FIFO order. \rev{The number of active flows is usually small (\S\ref{s:phpfcontrol}), but in certain settings, such as incast, it can exceed the number of queues. BFC's behavior is similar to stochastic fair queuing in such scenarios in that it incurs HoL blocking. 
BFC still outperforms existing protocols like DCQCN and HPCC except in the most extreme cases (see \App{limits}). Even during a large scale incast, BFC can leverage the large number of upstream queues feeding traffic to a bottleneck switch to (1) absorb larger bursts, and (2) limit congestion spreading.} In particular, when flows involved in an incast are spread among multiple upstream ports, BFC assigns these flows to separate queues at those ports. As long as the total number of flows does not exceed the total number of queues across {\em all} of the upstream ports, BFC will not incur HoL blocking at the upstream switches. As the size of the network increases and the fan-in to each switch gets larger, there will be even more queues at the upstream switches to absorb an incast, further reducing congestion spreading.

\smallskip
\noindent\textbf{Mechanism:}
To keep track of queue assignment, BFC maintains an array indexed by the egress port of a flow and the hash of the FID. All flows that map to the same index are assigned to the same queue. We maintain the following state per entry: the physical queue assignment (\texttt{\small qAssignment}), and the number of packets in the queue from the flows mapped to this entry (\texttt{\small size}). The pseudocode is as follows (we defer switch-specific implementation issues to \S\ref{s:tofino2}):

\begin{lstlisting}[basicstyle=\footnotesize]
On Enqueue(packet):
    key = <packet.egressPort, hash(packet.FID)>
    if flowTable[key].size == 0:
        reassignQueue = True:
    flowTable[key].size += 1
    if reassignQueue:
        if empty q available at packet.egressPort:
            qAssignment = emptyQ
        else:
            qAssignment = randomQ
        flowTable[key].qAssignment = qAssignment
    packet.qAssignment = flowTable[key].qAssignment

On Dequeue(packet):
    key = <packet.egressPort, hash(packet.FID)>
    flowTable[key].size -= 1
\end{lstlisting}

In the flow table, if two flows map to the same index they will use the same queue (collision). Since flows going through different egress ports cannot use the same queue, the index also includes the egress port. Index collisions in the flow table can hurt performance. These collisions decrease with the size of the table, but the flow table cannot be arbitrarily large as the switch has a limited stateful memory. In our design, we set the size of the flow table to 100 $\times$ the number of queues in the switch. This ensures that if the number of flows at an egress port is less than the number of queues, then the probability of index collisions is less than 1\%. If the number of flows exceeds the number of queues, then the index collisions do not matter as there will be collisions in the physical queues regardless. \rev{Tofino2} has 4096 queues in aggregate, and hence the size of the flow table is 409,600 entries, which is less than 10\% of the switch's dedicated stateful memory. 

While using an array is not memory efficient, accessing state involves simple operations. Existing solutions for maintaining flow state either involve slower control plane operations, or are more complex~\cite{pontarelli2019flowblaze, barbette2020high}. In the future, if the number of queues increases substantially, we can use these solutions for the flow table; however at the moment, the additional complexity is unnecessary.

\subsubsection{Backpressure mechanism}
\label{s:backpressure}
BFC pauses a flow if the occupancy of the queue assigned to that flow exceeds the pause threshold \emph{$Th$}. To pause/resume a flow, the switch could signal the flow ID to the upstream switch, which can then pause/resume the queue associated with the flow. While this solution is possible in principle, it is difficult to implement on today's programmable switches. The challenge is that, on receiving a pause, the upstream switch needs to perform a lookup to find the queue assigned to the flow and some additional bookkeeping to deal with cases when a queue has packets from multiple flows (some of which might be paused and some not).

We take a different approach. Switches directly signal to the upstream device to pause/resume a specific queue. Each upstream switch/source NIC inserts its local queue number in a special header field called \texttt{\small upstreamQ}. 

\smallskip
\noindent\textbf{Mechanism:} 
Recall that, in general, multiple flows can share a queue in rare cases. 
This has two implications. First,
we track the queue length (and not just the \texttt{\small flowTable.size}) and use that to determine
if the flow's upstream queue should be paused.  Second, each upstream queue can, in general, \rev{have flows sending packets}
to multiple queues at multiple egresses. We pause an upstream queue
if \emph{any} of its flows are assigned a congested queue, and we resume when \emph{none}
of its flows have packets at a congested queue (as measured at the time the packet arrived at
the switch).

We monitor this using a Pause Counter, an array
indexed by the ingress port and the \texttt{\small upstreamQ} of a packet. The upstream queue is paused if
and only if its Pause Counter at the downstream switch is non-zero. On enqueue of a packet, 
if its flow is assigned a queue that exceeds the pause threshold, 
we increment the pause counter at that index by 1. 
When this packet (the one that exceeded $Th$) leaves the switch we decrement the counter by 1. 
Regardless of the number of flows assigned to the \texttt{\small upstreamQ}, 
it will be resumed only once all of its packets that exceeded the pause threshold (when the packet arrived) 
have left the switch.

\begin{lstlisting}[basicstyle=\footnotesize,mathescape=true]
On Enqueue(packet):
    key = <packet.ingressPort, packet.upstreamQ>
    if packet.qAssignment.qLength > $Th$:
        packet.metadata.counterIncr = True
        pauseCounter[key] += 1
        if pauseCounter[key] == 1:
            //Pause the queue at upstream
            sendPause(key)
\end{lstlisting}

\begin{lstlisting}[basicstyle=\footnotesize,mathescape=true]
On Dequeue(packet):
    key = <packet.ingressPort, packet.upstreamQ>
    if packet.metadata.counterIncr == True:
        pauseCounter[key] -= 1
        if pauseCounter[key] == 0:
            //Resume the queue at upstream
            sendResume(key)
\end{lstlisting}

To minimize bandwidth consumed in sending pause/resumes, we only send a pause packet when the pause counter for an index goes from 0 to 1, and a resume packet when it goes from 1 to 0. For reliability against pause/resume packets being dropped, we also periodically send a bitmap of the queues that should be paused at the upstream (using the pause counter). Additionally, the switch uses a high priority queue for processing the pause/resume packets. This reduces the number of queues available for dynamic queue assignment by 1, but it eliminates performance degradation due to delayed pause/resume packets.

The memory required for the pause counter is small compared to the flow table. For example, if each upstream switch has 128 queues per egress port, then for a 32-port downstream switch, the pause counter is 4096 entries.

\smallskip
\noindent\textbf{Pause threshold.}
BFC treats any queue buildup as a sign of congestion. BFC sets the pause threshold $Th$ to 1-Hop BDP
at the queue drain rate.
Let $N_{active}$ be the number of {\em active queues} at an egress, i.e. queues with data to transmit that are not paused, $HRTT$ be the 1-Hop RTT to the upstream, and $\mu$ be the port capacity. Assuming fair queuing as the scheduling policy, the average drain rate for a queue at the egress is $\mu/N_{active}$. The pause threshold $Th$ is thus given by  $(HRTT) \cdot (\mu / N_{active})$. 
When the number of active queues increases, $Th$ decreases. In asymmetric topologies, egress ports can have different link speeds; as a result, we calculate a different pause threshold for every egress based on its speed. Similarly, ingress ports can have different 1-Hop RTTs.
Since a queue can have packets from different ingresses, 
we use the max of $HRTT$ across all the ingresses to calculate $Th$. 
We use a pre-configured match-action table indexed with $N_{active}$ and $\mu$ to compute $Th$.

BFC does not guarantee that a flow will never run out of packets due to pausing. First, a flow can be paused unnecessarily if it is sharing its upstream queue with other paused flows. 
Second, a switch only resumes an upstream queue once all its packets (that exceeded the pause threshold when they arrived) have left the downstream switch. Since the resume takes an $HRTT$ to take effect, a flow can run out of packets at the downstream switch for an $HRTT$, potentially hurting utilization. However, this scenario is unlikely\,---\,a pause only occurs when a queue builds up, typically because multiple flows are competing for the same egress port. In this case, the 
other flows at the egress will have packets to occupy the link, preventing under-utilization. 

We might reduce the (small) chance of under-utilization by resuming the upstream queue earlier, for example, when a flow's queue at the downstream drops below $Th$, or more precisely, when \emph{every} 
queue (with a flow
from the same upstream queue) drops below $Th$.
Achieving this would require 
extra bookkeeping, complicating the design. 

Increasing the pause threshold would reduce the number of pause/resumes generated,
but only at the expense of increased buffering (\Fig{qdepth}). In \App{impact_th}, we analyze the impact of $Th$ on under-utilization and peak buffer occupancy in a simple model, and we show that a flow runs out of packets at most 20\% of the time when $Th$ is set to 1-hop BDP. Our evaluation results show that BFC achieves much better throughput than this worst case in practice (Table~\ref{tab:mot_bfc_noincast}, \S\ref{s:eval}).

\smallskip
\noindent\textbf{Sticky queue assignment:} 
Using \texttt{\small upstreamQ} for pausing flows poses a challenge. Since a switch does not know the current queue assignment of a flow at the upstream, it uses the \texttt{\small upstreamQ} conveyed by the last packet of the flow to pause a queue. However, if a flow runs out of packets at the upstream switch (e.g., because it was 
bottlenecked at the downstream switch but not the upstream), then its queue assignment may change for subsequent packets, causing it to temporarily evade the pause signal sent by the downstream switch. Such a flow will be paused again when the downstream receives packets with 
the new \texttt{\small upstreamQ}. The old queue will likewise be unpaused
when its last packet (that exceeded $Th$) departs the downstream switch.

To reduce the impact of such queue assignment changes, we add a timestamp to the flow table state,
updated whenever a packet is enqueued or dequeued. A new queue assignment only happens if the \texttt{\small size} value in the flow table is 0, and the timestamp is older than a ``sticky threshold'' (\ie the entry in the flow table has had no packets in the switch for at least this threshold). Since with BFC's backpressure mechanism a flow can run out of packets for an $HRTT$, we set the sticky threshold to a small multiple of $HRTT$ (2 $HRTT$).

While sticky queue assignments reduce the chance that a backlogged flow will change queues, it doesn't completely eliminate it (\eg packets from the same flow may arrive slower than this interval due to an earlier bottleneck).
Such situations are rare, and we found that BFC performs nearly identically to an ideal (but impractical) variant that pauses flows directly using the flow ID without sticky queue assignments.

\section{Tofino2 implementation}
\label{switchx}
We implemented BFC in \rev{Tofino2}, a to-be-released P4-based programmable switch ASIC with a Reconfigurable Match Table (RMT) architecture~\cite{bosshart2014p4}. A packet in \rev{Tofino2} first traverses the ingress pipeline, followed by the traffic manager (TM) and finally the egress pipeline. \rev{Tofino2} has four ingress and four egress RMT pipelines. Each pipeline has multiple stages, each capable of doing stateful packet operations.
Ingress/egress ports are statically assigned to pipelines.

\noindent
\textit{Bookkeeping:} The flow table and pause counter are both maintained in the ingress pipeline. The flow table contains three values for each entry and is thus implemented as three separate register arrays (one for each value), updated one after the other. 

\noindent
\textit{Multiple pipelines:} The flow table is \emph{split} across the four ingress pipelines, and the size of the table in each ingress pipeline is 25 $\times$ the number of queues. During normal operation, packets of an active flow arrive at a single ingress pipeline (same ingress port). Since the state for a flow only needs to be accessed in a single pipeline, we can split the flow table. However, splitting can marginally increase collisions if the incoming flows are distributed unevenly among the ingress pipelines. Similarly, the pause counter is split among the ingress pipelines. An ingress pipeline contains the pause counter entries corresponding to its own ingress ports. 

\noindent
\textit{Gathering queue depth information:} We need queue depth information in the ingress pipeline for pausing and dynamic queue assignment. \rev{Tofino2} has an inbuilt feature tailored for this task. The TM can communicate the queue depth information for all the queues in the switch to all the ingress pipelines without consuming any additional ingress cycles or bandwidth. The bitmap of empty queues is periodically
updated with this data, with a different rotating starting point per pipeline to avoid new 
flows from being assigned to the same empty queue.

\noindent
\textit{Communicating from egress to ingress pipeline:} The enqueue operations described earlier are executed in the ingress pipeline when a packet arrives. Dequeue operations should happen at the egress but the bookkeeping data structures are at the ingress. To solve this, in the egress pipeline, we mirror packets
as they exit and recirculate the header of the mirrored packet back to the ingress pipeline it came from. The dequeue operations are executed on the recirculated packet header. 

Recirculating packets involves two constraints. First, the switch has dedicated internal links for recirculation, but the recirculation bandwidth is limited to 12\% of the entire switch capacity. 
Second, the recirculated packet consumes an additional ingress cycle. The switch has a cap on the number of packets it can process every second (pps capacity). 

Most workloads have an average packet size greater than 500 bytes~\cite{benson2010understanding}, 
and \rev{Tofino2} is designed with enough spare capacity in bandwidth and pps to handle header recirculation
for every packet for those workloads (with room to spare). 
If the average packet size is much smaller, we can reduce recirculations by sampling packets for recirculation (described in \App{incr_deploy}). 

Recirculation is not fundamental to BFC. For example, \rev{Tofino2} has native support for PFC bookkeeping
in the TM. Likewise, if BFC bookkeeping was implemented in the TM, it would not need recirculation.
Similarly, in switches with a disaggregated RMT architecture~\cite{chole2017drmt} where 
the same memory can be accessed at both the ingress and egress, there is no need for recirculation.

\section{Discussion}
\label{s:discussion}

\noindent\textbf{Guaranteed losslessness.} BFC does not guarantee losslessness. In particular, a switch in \bfc{} pauses an \texttt{\small upstreamQ} only after receiving a packet from it. This implies an \texttt{\small upstreamQ} can send packets for up to an $HRTT$ to the bottleneck switch before being paused, even if the switch is congested. In certain mass incast scenarios, this might be sufficient to trigger drops.
Using credits~\cite{anderson1993high,kung1995credit} could address this at the cost of added complexity. We leave an investigation of such prospective variants of BFC to future work. In our evaluation with realistic switch buffer sizes, BFC never incurred drops except under a 2000-to-1 incast (\S\ref{s:limits}) and even then only 0.007\% of the packets were dropped.

\noindent\textbf{Deadlocks:} Pushback mechanisms like PFC have been shown to be vulnerable to deadlocks in the presence of
cyclic buffer dependencies (CBD) or misbehaving NICs~\cite{hu2016deadlocks,rdmascale}. BFC NICs do not generate any backpressure and as a result cannot cause deadlocks. Since NICs always drain, in the absence of CBD, BFC cannot have deadlocks (see \App{deadlock} for a formal proof). A downstream switch in BFC \emph{will} resume an \texttt{\small upstreamQ} if it drains all the packets sent by the \texttt{\small upstreamQ}. If a downstream is not deadlocked, it will eventually drain packets from the upstream, and as a result, the corresponding upstream cannot be deadlocked.

To prevent CBD, we can reuse prior approaches for deadlock prevention. These approaches can be classified into two categories. The first is to redesign routing protocols to avoid installing routes that might cause CBD~\cite{bolt, stephens2016deadlock}. 
The other is to identify a subset of possible ingress/egress pairs that are provably CBD free, and only
send pause/resume along those pairs~\cite{hu2017tagger}.
For a fat-tree topology, this would allow up-down paths but not temporary loops or detour routes~\cite{f10}. 
In BFC, we use the latter approach. Given a topology, we pre-compute a match action table indexed by the ingress and egress port, and simply elide the backpressure pause/resume signal if it is disallowed. See \App{deadlock} for details.

\noindent\textbf{Incremental Deployment:} In a full deployment, BFC would not require end-to-end 
congestion control. In a partial deployment, 
we advocate some form of end-to-end congestion control, such as capping the number of inflight packets
of a flow. A common upgrade strategy is to upgrade switches more rapidly than server NICs. 
If only switches and not NICs are running BFC, capping inflight packets prevents a source 
NIC from overrunning the buffers of the first hop switch. The same strategy can be used
for upgrading one cluster's switches before the rest of the data center~\cite{minimalrewire}.
In our evaluation, we show incremental deployment would have some impact on buffer occupancy at the edge
but minimal impact on performance (\App{incr_deploy}). 
\section{Evaluation}
\label{s:eval}

We present a proof-of-concept evaluation of our \rev{Tofino2} implementation. To compare performance of BFC against existing schemes, we perform large scale ns-3~\cite{ns3} simulations.

\subsection{\rev{Tofino2} evaluation}
\label{s:tofino2}
\begin{figure}[t]
     \centering
    \includegraphics[width=0.7\columnwidth]{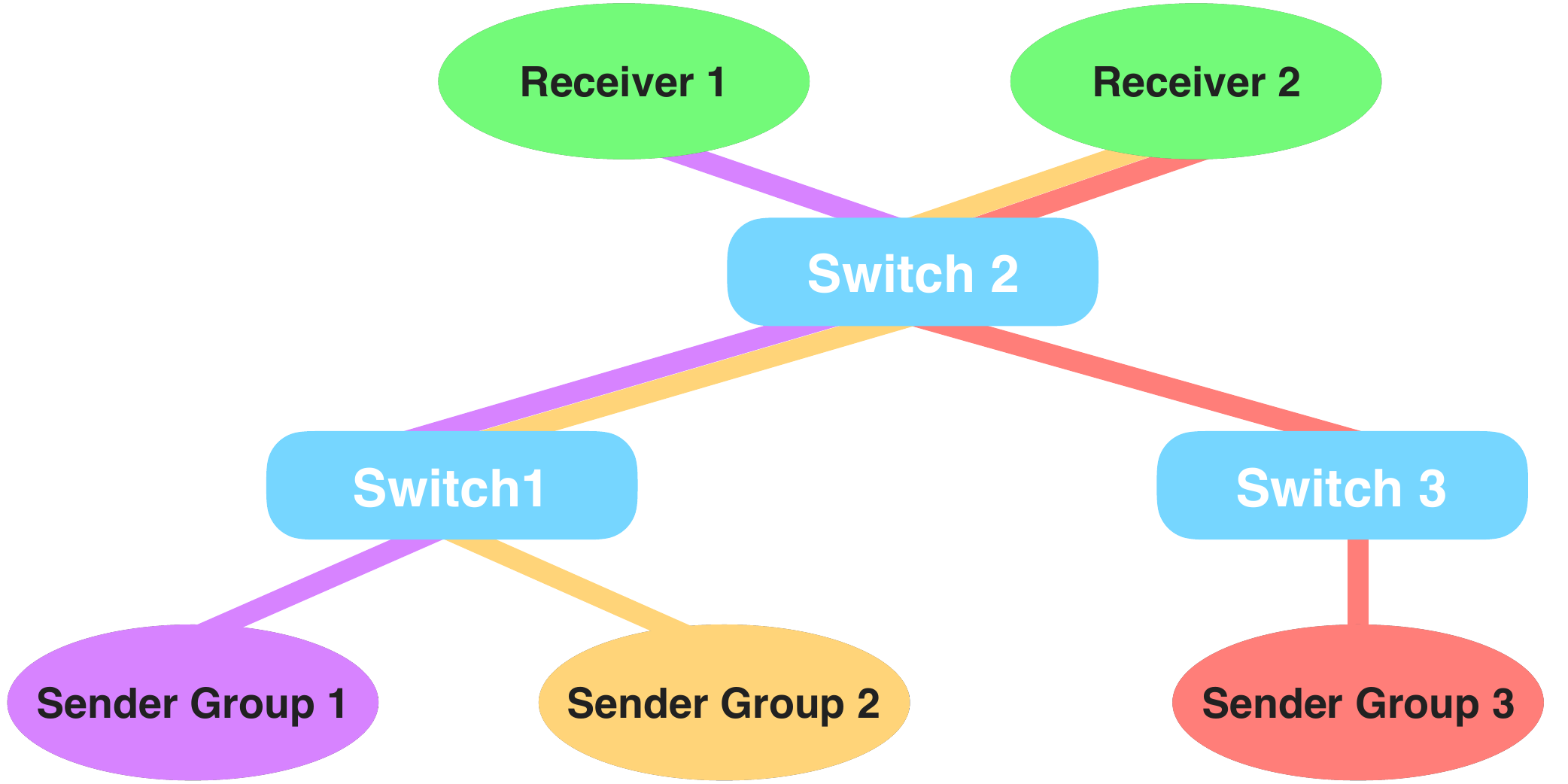}
    \vspace{-2mm}
    \caption{\small {\bf Testbed topology.} The colored lines show the path for different flow groups.}
    \label{fig:testbed}
    \vspace{-5mm}
 \end{figure}

\noindent\textbf{Testbed:} For evaluation, we were able to gain remote access to a \rev{Tofino2} switch. Using a single switch, we created a simple multi-switch topology (\Fig{testbed}) by looping back packets from the egress port back into the switch. All the ports are 100 Gbps, each port has 16 queues.\footnote{For 100 Gbps ports, \rev{Tofino2} has 32 queues, but in loopback mode only 16 queues are available.} The experiments include three groups of flows. 
\begin{CompactItemize}
\item Sender Group 1 $\rightarrow$ Switch 1 $\rightarrow$ Switch 2 $\rightarrow$ Receiver 1.
\item Sender Group 2 $\rightarrow$ Switch 1 $\rightarrow$ Switch 2 $\rightarrow$ Receiver 2.
\item Sender Group 3 $\rightarrow$ Switch 3 $\rightarrow$ Switch 2 $\rightarrow$ Receiver 2.
\end{CompactItemize}
To generate traffic we use the on-chip packet generator with no end-to-end congestion control.

  \begin{figure}[t]
    \centering
    \begin{subfigure}[tbh]{0.235\textwidth}
        \includegraphics[width=\textwidth]{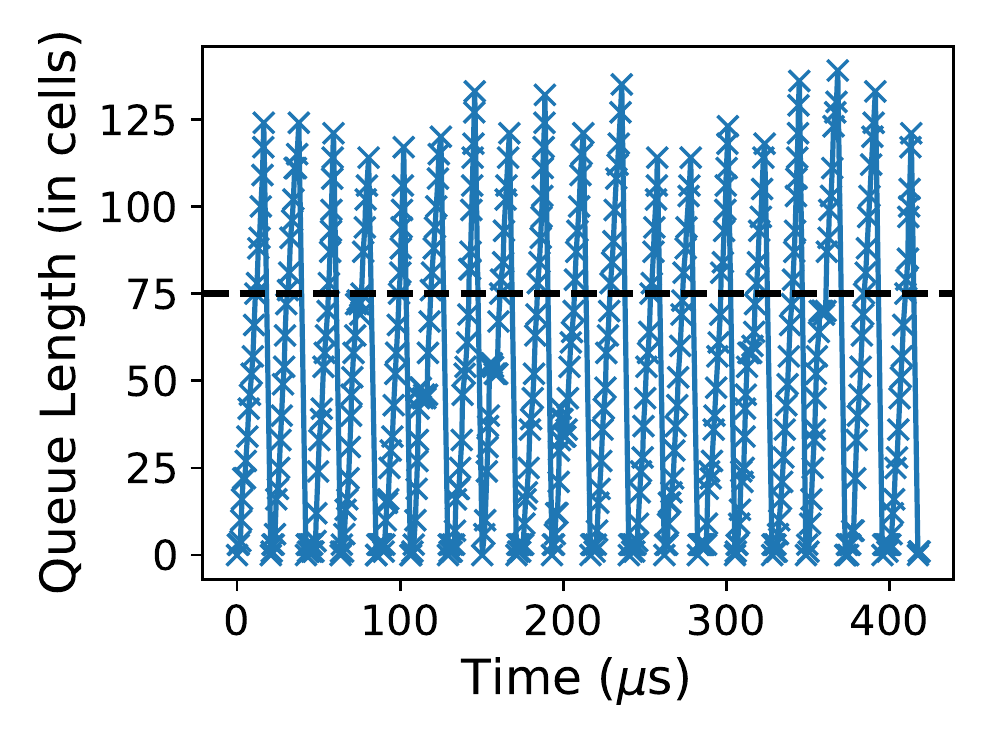}
        \vspace{-7mm}
        \caption{Queue Length}
        \label{fig:qdepth:qdepth}
    \end{subfigure}
    \begin{subfigure}[tbh]{0.235\textwidth}
        \includegraphics[width=\textwidth]{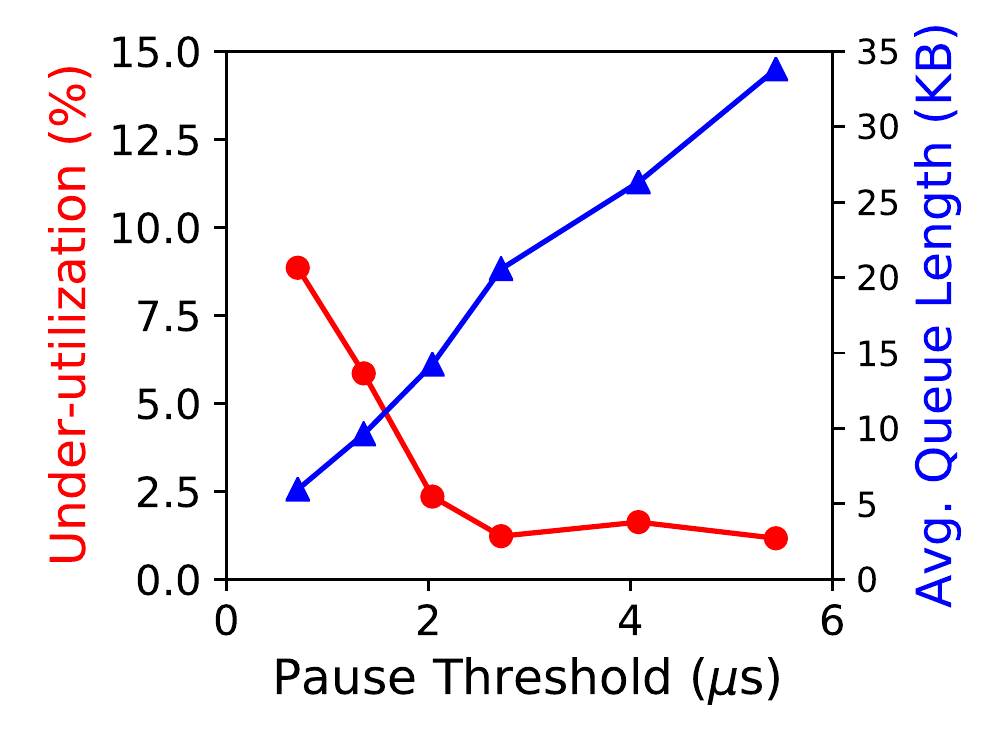}
        \vspace{-7mm}
        \caption{Under-utilization}
        \label{fig:qdepth:utilization}
    \end{subfigure}
    \vspace{-3.5mm}
    \caption{\small {\bf Queue length and under-utilization.} 2 flows are competing at a 100 Gbps link. \rev{Cell size is 176 bytes.} BFC achieves high utilization and low buffering. }
    \label{fig:qdepth}
    \vspace{-5mm}
\end{figure}

\noindent\textbf{Low buffering, high utilization:} \Fig{qdepth:qdepth} shows the queue length for a flow when two flows are competing at a link (a group 2 flow is competing with a group 3 flow at the switch 2 $\rightarrow$ receiver 2 link). The pause threshold is shown as a horizontal black line. BFC's pausing mechanism is able to limit the queue length near the pause threshold ($Th$). The overshoot from $Th$ is for two reasons. First, it takes an $HRTT$ for the pause to take effect. Second, \rev{Tofino2} has small hardware queues after the egress pipeline, and a pause from the downstream cannot pause packets already in these hardware queues. 

Notice that the queue length goes to 0 temporarily. Recall that a downstream switch only resumes the \texttt{\small upstreamQ} when it has drained all the packets from the \texttt{\small upstreamQ} that exceeded $Th$. As a result, a flow at the downstream can run out of packets for an $HRTT$.
This can cause under-utilization when the queues for the two flows go empty simultaneously. 
We repeat the above experiment but vary the pause threshold. \Fig{qdepth:utilization} shows the average queue length and the under-utilization of the congested link. With a pause threshold of 2 $\mu$s, BFC achieves close to 100\% utilization with an average queue length of 15 KB. 

\begin{figure}[t]
     \centering
    \includegraphics[width=0.8\columnwidth]{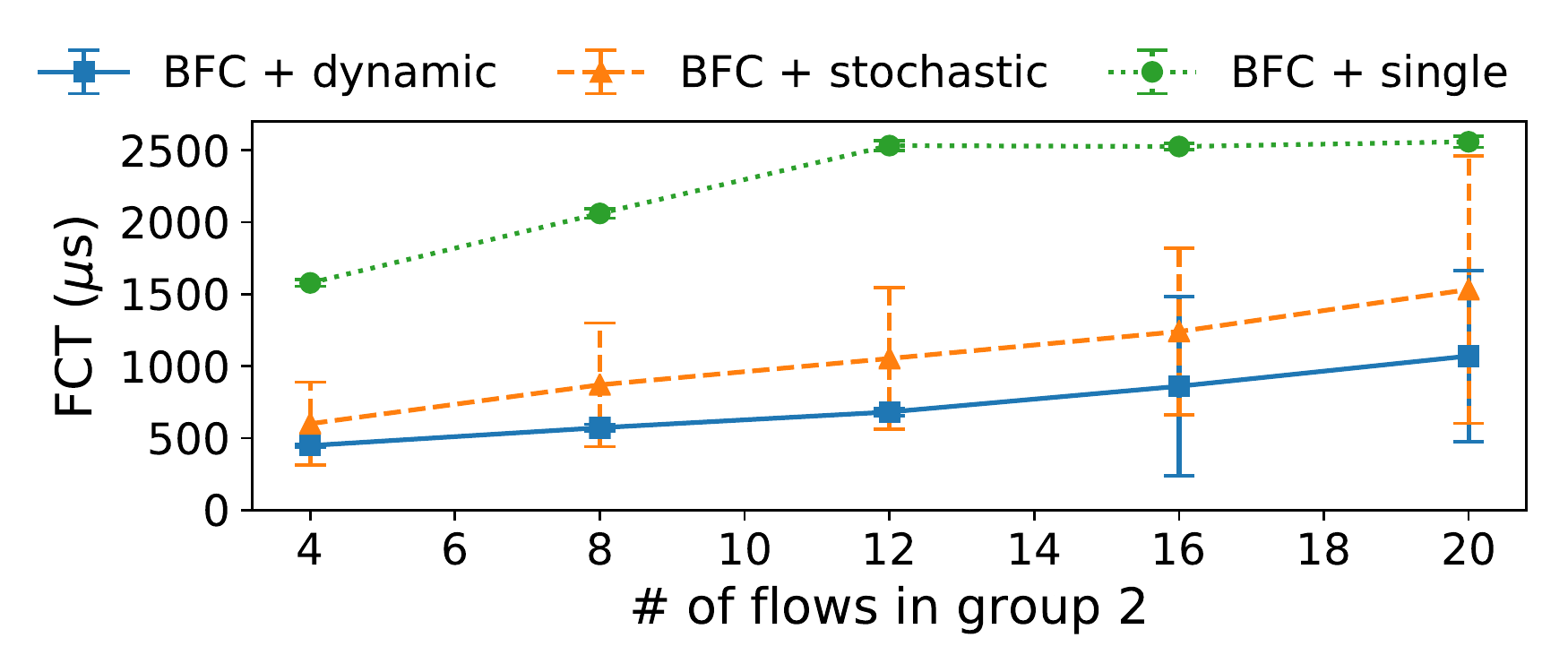}
    \vspace{-4mm}
    \caption{\small {\bf Congestion spreading.} Dynamic queue assignment reduces HoL blocking, improving FCTs on average and at the tail.}
    \label{fig:congestion_spreading}
    \vspace{-5.5mm}
 \end{figure}

\noindent\textbf{Queue assignment and congestion spreading:} We next evaluate the impact of queue assignment on HoL blocking and performance. We evaluate three different queue assignment strategies with BFC's backpressure mechanism: (1) ``BFC + single'': All flows are assigned to a single queue (similar to PFC); (2) ``BFC + stochastic'': Flows are assigned to queues using stochastic hashing; (3) ``BFC + dynamic'': Dynamic queue assignment as described in \S\ref{s:qassignment}. 

The setup consists of two group 1 flows, eight group 3 flows, and a number of group 2 flows varied between four to twenty. All flows are 1.5 MB in size. The experiment is designed such that for group 2 and 3 flows, the bottleneck is the switch 2 $\rightarrow$ receiver 2 link. The bottleneck for group 1 flows is the switch 1 $\rightarrow$ switch 2 link. Switch 2 will pause queues at switch 1 in response to congestion from group 2 flows. Notice that group 1 and group 2 flows are sharing the switch 1 $\rightarrow$ switch 2 link. If a group 1 flow shares a queue with a group 2 flow (a collision), the backpressure due to the group 2 flow can slow down the group 1 flow, causing HoL blocking and increasing its flow completion time (FCT) unnecessarily.

\Fig{congestion_spreading} shows the average FCT for group 1 flows across four runs. The whiskers correspond to one standard deviation in the FCT. BFC + single achieves the worst FCT as group 1 and 2 flows always share a queue. With stochastic assignment, the FCT is substantially lower, but the standard deviation in FCT is high. In some runs, group 1 and 2 flows don't share a queue and there is no HoL blocking. In other runs, due to the stochastic nature of assignment, they do share a queue (even when there are other empty queues), resulting in worse performance. With dynamic assignment, BFC achieves the lowest average FCT and the best tail performance. In particular, the standard deviation is close to 0 when the number of flows at the switch 1 $\rightarrow$ switch 2 link (group 1 + group 2 flows) is lower than the number of queues. In such scenarios, group 1 flows consistently incur no collisions. 
When the number of flows exceed the queues, collisions are inevitable, and the standard deviation in FCT increases.

\subsection{Simulation-based evaluation}
\label{s:simulation}
We also implemented BFC in ns-3~\cite{ns3}. \rev{For DCQCN we use~\cite{dcqcn_code}, for ExpressPass we use~\cite{expass_code}, and for all other schemes we use~\cite{hpcc_code}.} 

 \begin{figure*}[tbh]
    \centering
    \begin{subfigure}[b]{0.48\textwidth}
        \includegraphics[trim={0 0 0 4mm},width=\textwidth]{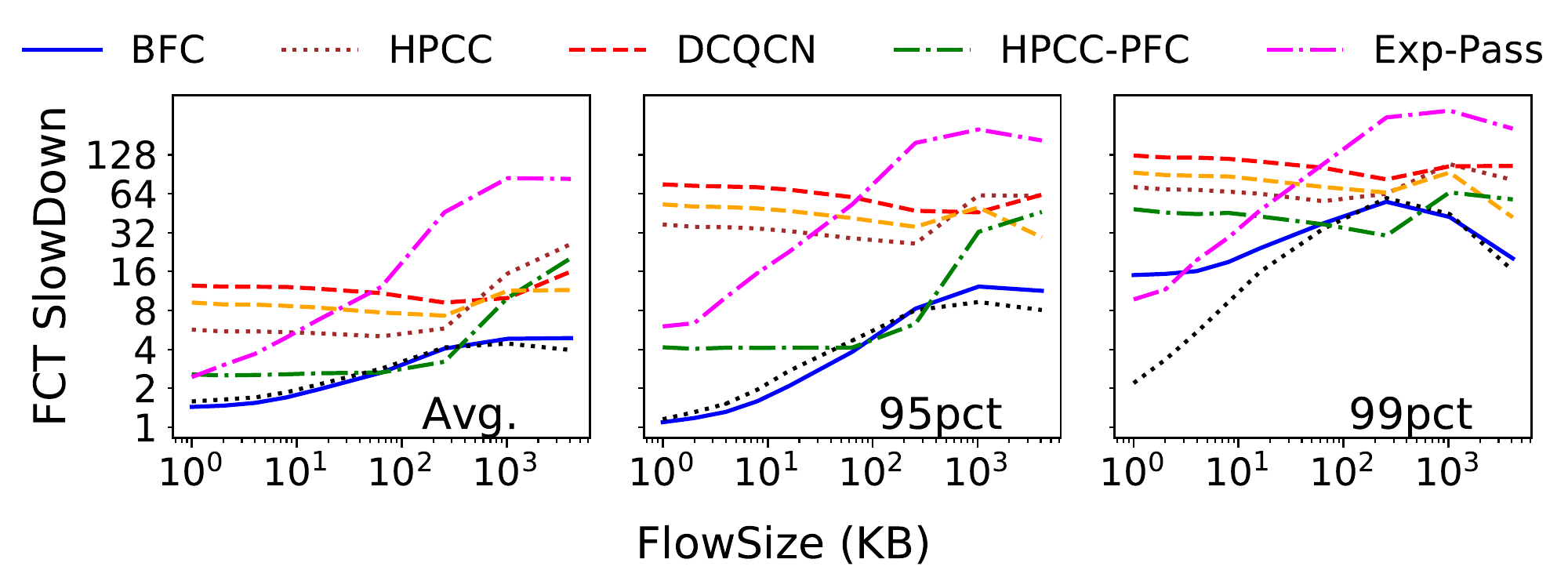}
        \vspace{-7mm}
        \caption{\small{FCT}}
        \label{fig:google_incast:fct}
    \end{subfigure}
    \begin{subfigure}[b]{0.25\textwidth}
        \includegraphics[trim={0 0 0 4mm},width=\textwidth]{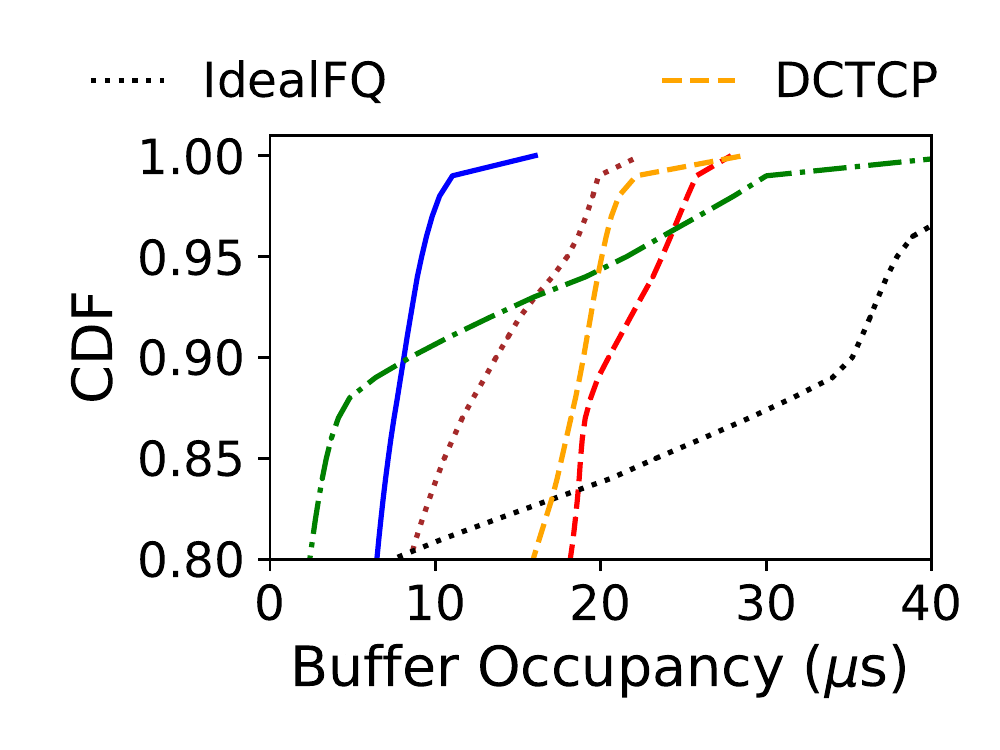}
        \vspace{-7mm}
        \caption{\small{Buffer occupancy}}
        \label{fig:google_incast:buffer}
    \end{subfigure}
    \begin{subfigure}[b]{0.18\textwidth}
        \includegraphics[width=\textwidth]{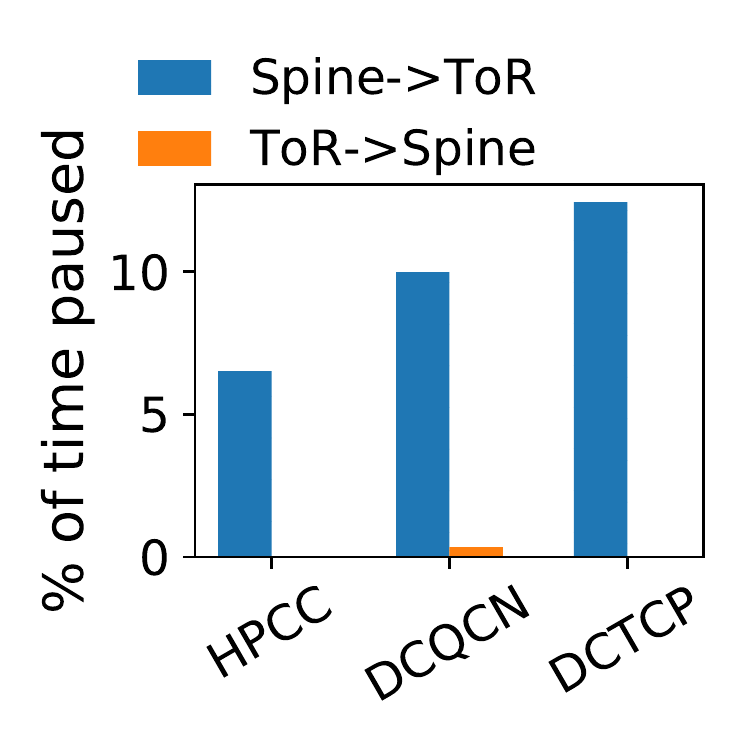}
        \vspace{-7mm}
        \caption{\small{PFC Time}}
        \label{fig:google_incast:pfc}
    \end{subfigure}
    \vspace{-3.5mm}
    \caption{\small \rev{Google distribution with 55\% load + 5\% 100-1 incast. BFC tracks the ideal behavior, improves FCTs, and reduces buffer occupancy. For FCT slowdown, both the x and y axis are log scaled.}}
    \label{fig:google_incast}
    \vspace{-5mm}
\end{figure*}

 \begin{figure*}
    \centering
    \begin{subfigure}[tbh]{0.48\textwidth}
        \includegraphics[width=\textwidth]{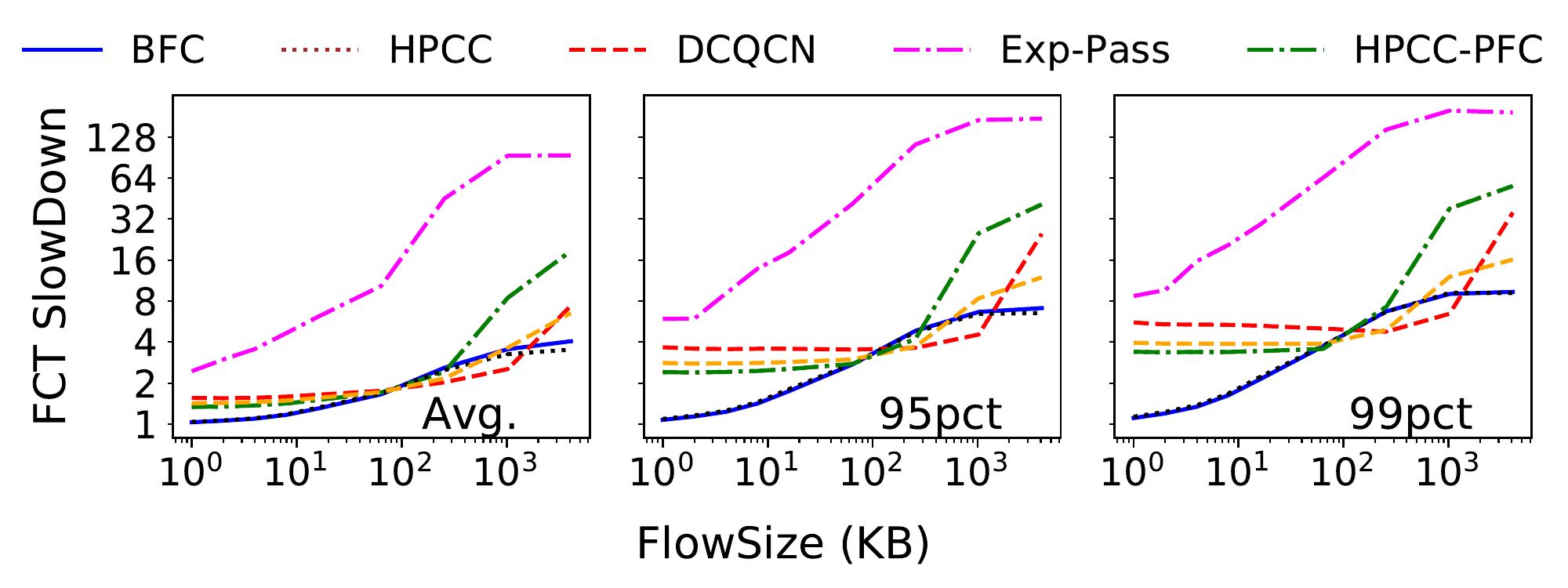}
        \vspace{-7mm}
        \caption{FCT}
        \label{fig:google:fct}
    \end{subfigure}
    \begin{subfigure}[tbh]{0.25\textwidth}
        \includegraphics[width=\textwidth]{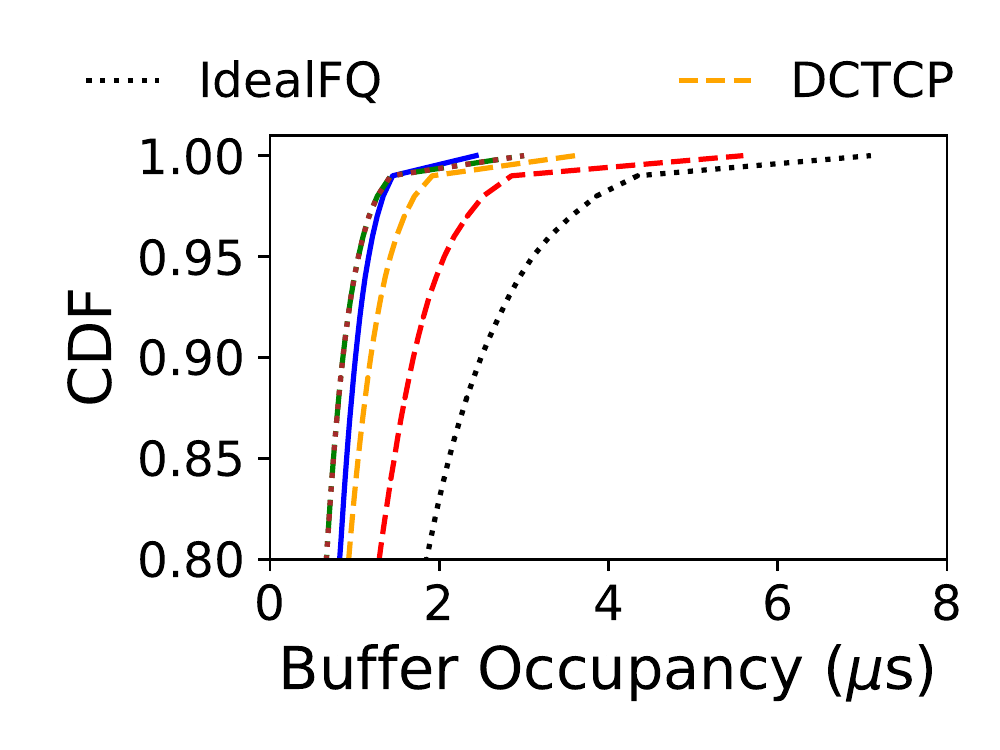}
        \vspace{-7mm}
        \caption{Buffer Occupancy}
        \label{fig:google:buffer}
    \end{subfigure}
        \begin{subfigure}[tbh]{0.24\textwidth}
        \includegraphics[width=\textwidth]{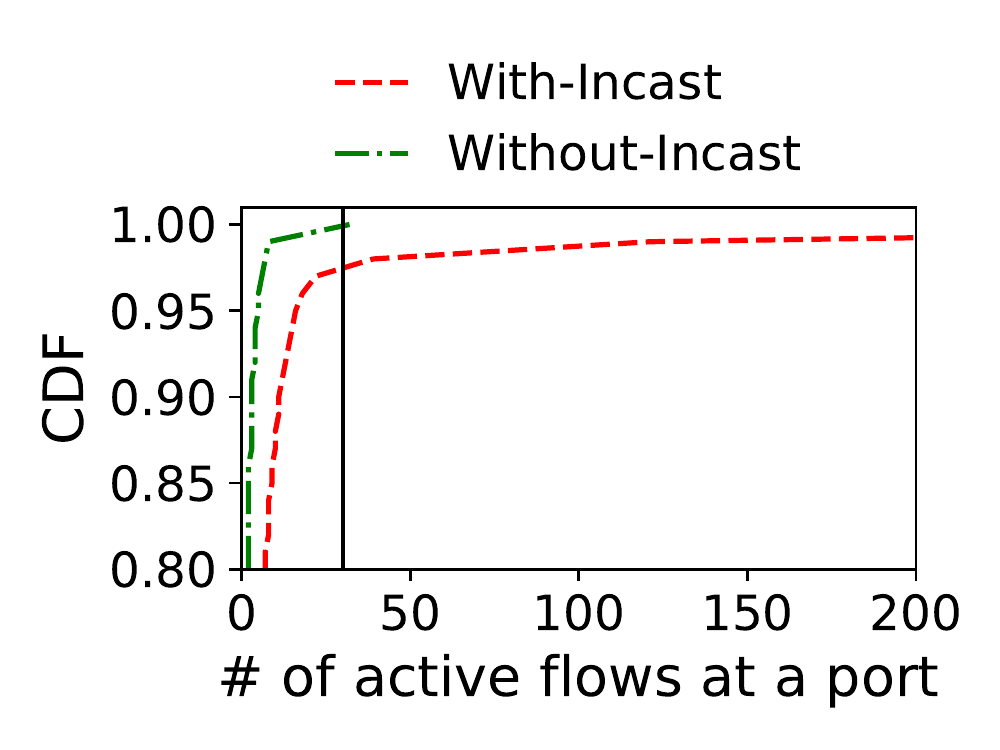}
        \vspace{-7mm}
        \caption{Active Flows}
        \label{fig:google:active}
    \end{subfigure}
    \vspace{-3.5mm}
    \caption{\small \rev{FCT slowdown and buffer occupancy for Google distribution with 60\% load. For all the schemes, PFC was never triggered. Part (c) shows the CDF of active flows at a port with and without incast, with the vertical bar showing the total number of queues per port.}}
    \label{fig:google}
    \vspace{-5mm}
\end{figure*}
 
 \subsubsection{Setup}
 \label{ss:setup}
 
 \noindent\textbf{Network Topology:}  We use a Clos topology with 128 leaf servers, 8 top of the rack (ToR) switches and 8 Spine switches (2:1 over subscription). Each Spine switch is connected to all the ToR switches, each ToR has 16 servers, and each server is connected to a single ToR. All links are 100\,Gbps with a propagation delay of 1\,us. The maximum end-to-end base round trip time (RTT) is 8\,$\boldsymbol{\mu}$s and the 1-Hop RTT is 2\,$\boldsymbol{\mu}$s. The switch buffer size is set to 12 MB. Relative to the ToR switch capacity of 2.4\,Tbps, the ratio of buffer size to switch capacity is 40\,$\boldsymbol{\mu}$s, the same as Broadcom's Tomahawk3 from \Fig{mot_broadcom}. \rev{We use an MTU of 1\,KB.
 Unless specified otherwise, we use Go-Back-N for retransmission, 
 flow-level  ECMP for load balancing, and the standard shared buffer memory model implemented in existing switches~\cite{broadcom}.} 
 
 \noindent\textbf{Comparisons:} \noindent\textit{HPCC:} HPCC uses explicit link utilization
information from the switches to reduce buffer occupancy and drops/PFCs at the congested switch. We use the parameters from the paper, $\eta = 0.95$ and $maxStage = 5$. The dynamic PFC threshold is set to trigger when traffic from an input port occupies more than 11\% of the free buffer (as in the HPCC paper). We use the same PFC thresholds for DCQCN and DCTCP.

\noindent
\rev{
\textit{HPCC-PFC:} This version replaces PFC with perfect retransmission. On a packet drop, the switch informs the sender directly, which then retransmits the dropped packet. We choose this (potentially impractical) strategy to provide a bound on the performance that can be achieved using any retransmission scheme.}

\noindent\textit{DCQCN:} DCQCN uses ECN bits and end-to-end control to manage buffer use at the congested switch. The ECN threshold triggers before PFC (K$_{min}$ = 100KB and K$_{max}$ = 400KB).
 
 \noindent\textit{DCTCP:} The ECN  threshold is same as DCQCN. Flows start at line rate to avoid degradation in FCTs from slow-start.
 
\noindent
\rev{
\textit{ExpressPass:} In ExpressPass, senders transmit data based on credits generated by the receiver. These credits are rate-limited at the switches to avoid congestion. We chose $\alpha = 0.5, w_{init} = 0.0625$ and a credit buffer size of 16 credits. The ExpressPass simulator does not follow a shared buffer model; instead it assumes dedicated per-port buffers. To eliminate drops, we supplied a high per-port buffer value of 75\,MB. There is no PFC.
}
 
 \noindent\textit{\bfc{}:} We use 32 physical queues per port (consistent with \rev{Tofino2}) and our flow table has 76K entries.  The flow table takes 400 KB of memory. We chose per-flow fair queuing as our scheduling mechanism; all the comparison schemes strive for per-flow fairness, thus, fair queuing provides for a just comparison.
 
 \noindent\textit{Ideal-FQ:} To understand how close \bfc{} comes to optimal performance, we simulate ideal fair queuing with infinite buffering at each switch. The NICs cap the in-flight packets of a flow to 1 BDP. Note that infinite buffering is not realizable in practice; its role is to bound how well we could possibly do.
 
 \noindent
 \rev{\textbf{Sensitivity to parameters:} All systems were configured to 
 achieve full throughput for a single flow on an
 unloaded network. For end-to-end schemes, the choice of parameters 
 governs the trade-off between the performance of short flows (through reduced queuing) and long flows (higher link utilization). 
  We perform parameter sensitivity analysis for HPCC, DCTCP and ExpressPass in \App{par_sense_comp}.}
 
\noindent\textbf{Performance metrics:} We consider three performance metrics: (1) FCT normalized to the best possible FCT for the same size flow, running at link rate (referred as the FCT slowdown); (2) Overall buffer occupancy at the switch; (3) Throughput of individual flows. 

 \noindent\textbf{Workloads:}  We synthesized a trace to match the flow size distributions from the industry workloads discussed in \Fig{mot_flowsize}: (1) Aggregated workload from all applications in a Google data center; (2) a Hadoop cluster at Facebook (FB\_Hadoop). The flow arrival pattern is open-loop and follows a \emph{bursty} log-normal inter-arrival time distribution with $\sigma$ = 2.\footnote{\rev{Most prior work evaluates using Poisson flow arrivals~\cite{expresspass,homa}, 
 but we use the more bursty Lognormal as it provides a more challenging case for BFC.} 
 } For each flow arrival, the source-destination pair is derived from a uniform distribution. We consider scenarios with and without incast, different traffic load settings, and incast ratios. Since our topology is oversubscribed, on average links in the core (Spine-ToR) will be more congested than the ToR-leaf server links. In our experiments, by X\% load we mean X\% load on the links in the core.

 \subsubsection{Performance}
 \label{s:eval_performance}

 \Fig{google_incast} and ~\ref{fig:google} show our principal results. The flow sizes are drawn from the Google distribution and 
the average load is set to 60\% of the network capacity. For \Fig{google_incast} (but not \Fig{google}), 5\% of the traffic (on average) is from incast flows. The incast degree is 100-to-1 and the size is 20\,MB in aggregate. A new incast event starts every 500\,$\mu$s. Since the best-case completion time for an incast is 1.6 ms (20\,MB/100\,Gbps), multiple incasts coexist simultaneously in the network. We report the FCT slowdowns at the average, 95$^{\text{th}}$ and 99$^{\text{th}}$ percentile, the tail buffer occupancy \rev{(except for ExpressPass simulations which do not follow the shared  buffer model)},  and the fraction of time links were paused due to PFC. We report the FCT slowdowns for the incast traffic separately in \App{incast_fct}.  
 
Out of all the schemes, DCQCN is worst on latency for small flow sizes, both at the average and the tail. Compared to DCQCN, DCTCP improves latency as it uses per-ACK feedback instead of periodic feedback via QCN.
However, the frequent feedback is not enough, and the performance is far from optimal (Ideal-FQ). The problem is that both
 DCQCN and DCTCP are slow in responding to congestion. 
 Since flows start at line rate, a flow can build up an entire end-to-end bandwidth-delay product 
 (BDP) of buffering (100 KB) at the bottleneck
 before there is any possibility of reducing its rate.
 The problem is aggravated during incast events. 
The bottleneck switch can potentially accumulate one BDP of packets per incast flow (10\,MB in aggregate for 100-to-1 incast).

Both protocols have low throughput for long flows. When capacity becomes available, a long flow may fail 
to ramp up quickly enough, reducing throughput and shifting its work to busier periods where it can impact other flows. Moreover, on sudden onset of congestion, a flow may not reduce its rate fast enough, 
slowing short flows.

HPCC improves on DCQCN and DCTCP by using link utilization instead of ECN and a better control algorithm. 
Compared to DCQCN and DCTCP, HPCC reduces tail latency, tail buffer occupancy, and PFC pauses (in case of incast). Compared to \bfc{}, however, HPCC has 5-30$\times$ worse tail latency for short flows with incast, and 2.3-3$\times$ worse without. Long flows do worse with HPCC than DCQCN and DCTCP \rev{since HPCC deliberately targets 95\% utilization and very small queues to improve tail latency for short flows}. 

\rev{With ideal retransmission, HPCC performance improves, especially for short and medium flows. However, HPCC without PFC has higher tail buffer occupancy and suffers packet loss. Compared to BFC, overall performance is still worse for both long and short flows.}
 
\rev{Across all systems, ExpressPass achieves the worst throughput for long flows. In ExpressPass, the receiver can generate unnecessary credits for an additional RTT before learning that a flow is finished. These credits are considered ``wasted'' as the sender cannot transmit packets in response, and can therefore cause link under-utilization. Credit waste and the corresponding under-utilization increase with faster link speeds and/or when the flow sizes get shorter (see \S6.3 and \S7 in \cite{expresspass}).}
 
Ideal-FQ achieves lower latency than all the schemes,  but its buffer occupancy can grow to an unfeasible level.

BFC achieves the best FCTs (both average and tail) among all the schemes.
Without incast, BFC performance
closely tracks optimal.  With incast, incoming flows exhaust the number of physical queues, triggering
HoL blocking and hurting tail latency.
This effect is largest for the smallest flows at the tail. \Fig{google:active} shows the CDF of the number of active flows at a port. In the absence of incast, the number of active flows is smaller than the total queues 99\% of the time, and collisions are rare. 
With incast, the number of active flows increases, causing collisions. However,
the tail latency for short flows with BFC is still 5-30$\times$ better than existing schemes. BFC also improves the performance of incast flows, achieving 2$\times$ better FCTs at the tail compared to HPCC (see \App{incast_fct}). 

\cut{
In the absence of incast, the buffer occupancy for HPCC is similar to that of BFC, but this low buffering comes at the cost of link under-utilization and high FCT for long flows. \ma{can be cut}
}

Note that, compared to \bfc{} and Ideal-FQ, latency for medium flows (200-1000KB) is slightly better with existing schemes. Because they slow down long flows relative to perfect fairness, medium flows have room to get through more quickly. Conversely, tail slowdown is better for long flows than medium flows with \bfc{} and Ideal-FQ. Long flows achieve close to the long term average available bandwidth, while medium flows are more affected
by transient congestion. 

  \begin{figure}[t]
    \centering
    \begin{subfigure}[tbh]{0.235\textwidth}
        \includegraphics[trim={0 0 0 4mm}, clip, width=\textwidth]{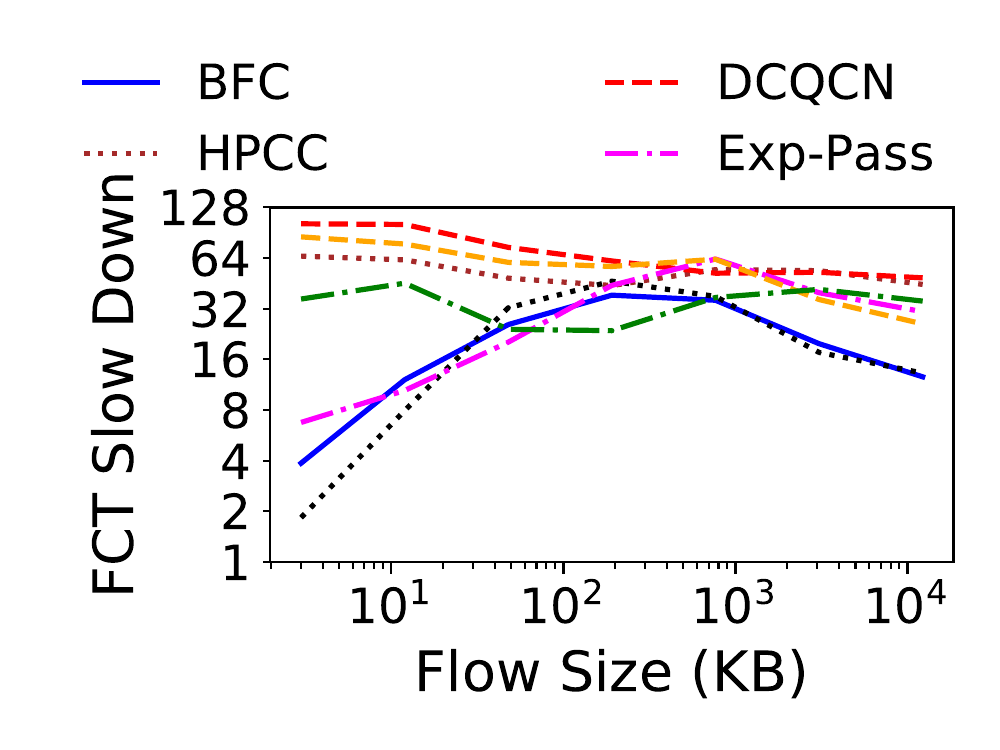}
        \vspace{-7mm}
        \caption{55\% + 5\% 100-1 incast}
        \label{fig:fb:incast}
    \end{subfigure}
    \begin{subfigure}[tbh]{0.235\textwidth}
        \includegraphics[trim={0 0 0 4mm}, clip,width=\textwidth]{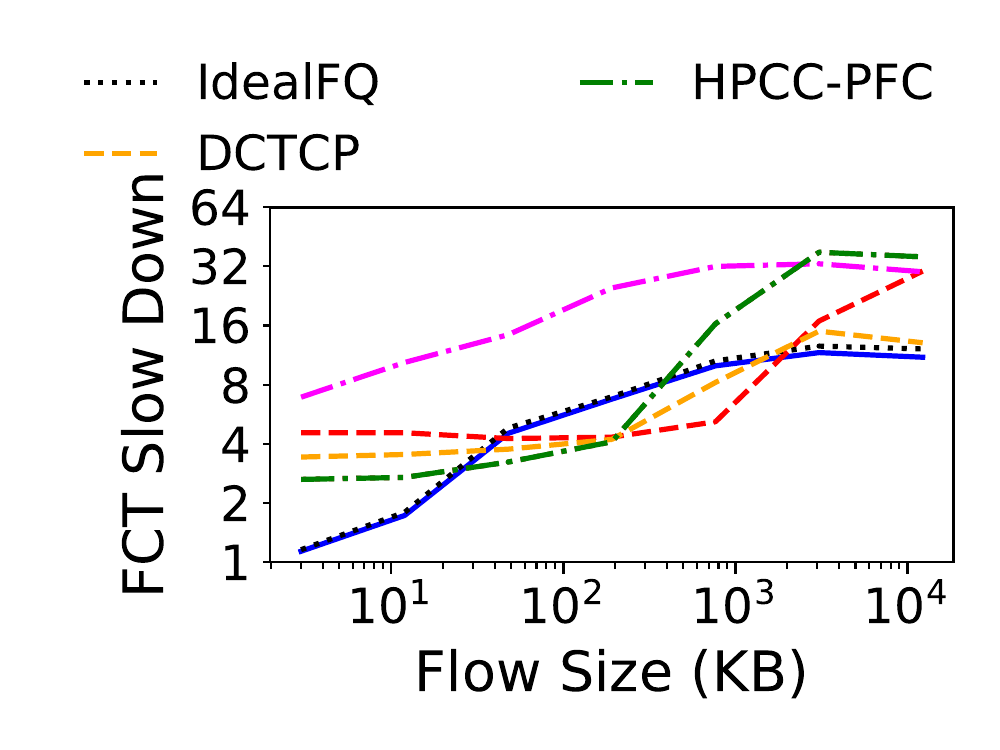}
        \vspace{-7mm}
        \caption{60\%}
        \label{fig:fb:no_incast}
    \end{subfigure}
    \vspace{-3mm}
    \caption{\small \rev{FCT slowdown (99$^{th}$ percentile) for Facebook distribution with and without incast. }}
    \label{fig:fb}
    \vspace{-4.5mm}
\end{figure}
 
\noindent\rev{{\bf Another workload:}} We repeated the experiment in \Fig{google_incast} and ~\Fig{google} with the Facebook distribution. \Fig{fb} shows the 99$^{th}$ percentile FCT slowdown. The trends in the FCT slowdowns are similar to that of the Google distribution, \rev{except that ExpressPass performs better since it incurs fewer wasted credits (as a percentage) for the Facebook workload, which has larger flows}. We omit other statistics presented earlier in the interest of space, but the trends are similar to \Fig{google_incast} and ~\ref{fig:google}. 
Henceforth, all the experiments use the Facebook workload.

\subsection{Stress-testing BFC}
\label{s:limits}
In this section we stress-test BFC under high load and large incast degree. Flow arrivals follow a bursty log-normal distribution ($\sigma=2$). We evaluate BFC under two different queue configurations: (1) 32 queues per port (BFC 32); (2) 128 queues per port (BFC 128). We show the average slowdown for long flows (> 3MB) and 99$^{th}$ percentile slowdown for short flows (< 3KB).

 \begin{figure}[t]
    \centering
    \begin{subfigure}[tbh]{0.22\textwidth}
        \includegraphics[trim={0 0 0 4mm}, clip,width=\textwidth]{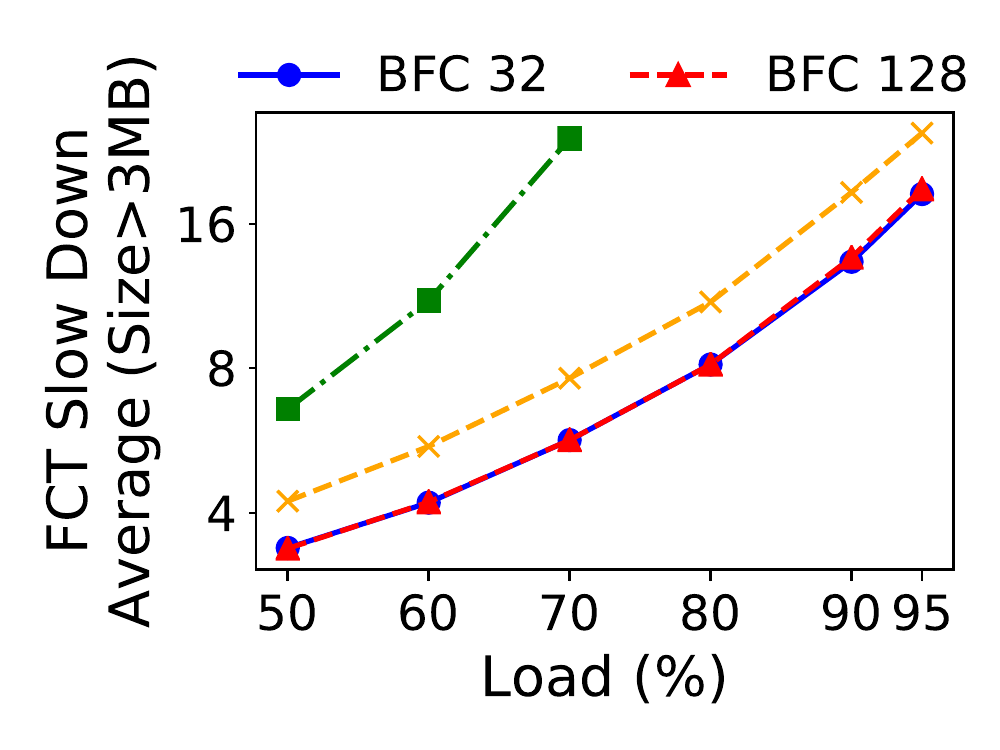}
         \vspace{-7mm}
        \caption{Average FCT for long flows}
        \label{fig:load_var:long}
    \end{subfigure}
    \begin{subfigure}[tbh]{0.22\textwidth}
        \includegraphics[trim={0 0 0 4mm}, clip,width=\textwidth]{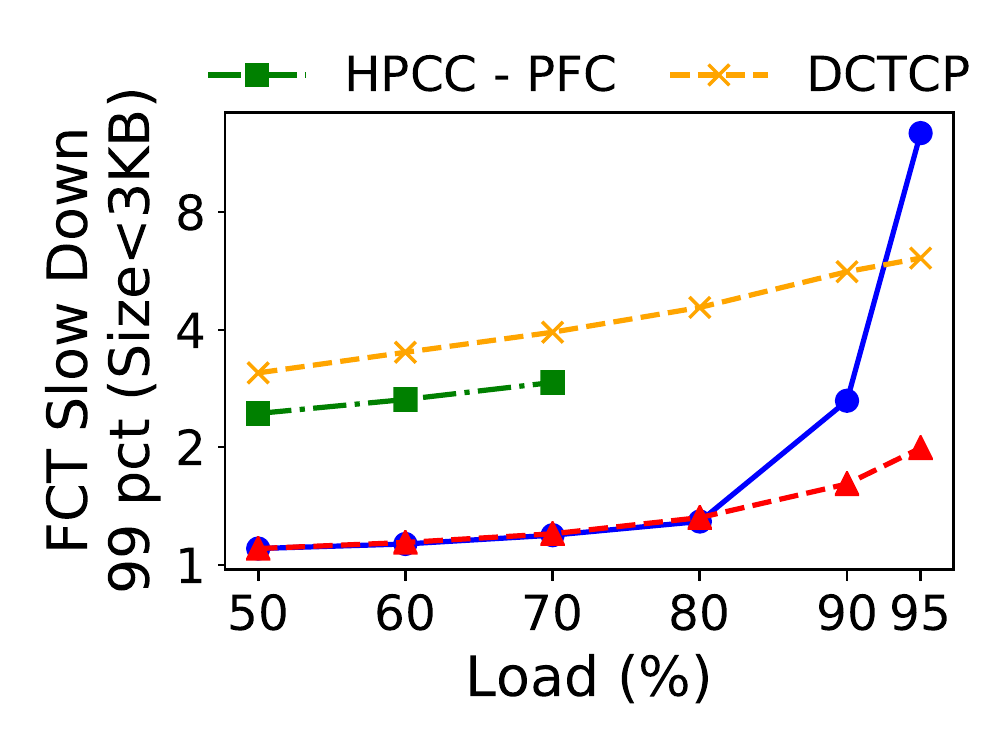}
        \vspace{-7mm}
        \caption{Tail FCT for short flows}
        \label{fig:load_var:short}
    \end{subfigure}
    \vspace{-3mm}
    \caption{\small \rev{Average FCT slowdown for long flows, and 99$^{th}$ percentile tail FCT slowdown for small flows, as a function of load.}}
    \label{fig:load_var}
    \vspace{-5.5mm}
\end{figure}

\noindent
\textbf{Load:}
\Fig{load_var} shows the performance as we vary the average load from 50 to 95\% (without incast). HPCC only supports loads up to 70\%. At higher loads, it becomes unstable (the number of outstanding flows  grows without bound), in part due to the overhead of the INT header (80\,B per-packet). 
All other schemes were stable across all load values.

At loads $\leq$ 80\%, BFC 32 achieves both lower tail latency (\Fig{load_var:short}) for short flows and higher throughput for long flows (\Fig{load_var:long}). The tail latency for short flows is close to the perfect value of 1. At higher loads, flows remain queued at the bottleneck switch for longer periods of time, raising the likelihood that we run out of physical queues, leading to head of line blocking. This particularly hurts tail performance for short flows as they might be delayed for an extended period if they are assigned to the same queue as a long flow.
At the very high load of 95\%, the HoL blocking degrades tail latency substantially for BFC 32. However, it still achieves good link utilization, and the impact of collisions is limited for long flows. 

Increasing the number of queues reduces collisions and the associated HoL blocking. BFC 128 achieves better tail latency for short flows at load $\geq 90\%$.

 \begin{figure}[t]
    \centering
    \begin{subfigure}[tbh]{0.22\textwidth}
        \includegraphics[trim={0 0 0 4mm}, clip,width=\textwidth]{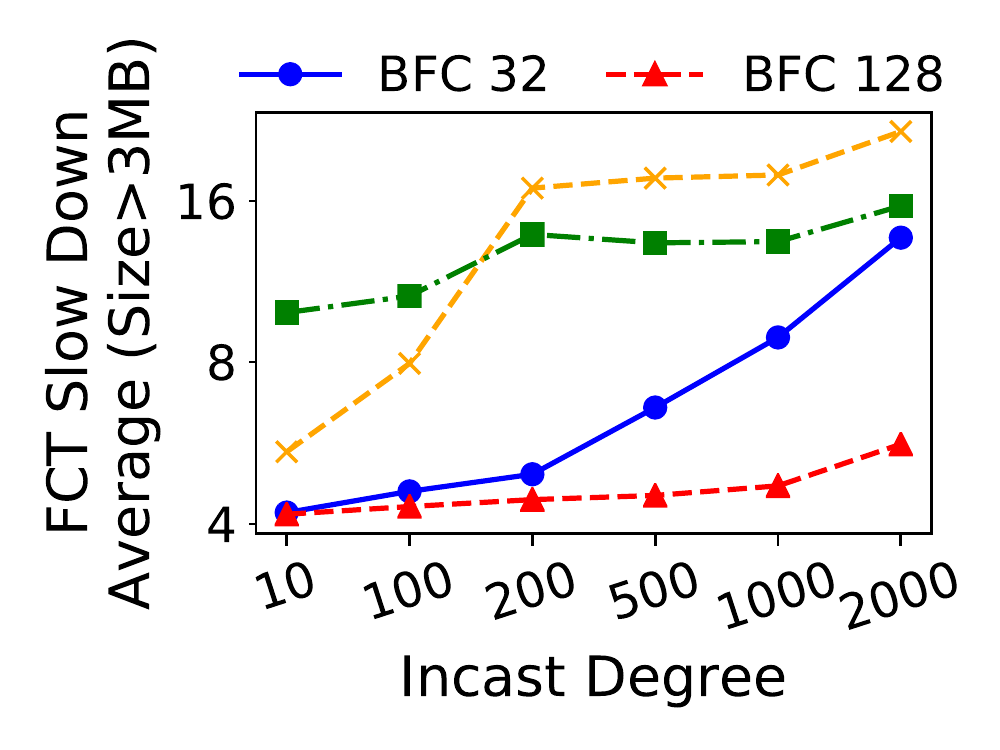}
         \vspace{-7mm}
        \caption{Average FCT for long flows}
        \label{fig:incast_var:long}
    \end{subfigure}
    \begin{subfigure}[tbh]{0.22\textwidth}
        \includegraphics[trim={0 0 0 4mm}, clip,width=\textwidth]{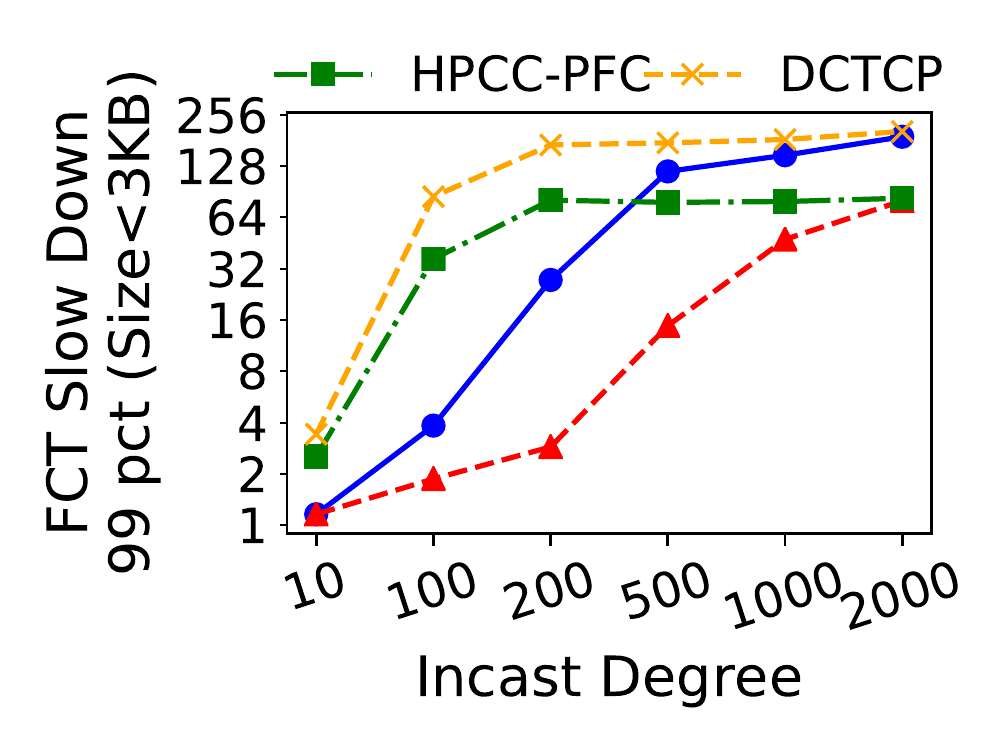}
        \vspace{-7mm}
        \caption{Tail FCT for short flows}
        \label{fig:incast_var:short}
    \end{subfigure}
    \vspace{-3mm}
    \caption{\small \rev{Average FCT slowdown for long flows, and 99$^{th}$ percentile tail FCT slowdown for small flows, as a function of incast degree.}}
    \label{fig:incast_var}
    \vspace{-5.5mm}
\end{figure}

\noindent
\textbf{Incast degree:}
If the size of an incast is large enough, it can
exhaust physical queues and hurt performance. \Fig{incast_var} shows the effect of varying the degree of incast on performance. The average load is 60\% and includes a 5\% incast. The incast size is 20\,MB in aggregate, but we vary the degree of incast from 10 to 2000. 

For throughput, both BFC 32 and BFC 128 perform well as long as the incast degree is moderate compared
to the number of queues. \rev{Both start to degrade when the incast degree exceeds 8$\times$ the number of queues per port.
Till this point, BFC can leverage the FanIn from the larger number of upstream queues (and greater
aggregate upstream buffer space) to keep the incast
from impeding unrelated traffic.}
As the incast degree scales up further, 
BFC 32 is able to retain some of its advantage relative to HPCC and DCTCP.

\rev{For high incast degree, the tail latency for short flows becomes worse than HPCC.} The tail is skewed by the few percent of small requests
that happen to go to the same destination as the incast. (Across the 128 leaf servers in our setup, several servers are the target of an incast at any one time, and these also receive their share 
of normal traffic.) 
As the incast degree increases, more small flows share physical queues with incast flows, leading
to more HoL blocking.

\rev{In \App{limits}, we further explore this issue with microbenchmarks
designed to trigger a variable number of active flows at the bottleneck 
switch. We show that by adding a very simple end-to-end control mechanism to BFC, 
we can ameliorate the impact of HoL blocking while still fully utilizing the link.}

 \begin{figure}[t]
    \centering
    \begin{subfigure}[tbh]{0.235\textwidth}
        \includegraphics[trim={0 0 0 4mm}, clip,width=\textwidth]{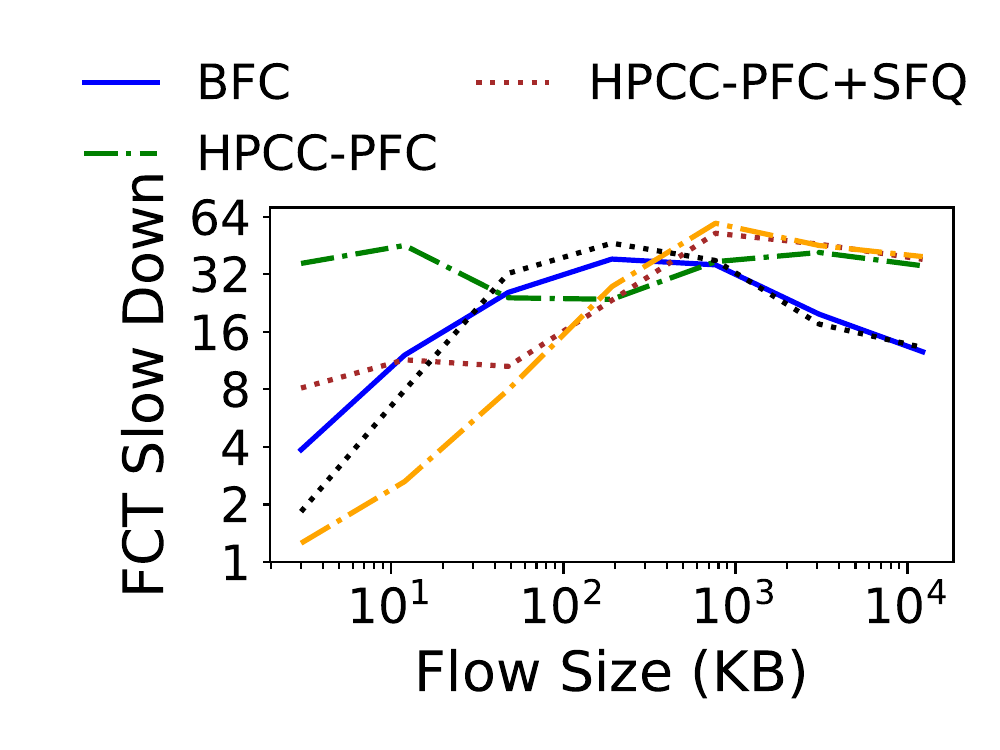}
         \vspace{-7mm}
        \caption{FCT SlowDown}
        \label{fig:hpcc_dqa:fct}
    \end{subfigure}
    \begin{subfigure}[tbh]{0.235\textwidth}
        \includegraphics[trim={0 0 0 4mm}, clip,width=\textwidth]{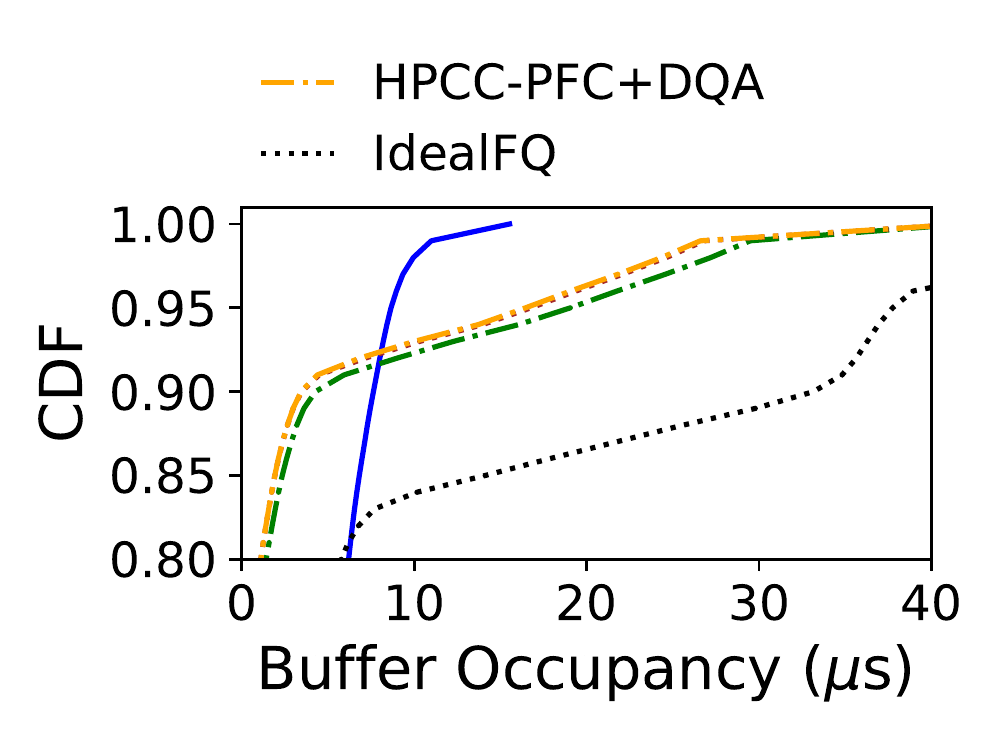}
        \vspace{-7mm}
        \caption{Buffer Occupancy}
        \label{fig:hpcc_dqa:buf}
    \end{subfigure}
    \vspace{-3mm}
    \caption{\small \rev{FCT slowdown (99$^{th}$ percentile) and buffer occupancy of HPCC variants, using the setup in \Fig{fb:incast}.}}
    \label{fig:hpcc_dqa}
    \vspace{-6.5mm}
\end{figure}

\subsection{Dynamic Queue Assignment}
\rev{We next consider the effect of applying BFC's dynamic queue assignment separately
from the backpressure mechanism. For this, we modified HPCC with
idealized retransmission (HPCC-PFC) to add stochastic fair queuing (HPCC-PFC+SFQ) and dynamic queue assignment (HPCC-PFC+DQA). To match BFC, we use 32 physical queues with HPCC.}
\rev{We repeat the experiment from \Fig{fb:incast},
showing tail slowdown and buffer occupancy for the HPCC variants, BFC, and IdealFQ
in \Fig{hpcc_dqa}.}

\rev{Adding SFQ to HPCC improves short flow latency by isolating them from long flows in different queues, but it still suffers from more collisions (and thus higher tail latency for short flows) than DQA. DQA on its own, however, has no benefit for long flows: since HPCC is unable to adapt to rapid changes in the number of flows (and the fair-share rate), it is unable to fully
utilize the link for long flows, even with DQA. Moreover, both HPCC-PFC+SFQ and HPCC-PFC+DQA build deep buffers and experience drops at the same rate as HPCC-PFC.  Notice  that HPCC's lower throughput for long flows favors short flows to such an extent  that HPCC-PFC+DQA achieves better tail latency for short flows than both BFC and IdealFQ.}   

\subsection{Additional Experiments}
\rev{In \App{additional}, we 
use our simulation framework to 
further characterize the limits of BFC, compare BFC to Homa, 
as well as study the impact of
priority scheduling, parameter selection, locality in the traffic matrix, 
slow start, incast labelling, 
 and other factors.}
\section{Conclusion}
\label{s:concl}
In this paper, we present Backpressure Flow Control (BFC), a practical congestion control architecture for data center networks. BFC provides per-hop per-flow flow control, but with bounded state, constant-time switch operations, and careful use of buffers. 
Switches dynamically assign flows to physical queues, allowing fair scheduling among competing flows and use selective backpressure to reduce buffering with minimal head of line blocking. 
Relative to existing end-to-end congestion control schemes, BFC improves
short flow tail latency and long flow utilization for networks with high bandwidth links and bursty traffic. 
We demonstrate BFC’s feasibility by implementing it on \rev{Tofino2}, a state-of-art P4-based programmable hardware switch. 
In simulation, compared to several deployed end-to-end schemes, BFC achieves 2.3\,-\,60$\times$ lower tail latency for short flows and 1.6\,-\,5$\times$ better average completion time for long flows.

\bibliographystyle{plain}
\bibliography{bfc}



\clearpage
\appendix
\section{Additional Experiments}
\label{app:additional}
\rev{In this appendix, we present a more complete set of simulation results
for BFC. We first summarize those results, and then present them.}

\noindent\rev{\textbf{Understanding the Limits of BFC:}
A limitation of BFC is that performance can degrade when collisions occur.
The worst case is when many long-running flows share a bottleneck
link with bursty traffic. We synthetically create this scenario and show that
by adding a very simple end-to-end control system to BFC, we can largely
ameliorate the impact of long flows, while still fully utilizing the link.
See \App{limits} for details.}

\noindent\rev{\textbf{Comparison with Homa:} Homa is a receiver driven data center transport that uses network priorities to achieve an approximation of the shortest
remaining flow first (SRF) scheduling to provide low latency for short flows while
still using the full bandwidth of the bottleneck for long flows. Homa also uses
packet spraying. In \App{homa_comp}, we configure BFC with a similar scheduling policy.
We show that Homa with packet spraying outperforms BFC, but when we turn off packet spraying, BFC outperforms Homa.}

\noindent\rev{\textbf{Priority Scheduling:} Data center operators often classify traffic into multiple classes and use scheduling priorities to ensure performance for the most
time-sensitive traffic. We repeat the experiment in \Fig{fb:no_incast} but with traffic split equally among four priority traffic classes, and show that BFC performs
well in this case. See \App{multi_tfk_classes} for details.}

\noindent\rev{\textbf{Parameter Sensitivity:}  We perform parameter sensitivity analysis for HPCC, DCTCP and ExpressPass. See \App{par_sense_comp} for details.}

\noindent\rev{\textbf{Spatial Locality:}
We repeat the experiment in \Fig{fb} with spacial locality in source-destination pairs such that the average load on all links across the network is same. The trends in performance are similar. See \App{locality} for details.}

\noindent\textbf{Slow-start:} We evaluate the impact of using TCP slow-start instead of starting flows at line rate. We repeat the experiment in \Fig{fb} and compare the original DCTCP with slow start (DCTCP + SS) and our modified DCTCP where flows start at the line rate. With incast, DCTCP + SS reduces buffer occupancy by reducing the intensity of incast flows, improving tail latency. However, it also increases median FCTs by up to 2$\times$. Flows start at a lower rate, taking longer to ramp up to the desired rate. In the absence of incast, it increases both the tail and median FCT for short flows. See \App{ss} for details.

\noindent\textbf{Reducing Contention for Queues:} We tried a variant of BFC where the sender labels incast flows explicitly (similar to the potential optimization in \cite{homa}). 
All the incast flows at an egress port are assigned to the same queue. This frees up queues for non-incast traffic and reduces collisions substantially under large incasts. (see \App{isolate_incast}).

\noindent\textbf{Incremental Deployment:} 
We repeated the experiment in \Fig{fb:incast} in the scenario where (i) BFC is deployed in part of the network; (ii) The switch doesn't have enough capacity to handle all the recirculations. The impact on FCTs is minimal under these scenarios (see \App{incr_deploy}).

\noindent\textbf{Performance in Asymmetric Topologies:} BFC makes no assumption about the topology, link speeds and link delays. We evaluate the performance of BFC in a multi-data-center topology. BFC achieves low FCT for flows within the data center, and high link utilization for the inter-data-center links (see \App{cross}). 

\noindent\textbf{Dynamic vs. Stochastic Queue Assignment in BFC:} We  repeat  the  experiment in \Fig{fb:incast} but use stochastic hashing to statically assign flows to physical queue instead. With stochastic assignment, the number of collisions in physical queues increases, hurting FCTs (see \App{qassign}).

\noindent\textbf{Size of Flow Table:} Reducing the size of the flow table can increase index collisions in the flow table, potentially hurting FCTs. We repeat the experiment in \Fig{fb:incast} and evaluate the impact of size of flow table. Reducing the size partly impacts the short flow FCTs (see \App{sensitivity}).

\noindent
\textbf{Incast Flow Performance:} \App{incast_fct} shows the slowdown for incast flows for the Google workload used in \Fig{google_incast}. BFC reduces the FCT for incast flows compared to other feasible schemes.

\begin{figure}[t]
     \centering
    \includegraphics[trim={0 0 0 4mm},clip,width=\columnwidth]{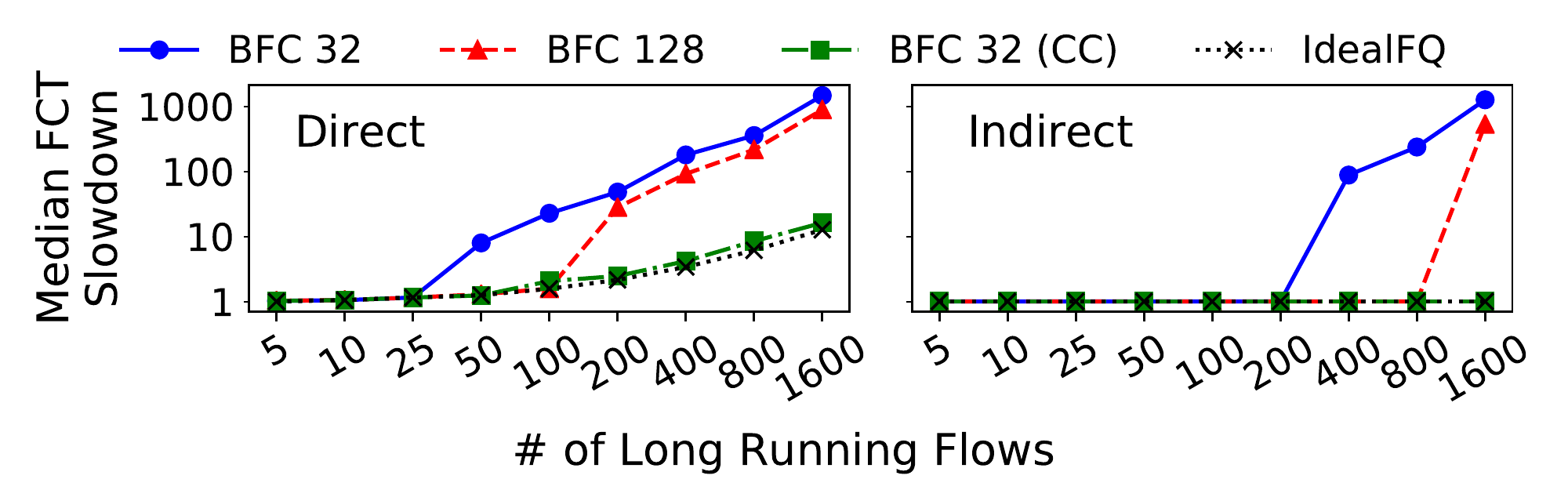}
    \vspace{-7mm}
    \caption{\small \rev{Median FCT slowdown for mice flows in the presence of long-running flows.}}
    \label{fig:long_running}
    \vspace{-4mm}
 \end{figure}
 
\subsection{\rev{Understanding the limits of BFC}}
\label{app:limits}
\rev{This section investigates the impact of large numbers of active flows on BFC's performance through controlled microbenchmarks. We also show that adding a simple end-to-end flow control mechanism on top of pure BFC helps alleviate the performance impairments caused by large numbers of flows.}

\rev{Collisions hurt performance in two ways. Consider a congested port $X$. First, at $X$, the packets of a short flow can get stuck behind the packets of a long flow sharing the same queue, increasing the FCT. Such performance degradation occurs when the number of active flows exceeds the number of queues at $X$. Second, $X$ can pause an upstream queue. Unrelated flows sharing this upstream queue will get paused even though they are not going through the congested port $X$ (congestion spreading). 
BFC can leverage the larger number of upstream queues at the upstream switches to limit congestion spreading (\S\ref{s:qassignment}).
Typically, congestion spreads only once the number of flows at the congested port exceeds the total number of upstream queues. As a result, in larger topologies with more upstream switches, congestion spreading is harder to create.}

\rev{
To illustrate these issues, we conduct experiments on our standard topology (\S\ref{ss:setup}) where we create different numbers of long-running elephant flows destined to the same receiver (Receiver $A$). All elephant flows start at the beginning of the experiment. We then create two groups of short flows: (1) destined to the same receiver $A$ (referred as ``direct'' mice flows), and (2) destined to a different receiver $B$ in the same rack as receiver $A$ (referred to as ``indirect'' mice flows). The aggregate load for each group of mice flows is 3\% of the link capacity, and the size of the mice flows is 1~KB. \Fig{long_running} shows the median FCT slowdown for mice flows as we vary the number of long-running flows. We show results for BFC with 32 and 128 queues, and also IdealFQ (described in \S\ref{ss:setup}) for reference. As expected, for direct mice flows, the FCT degrades when the number of long-running flows exceeds the number of queues. For indirect flows, the degradation only happens when long flows exceed 8$\times$ the number of queues, since the topology has 8 spine switches connected to each ToR switch. In this case, some indirect mice flows get paused unnecessarily because they share an upstream queue with a paused long-running flow.}

\noindent \rev{\textbf{Combining end-to-end congestion control with BFC:} 
In the previous experiment, each long-running flow can build up to 1 Hop-BDP of buffering before getting paused. With $N$ long-running flows, in the worst case, a mice flow experiencing a collision can get stuck behind $N \times$ 1-Hop BDP of buffering. BFC can use a simple end-to-end congestion control mechanism to reduce this buffering and limit HoL blocking. This mechanism is helpful in scenarios with persistently large numbers of active flows. As our evaluations showed (\S\ref{s:limits}), even in workloads with high load and occasional large-scale incast, pure BFC (with no end-to-end control) performs well except in extreme cases.}   

\rev{Augmenting BFC with end-to-end control is simple. The main goal of the end-to-end control is to prevent flows from sending an excessively large number of packets into the network. Importantly, the end-to-end mechanism need not try to accurately control queuing, react quickly to bursts, or achieve fairness\,---\,typical requirements for low-latency data center congestion control protocols\,---\,since BFC already achieves these goals.}

\rev{As an example, we implemented a simple delay-based congestion control that tries to maintain the end-to-end RTT at a certain threshold (\texttt{RTT}$_{Target}$). We chose a high \texttt{RTT}$_{Target}$ value of $2.5 \times$ base RTT to avoid hurting the throughput of long flows, exploiting the fact that it isn't necessary to tightly control queuing in BFC. The algorithm adjusts the sender's window ($w$) as follows.}

\begin{algorithm}
\small
\texttt{RTT}$_{Target}$ $=$ $2.5 \times$ Base RTT\;
$w = $1 BDP\;
\For{each Acknowledgement}{
    \eIf{RTT $>$ \texttt{RTT}$_{Target}$}{
        $w = w - \frac{\texttt{RTT - RTT}_{Target}}{RTT}$
    }{
        $w = w + \frac{\texttt{RTT}_{Target} - RTT}{RTT}$
    }
}
\vspace{4mm}
\caption{\rev{Simple end-to-end congestion control}}
\label{alg:bfc_cc}
\end{algorithm}

\begin{figure}[t]
    \centering
    \begin{subfigure}[tbh]{0.235\textwidth}
        \includegraphics[trim={0 0 0 4mm}, clip,width=\textwidth]{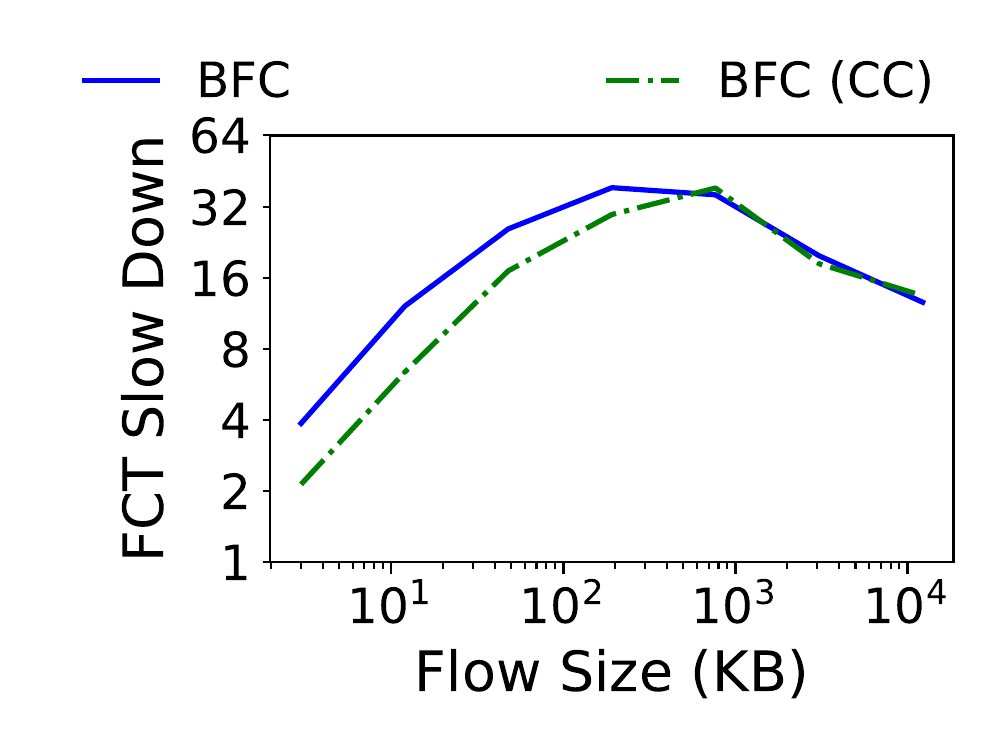}
         \vspace{-7mm}
        \caption{55\% + 5\% 100-1 incast}
        \label{fig:bfc_cc:incast}
    \end{subfigure}
    \begin{subfigure}[tbh]{0.235\textwidth}
        \includegraphics[trim={0 0 0 4mm}, clip,width=\textwidth]{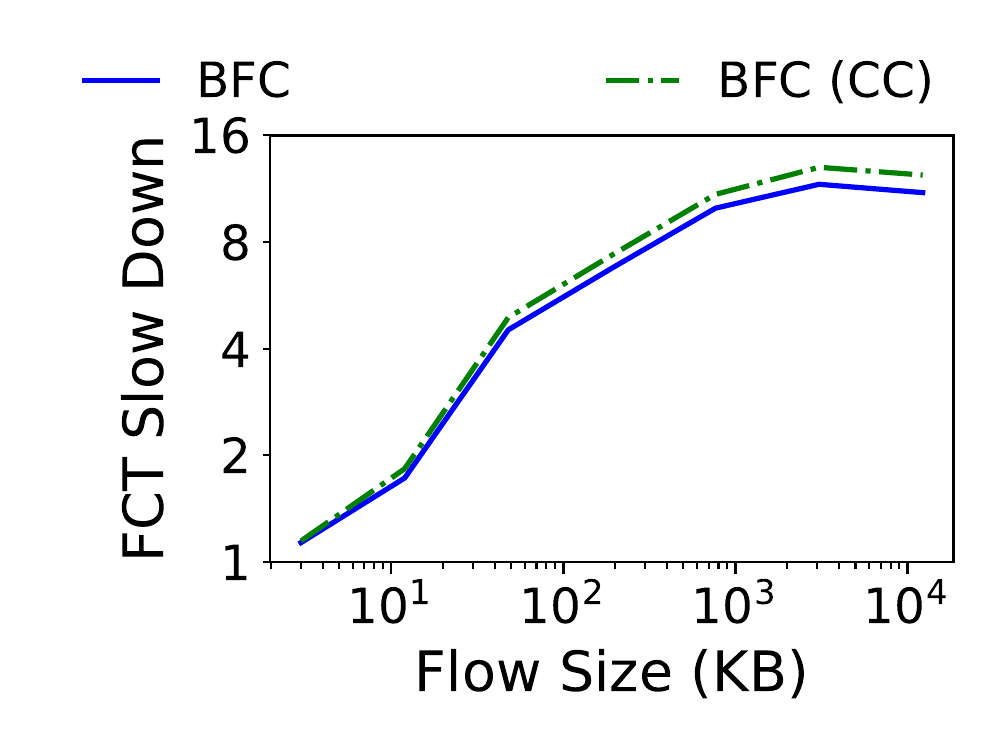}
        \vspace{-7mm}
        \caption{60\%}
        \label{fig:bfc_cc:noincast}
    \end{subfigure}
    \vspace{-3mm}
    \caption{\small \rev{99$^{th}$ percentile FCT slowdown when combined with congestion control. Facebook workload, same setup as \Fig{fb}.}}
    \label{fig:bfc_cc}
    \vspace{-3.5mm}
\end{figure}

\rev{With the above rule, the window of a sender roughly goes from $w \rightarrow w \times \frac{\texttt{RTT}_{Target}}{RTT}$ within an RTT. \Fig{long_running} shows the performance with this variant (BFC 32 (CC)). The performance is close to IdealFQ in all the cases. To check if this change negatively affected the overall behavior
of BFC, we repeat the principle experiment in \Fig{fb} (Facebook workload) 
with BFC 32 (CC). 
\rev{\Fig{bfc_cc} shows the 99$^{th}$ percentile FCT slowdowns.}
The FCTs of long flows are similar to that of the original BFC (within 10\%). 
However, in the presence of incast, adding congestion control improves the 99$^{th}$ percentile FCT of short flows and the peak buffer occupancy by 30\%.
While using end-to-end congestion control can improve performance under frequent collisions (and we advocate supplementing BFC with such a mechanism in practice), in this paper we focus on BFC without any such mechanism to better understand the core benefits and limitations of BFC in its purest form.}

\rev{In \App{isolate_incast}, we experiment with a variant of BFC where the sender labels incast flows explicitly (similar to the potential optimization in ~\cite{homa}). All the incast flows at an egress port are assigned to the same queue. This frees up queues for non-incast traffic and reduces collisions substantially under large incasts.}

\subsection{\rev{Comparison with Homa}}
\label{app:homa_comp}
\rev{Homa is a receiver driven data center transport that uses network priorities to achieve an approximation of shortest-remaining-flow-first (SRF) scheduling. Homa divides a flow's data into unscheduled (first BDP of traffic) and scheduled categories. The sender assigns a fixed priority level to a flow's unscheduled bytes based on its size and the flow size distribution of the workload. The unscheduled bytes are transmitted at line rate. The receiver assigns priority levels to the scheduled bytes and issues grants (credits) for them. Homa assumes per-packet spraying to ensure load balancing across core links, and sufficient core capacity to guarantee minimal congestion in the core.}

\rev{While we focus on fair queuing in this paper, BFC's design is applicable to other scheduling policies. In this section, we evaluate a variant of BFC, BFC-SRF, that aims to approximate SRF. Flows insert their remaining size into a header field in each packet transmitted, and the switch schedules queues in order of remaining size of the packet at the head of the queue. Similarly to Homa, NICs also follow SRF scheduling. We ran Homa using its OMNet++simulator~\cite{homa_code}. The Homa simulator assumes unbounded buffers at the switch. For BFC, we use a 12\,MB shared buffer. We use 32 queues for both Homa and BFC. For Homa, the 32 priority levels are divided between unscheduled and scheduled priorities based on the ratio of unscheduled and scheduled traffic; the overcommitment level is equal to the number of scheduled priorities~\cite{homa}. We use our default topology with 128 servers and 2:1 oversubscription at the ToR uplinks (\S\ref{ss:setup}).} 

\rev{Two differences between Homa and BFC-SRF are worth highlighting. First, BFC-SRF uses flow-level ECMP rather than packet spraying for enforcing per-flow backpressure. Second, BFC-SRF uses dynamic queue assignment and performs SRF scheduling directly on the switch, as opposed to Homa's priority assignment from the end-points. To understand the impact of these aspects separately, we also evaluate a variant of Homa with ECMP, and report results for IdealSRF+ECMP, an idealized SRF scheme with unlimited queues and unbounded buffers at each switch with ECMP load balancing.}


 \begin{figure}[t]
    \centering
    \begin{subfigure}[tbh]{\columnwidth}
        \includegraphics[width=\textwidth]{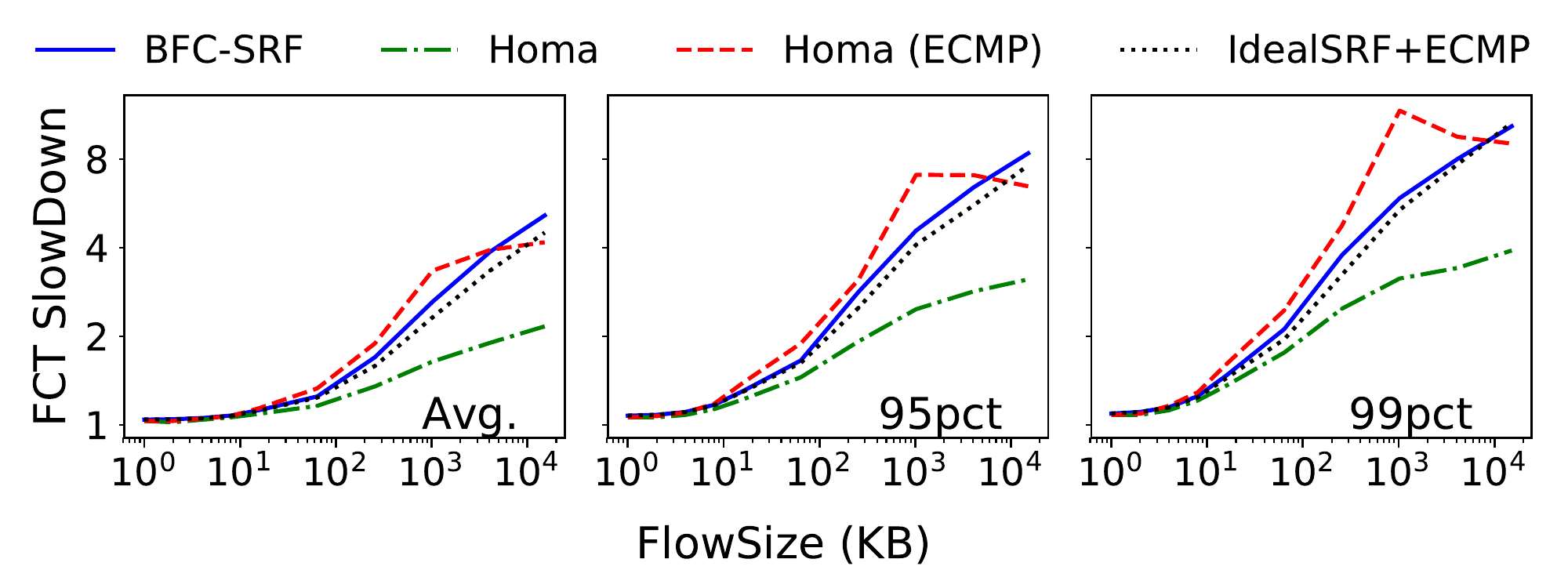}
         \vspace{-7mm}
        \caption{Google, 60\%}
        \label{fig:homa_comp_os:google}
    \end{subfigure}
    \begin{subfigure}[tbh]{\columnwidth}
        \includegraphics[width=\textwidth]{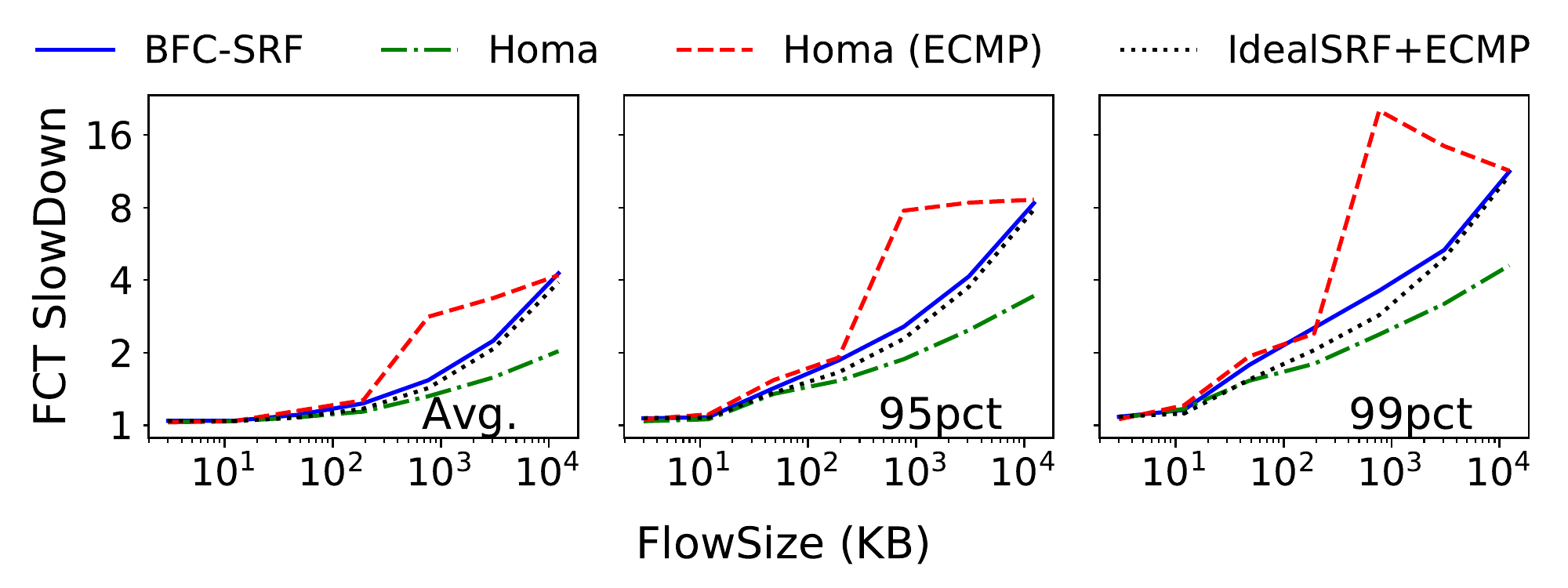}
         \vspace{-7mm}
        \caption{Facebook Hadoop, 60\%}
        \label{fig:homa_comp_os:fb}
    \end{subfigure}
    \vspace{-3mm}
    \caption{\small \rev{FCT slowdown on an oversubscribed clos topology. With packet spraying, Homa encounters minimal congestion in the core and outperforms other schemes.}}
    \label{fig:homa_comp_os}
    \vspace{-4mm}
\end{figure}

\rev{We repeat the experiments in \Fig{google} and \Fig{fb:no_incast} for the Google and Facebook workloads at 60\% load (log-normal flow arrivals without incast). \Fig{homa_comp_os} reports the FCTs. Homa performs the best out of all schemes, achieving up to 2$\times$ better FCTs for long flows. With packet spraying, flows encounter minimal congestion in the core, and compete for bandwidth primarily at the last-hop. In contrast, ECMP is prone to path collisions~\cite{conga} and flows encounter congestion in the core. Notice that a last-hop link carries half the load of a core link (30\% vs 60\%) in this experiment on average (\S\ref{ss:setup}). Since packet spraying essentially eliminates congestion on the core links, with Homa flows experience congestion only on the last-hop links. But with the ECMP-based schemes, flows contend at the core links (with $2\times$ the load). As a result, Homa even outperforms IdealSRF+ECMP. This result illustrates the benefits of packet spraying; nevertheless, packet spraying is rarely deployed in practice because it can cause packet reordering, increasing CPU overhead at endpoints\footnote{\rev{Packet reordering makes hardware offloads such as Large Receiver Offload (LRO) ineffective~\cite{juggler}}.}, and it can hurt performance in asymmetric topologies (e.g., caused by rolling upgrades or link failures)~\cite{letflow}.}    


\begin{table}[t]
\small
\begin{center}
\begin{tabular}{c|c|c|c}
Scheme & Link & 95\% Delay ($\boldsymbol{\mu}$s) & 99\% Delay ($\boldsymbol{\mu}$s)\\ 
 \hline
 \hline
 Homa & Agg-ToR &2.4 & 6.7 \\
  \hline
  Homa & ToR-Agg &2.1 & 6.0 \\
\hline
 Homa ECMP & Agg-ToR &{40.8} & {87.2} \\

  \hline
 Homa ECMP & ToR-Agg &{43.7} & {93.3} \\

\end{tabular}
\end{center}
\vspace{-5mm}
\caption{\small\rev{Per-packet queuing delay for scheduled traffic in the core.}}
\label{tab:homa_sched+qdelay}
\vspace{-4.5mm}
\end{table}

\rev{Among the ECMP approaches, BFC-SRF is close to IdealSRF+ECMP and Homa is worse.
In Homa, receivers have no visibility into congestion in the core and don't react to queue buildup in the core (though each flow limits its total in-flight data to 1 BDP). Also, Homa's receiver-set priorities are only based on contending flows at the last hop, and can violate SRF scheduling when congestion occurs in the core. Table~\ref{tab:homa_sched+qdelay} shows that with ECMP, the scheduled traffic encounters significantly higher queuing in the core.}

\begin{figure}[t]
    \centering
    \begin{subfigure}[tbh]{\columnwidth}
        \includegraphics[width=\textwidth]{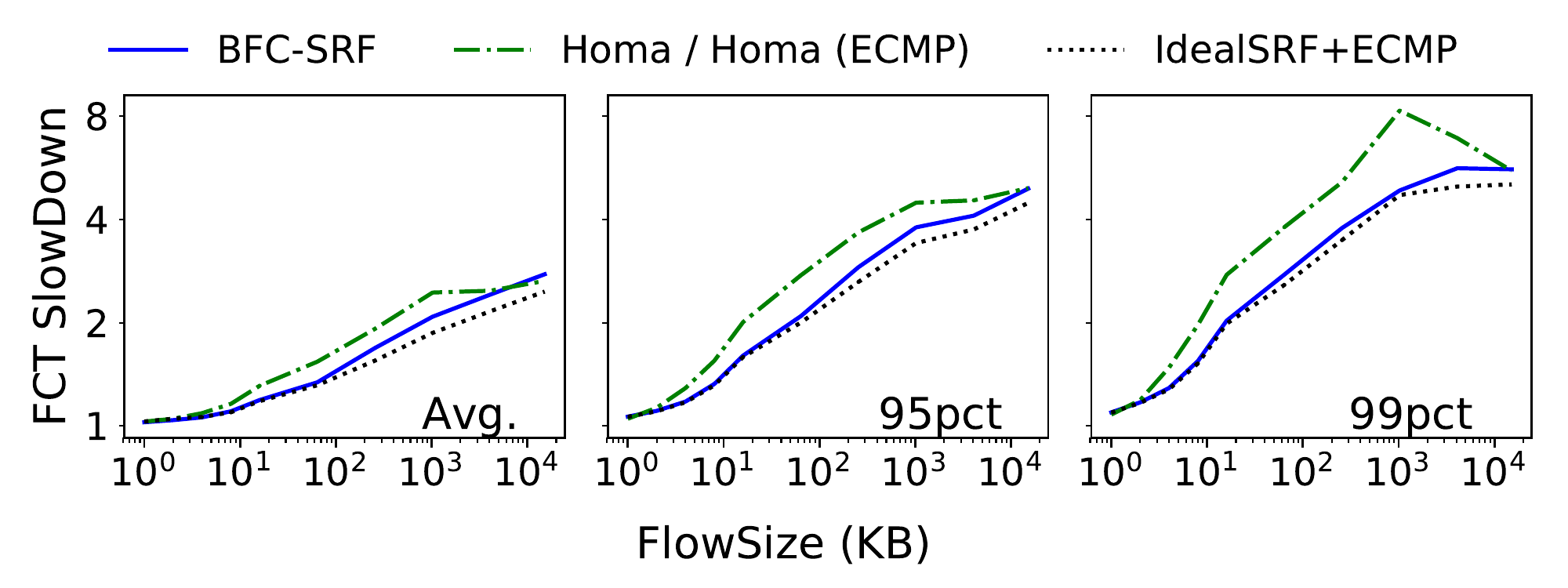}
         \vspace{-7mm}
        \caption{Google, 60\%}
        \label{fig:homa_comp_sl:google}
    \end{subfigure}
    \begin{subfigure}[tbh]{\columnwidth}
        \includegraphics[width=\textwidth]{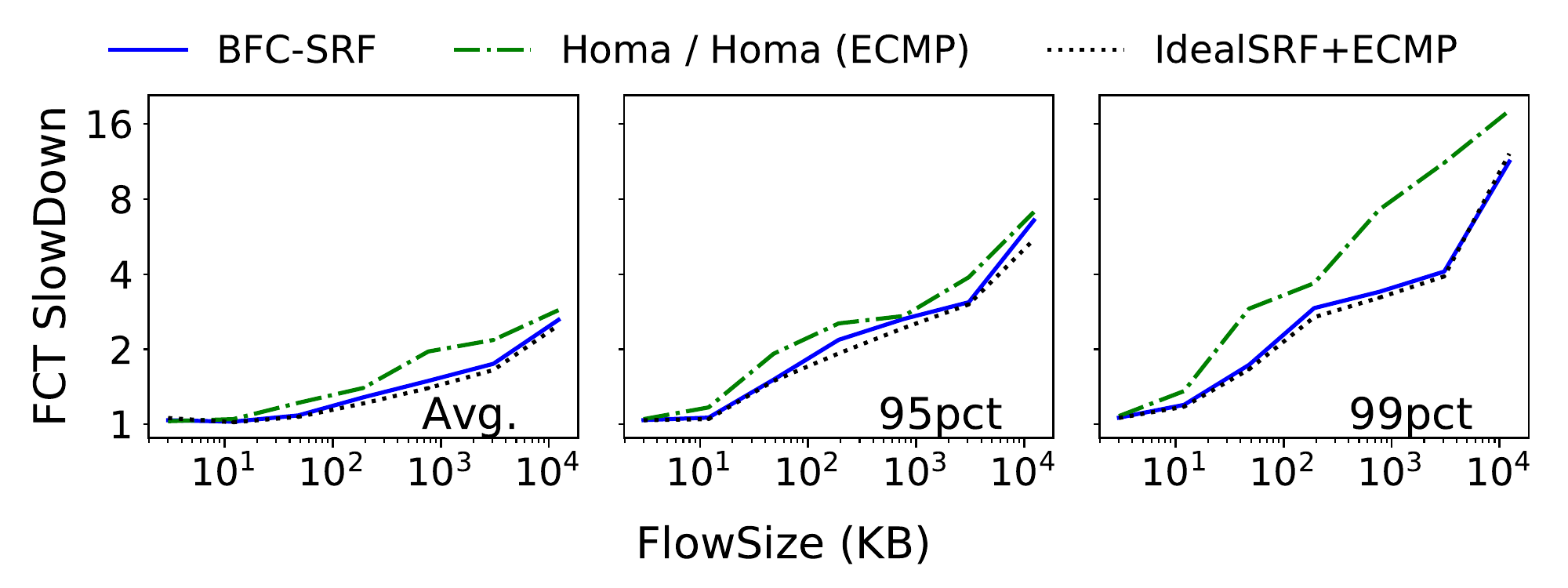}
         \vspace{-7mm}
        \caption{Facebook Hadoop, 60\%}
        \label{fig:homa_comp_sl:fb}
    \end{subfigure}
    \vspace{-3mm}
    \caption{\small \rev{BFC's dynamic queueue assignment achieves a better approximation of the SRF scheduling policy. BFC-SRF achieves close to optimal FCTs.}}
    \label{fig:homa_comp_sl}
    \vspace{-5mm}
\end{figure}

\noindent
\rev{\textbf{Benefits of BFC's dynamic queue assignment over Homa.} BFC makes queue assignment and scheduling decisions at the switch, based on an instantaneous view of competing flows. In principle, this should allow BFC to more accurately approximate SRF compared to Homa. To understand if this is actually the case, we conduct an experiment with the same Google and Facebook workloads but with all flows destined to a single receiver, and the senders located within the same rack as the receiver. Since there is no traffic in the core, load balancing (ECMP vs. packet spraying) does not matter in this case. Flow arrivals are log-normal and the load on the receiver's link is 60\%. \Fig{homa_comp_sl} shows the results. BFC-SRF achieves better FCTs primarily at the tail.} 

\rev{We give two examples of priority inversions in Homa which BFC avoids. First, the Homa sender assigns priorities to unscheduled traffic based on flow size distributions rather than using the current set of flows competing at the switch due to lack of visibility for the first RTT. As a result with Homa, short flows (< 1 BDP) with similar flow sizes can end up sharing unscheduled priority queues unnecessarily, even when there are sufficient queues at the switch to assign each flow a unique queue. 
Second, in Homa the unscheduled bytes of a flow are always scheduled ahead of the scheduled bytes of competing flows. This implies that the unscheduled bytes of a new long flow will be \emph{incorrectly} scheduled ahead of the scheduled bytes of a shorter flow. This also violates SRF and increases FCT for flows larger than a BDP.}

\begin{figure}[t]
    \centering
    \begin{subfigure}[tbh]{\columnwidth}
        \includegraphics[width=\textwidth]{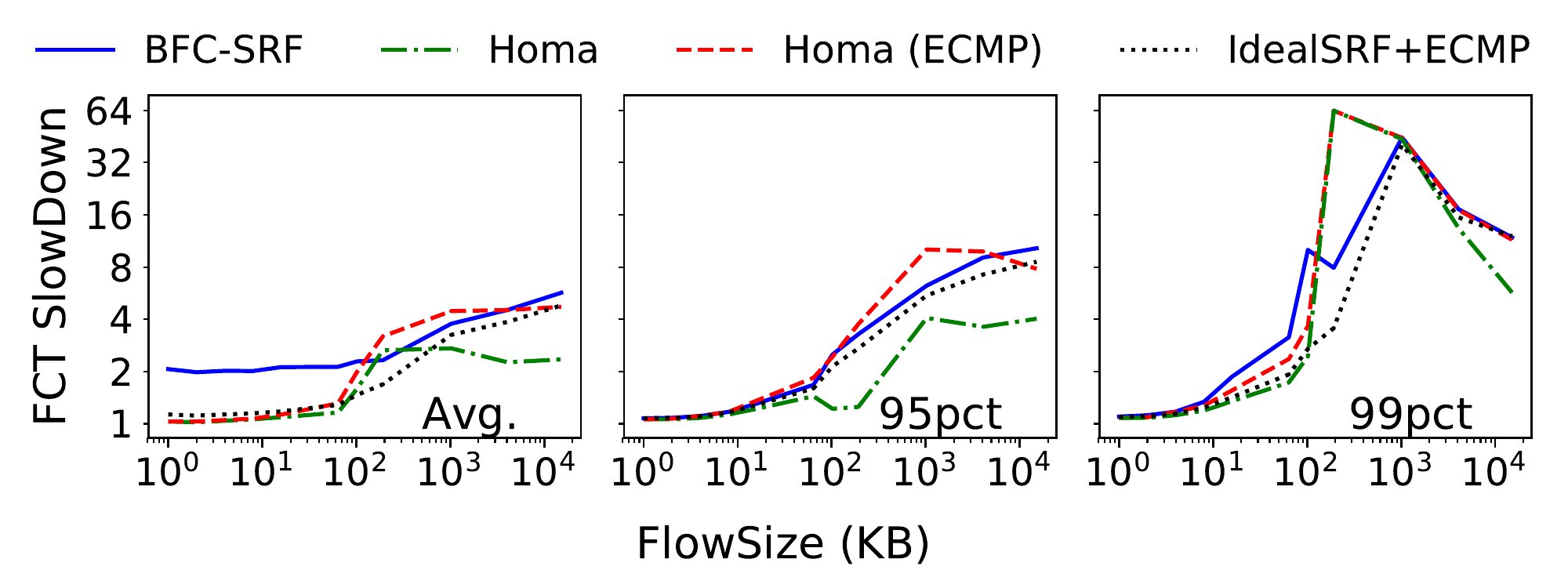}
         \vspace{-7mm}
        \caption{Google, 55\% + 5\% 100-1 incast}
        \label{fig:homa_comp_incast_os:google}
    \end{subfigure}
    \begin{subfigure}[tbh]{\columnwidth}
        \includegraphics[width=\textwidth]{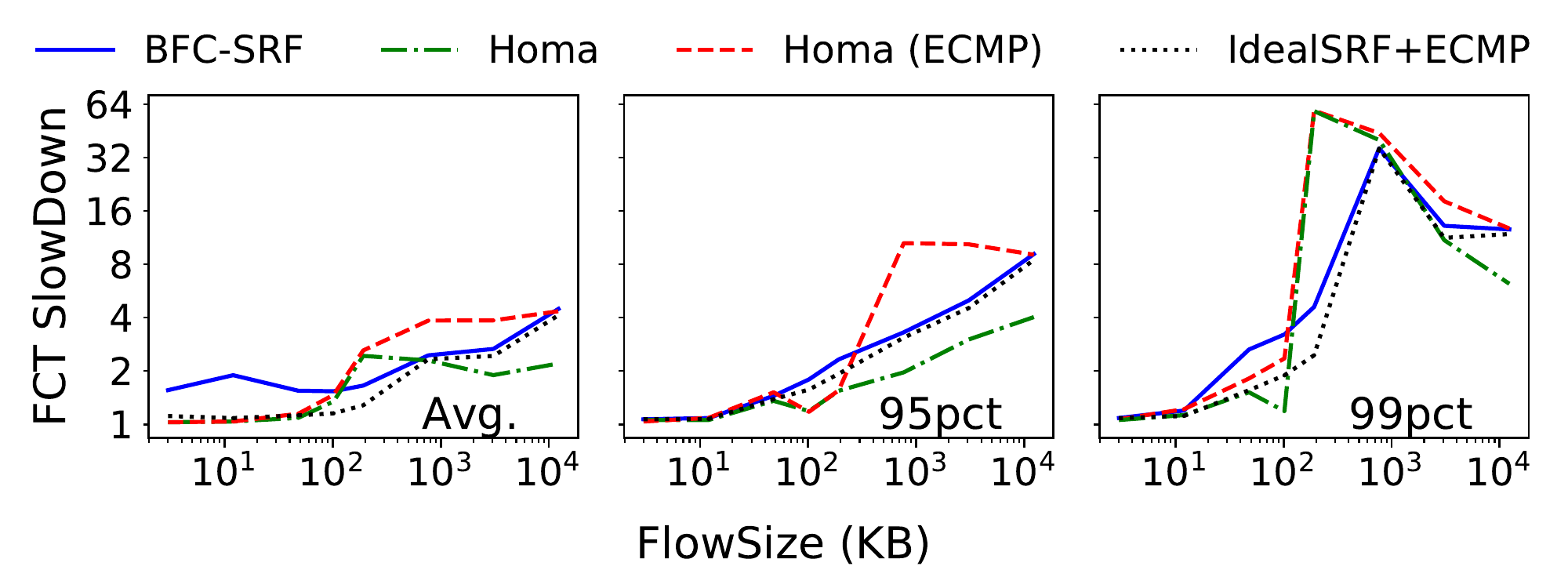}
         \vspace{-7mm}
        \caption{Facebook Hadoop, 55\% + 5\% 100-1 incast}
        \label{fig:homa_comp_incast_os:fb}
    \end{subfigure}
    \vspace{-3mm}
    \caption{\small \rev{FCT slowdown with 100-1 incast. Collisions in BFC-SRF can cause priority inversions hurting FCTs}}
    \label{fig:homa_comp_incast_os}
    \vspace{-5mm}
\end{figure}

\noindent\rev{{\bf Impact of collisions on BFC-SRF.} Recall that with large incast, BFC can experience collisions. For BFC-SRF, such collisions can cause priority inversions that hurt FCTs. To illustrate this, we repeat the experiments in \Fig{google_incast} and \Fig{fb:incast} (55\% load plus 5\% 100-1 incast traffic). \Fig{homa_comp_incast_os} shows that the average FCT for short flows is higher with BFC-SRF. This is because of high completion times for a (small) fraction of short flows sharing queues with longer flows. To understand why, consider the following situation. An incoming short flow arrives when there are no free queues, and ends up sharing the queue with a long flow. Let's say the remaining size of the long flow is greater than the incast flow size (200~KB in this experiment). In case there are competing incast flows present in other queues, the incast flows will be scheduled ahead of this long flow. Therefore, the short flow will have to wait for {\em all} the traffic from the incast flows to finish to make any progress. This can severely degrade its completion time. The core of this problem is that when a port runs out of queues, the BFC switch assigns the new flow to a queue randomly. This is fine for fair queuing but with SRF, a more sophisticated strategy may improve performance (e.g., assign the new flow to a queue with similar remaining flow sizes).} 

\rev{As explained earlier, Homa is not immune to priority inversions. \Fig{homa_comp_incast_os} shows that with Homa, flows with size greater than 1 BDP but less than 2 BDP have high FCTs at the tail. This is because unscheduled bytes of the the incast flows are incorrectly scheduled ahead of the scheduled bytes of such flows. 
}

\rev{These experiments suggest an interesting possibility to try to get the best of both schemes: we could combine BFC's dynamic queue assignment for unscheduled traffic with Homa's grant mechanism for controlling scheduled traffic. We leave exploration of such a design to future work.} 


 \begin{figure*}[t]
    \centering
    \begin{subfigure}[tbh]{0.235\textwidth}
        \includegraphics[width=\textwidth]{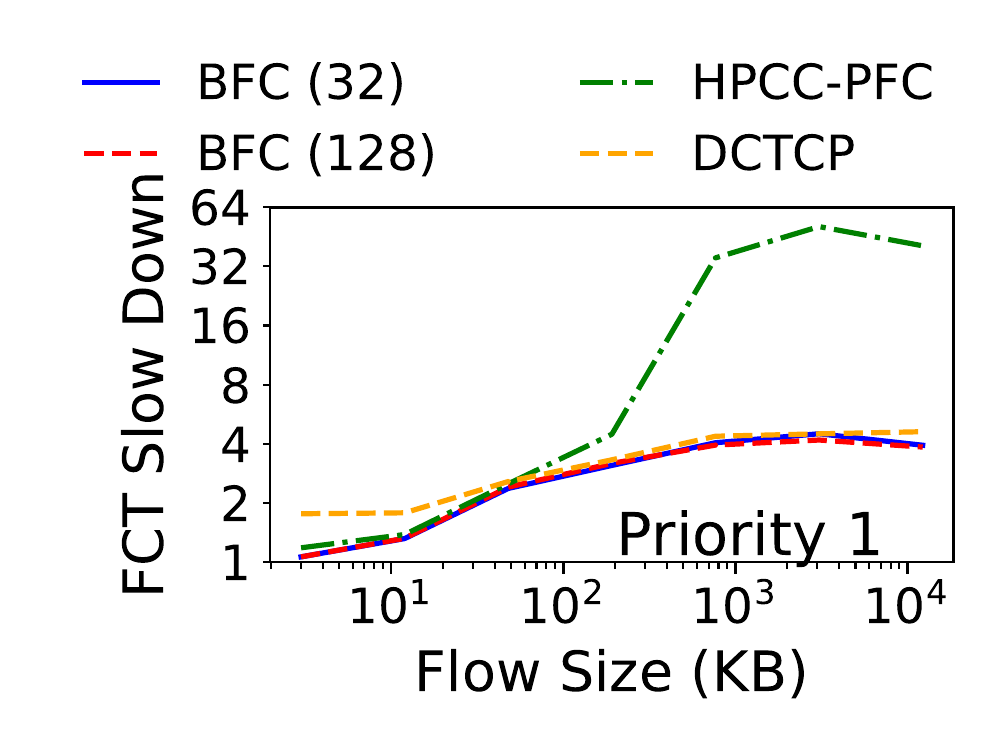}
        \vspace{-7mm}
        \caption{Priority Class 1 (highest)}
        \label{fig:pgroup:1}
    \end{subfigure}
    \begin{subfigure}[tbh]{0.235\textwidth}
        \includegraphics[width=\textwidth]{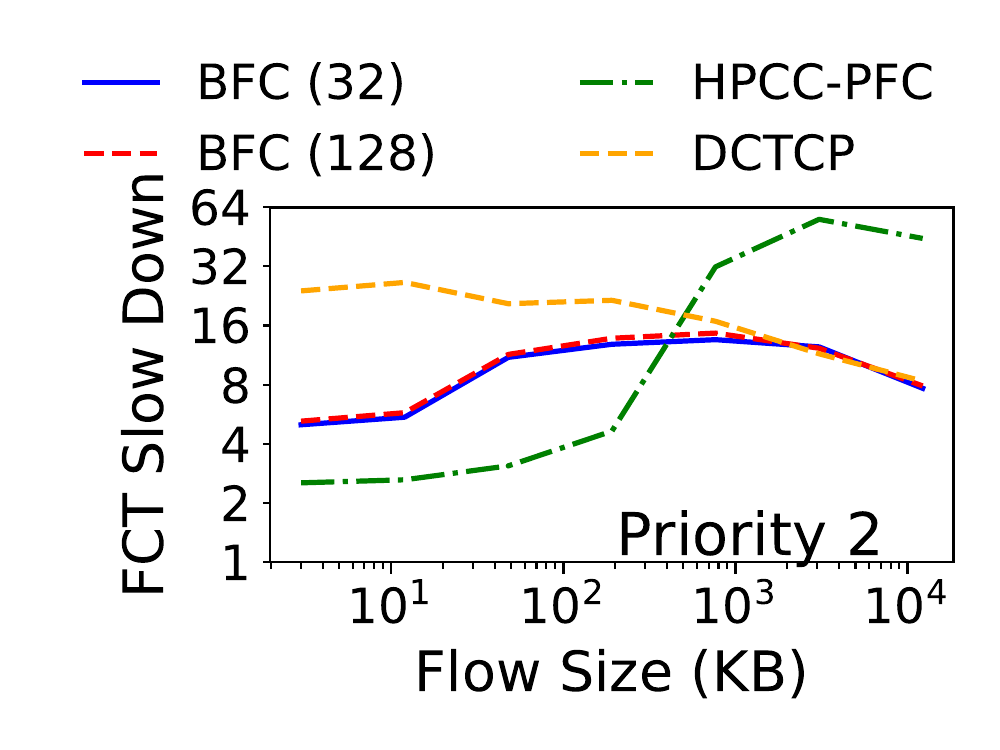}
        \vspace{-7mm}
        \caption{Priority Class 2}
        \label{fig:pgroup:2}
    \end{subfigure}
    \begin{subfigure}[tbh]{0.235\textwidth}
        \includegraphics[width=\textwidth]{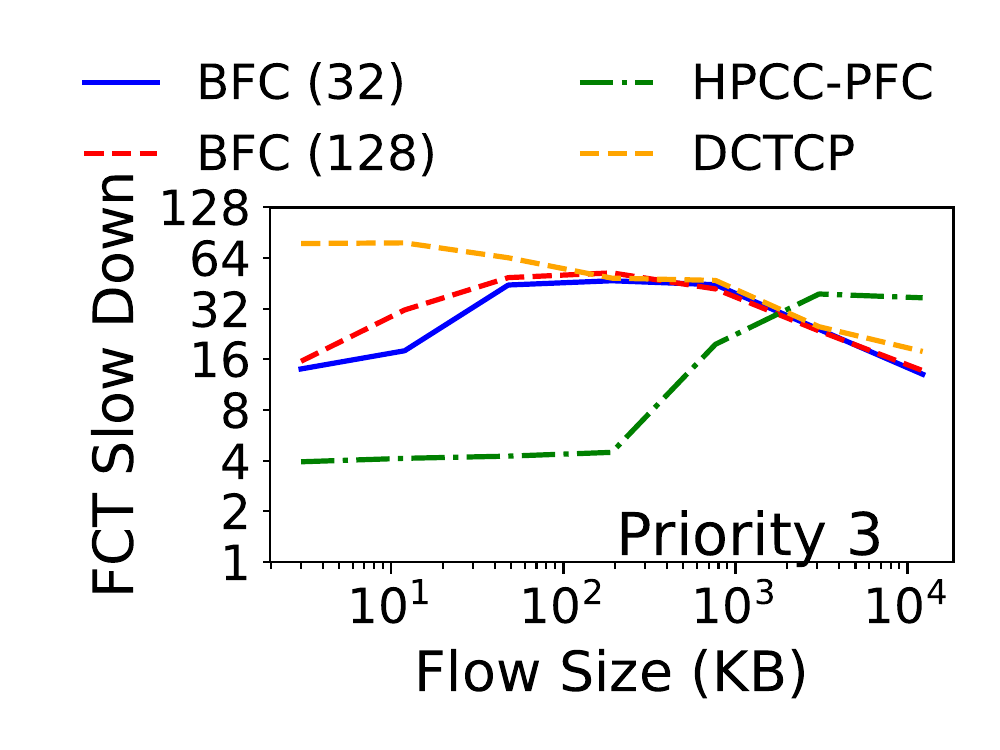}
        \vspace{-7mm}
        \caption{Priority Class 3}
        \label{fig:pgroup:3}
    \end{subfigure}
    \begin{subfigure}[tbh]{0.235\textwidth}
        \includegraphics[width=\textwidth]{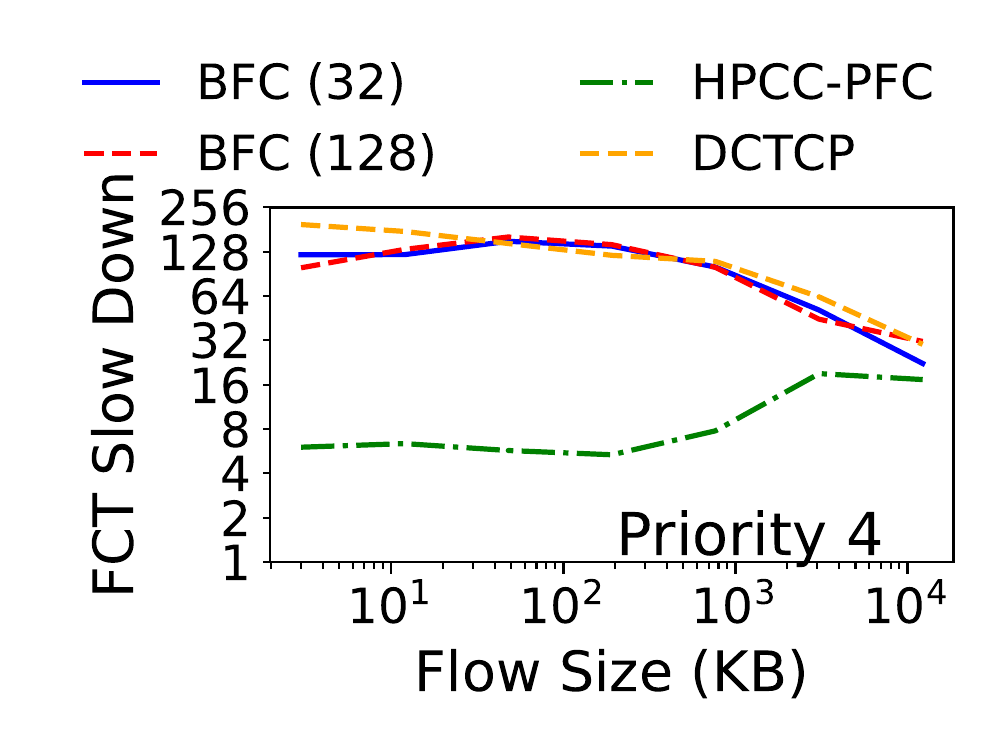}
        \vspace{-7mm}
        \caption{Priority Class 4 (lowest)}
        \label{fig:pgroup:4}
    \end{subfigure}
    \vspace{-3mm}
    \caption{\small \rev{Multiple traffic classes with BFC, reporting 99$^{th}$ percentile FCT slowdown for the Facebook workload, 60\% load, and no incast.}}
    \label{fig:pgroup}
    \vspace{-4mm}
\end{figure*}


\subsection{\rev{Multiple traffic classes}}
\label{app:multi_tfk_classes}

\rev{Many data center operators allocate network traffic into a small number
of priority traffic classes to ensure that mission critical traffic 
is delivered with low tail latency, while other traffic is delivered according
to its quality of service needs. BFC has a simple extension to support priority groups. 
To avoid priority inversion where a flow at one priority can be stalled behind
a flow of a lower priority, we assume queues at a port are statically 
assigned to different priority levels. 
The switch performs dynamic queue assignment for each class independently. A flow with priority $X$ is only assigned to physical queues associated with that priority. Queues at the same priority level follow fair scheduling.}

\rev{Statically partitioning physical queues among traffic classes could make
it more likely for traffic within a class to run out of queues and suffer 
degraded performance with collisions and HoL blocking. 
On the other hand, high priority traffic is preferentially scheduled, leading to
short queues and few active flows. Collisions will be more likely at 
lower priority traffic classes, where performance is already degraded.
Priority scheduling results in rapid and extreme changes 
in the available rate for these background classes. Relative to end-to-end control,
per-hop backpressure can more easily utilize rapidly changing spare capacity.}

\rev{To test how BFC behaves with multiple traffic classes, we repeat the experiment in \Fig{fb:no_incast}: Facebook workload, 60\% load, and no incast. We configure
the system with 4 priority classes, each with equal load (15\% each, 60\% in aggregate). We allocate physical queues evenly to each traffic class. We consider configurations
with 32 and 128 queues per port (8 or 32 queues per class). 
We also show results for HPCC and DCTCP. In this study, DCTCP marks packets based on per-class queueing, while HPCC uses switch aggregates. 
\Fig{pgroup} shows the 99$^{th}$ percentile FCT slowdown for different priority classes. BFC achieves good performance across
all traffic classes and flow sizes. In particular, BFC achieves up to 5$\times$ better tail latency for short flows than DCTCP. At the lowest priority level, DCTCP's short flow tail latency converges to that of BFC. For low priority flows, tail latency is primarily governed by time spent waiting to be scheduled at the switch.}

\rev{HPCC's performance is somewhat anomalous. Long flows suffer priority inversion, where long flows at high priority achieve significantly worse service than
short flows at lower priority.
In HPCC,
long flows back off in an attempt to keep queues empty. 
The (transient) extra capacity left by such long flows can be used by 
short flows traffic at all priority levels, improving performance for these short
flows.} 

\rev{BFC has only slightly better performance with 32 vs. 8 queues per priority level, indicating that collisions did not have much impact. For high priority traffic,
the setup is equivalent to running our experiment with just one traffic class at 15\% load and a small number of queues---even modest numbers of active queues are unlikely
at such low load.  Lower priority traffic can run out of queues, but they gain the benefit of being able to take immediate advantage when the high priority queues are empty. In other words, work conserving behavior is more important for background traffic than the number of queues. We acknowledge this is just one study, and there are likely scenarios where BFC’s performance could suffer when using multiple traffic classes.} 

\rev{One obvious improvement is to split queues dynamically among classes rather than statically. But in the long run, we strongly believe that the number of queues per port is likely to continue to grow to whatever is needed to deliver good performance.}



\subsection{\rev{Parameter sensitivity for comparison schemes}}
\label{app:par_sense_comp}

 \begin{figure}[t]
    \centering
    \begin{subfigure}[tbh]{\columnwidth}
        \includegraphics[width=\textwidth]{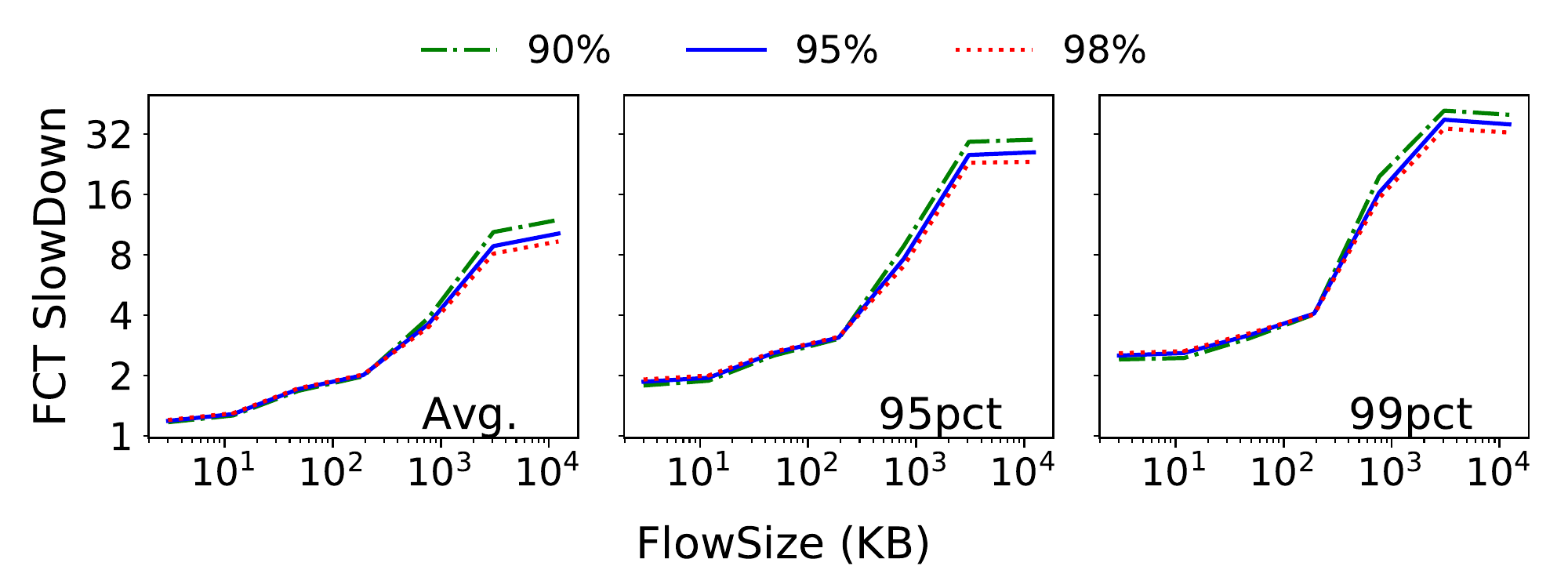}
         \vspace{-7mm}
        \caption{HPCC ($\eta$)}
        \label{fig:par_sense:hpcc}
    \end{subfigure}
    \begin{subfigure}[tbh]{\columnwidth}
        \includegraphics[width=\textwidth]{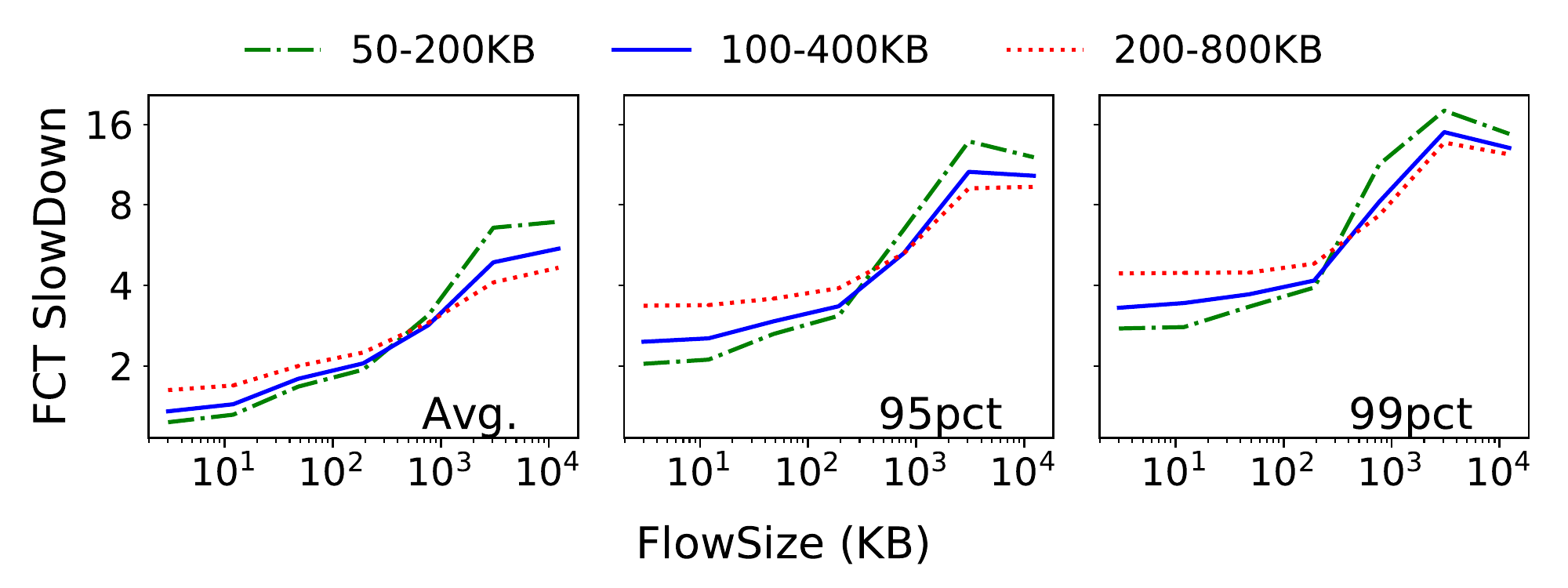}
         \vspace{-7mm}
        \caption{DCTCP (ECN marking threshold: $K_{min}$-$K_{max}$)}
        \label{fig:par_sense:dctcp}
    \end{subfigure}
    \begin{subfigure}[tbh]{\columnwidth}
        \includegraphics[width=\textwidth]{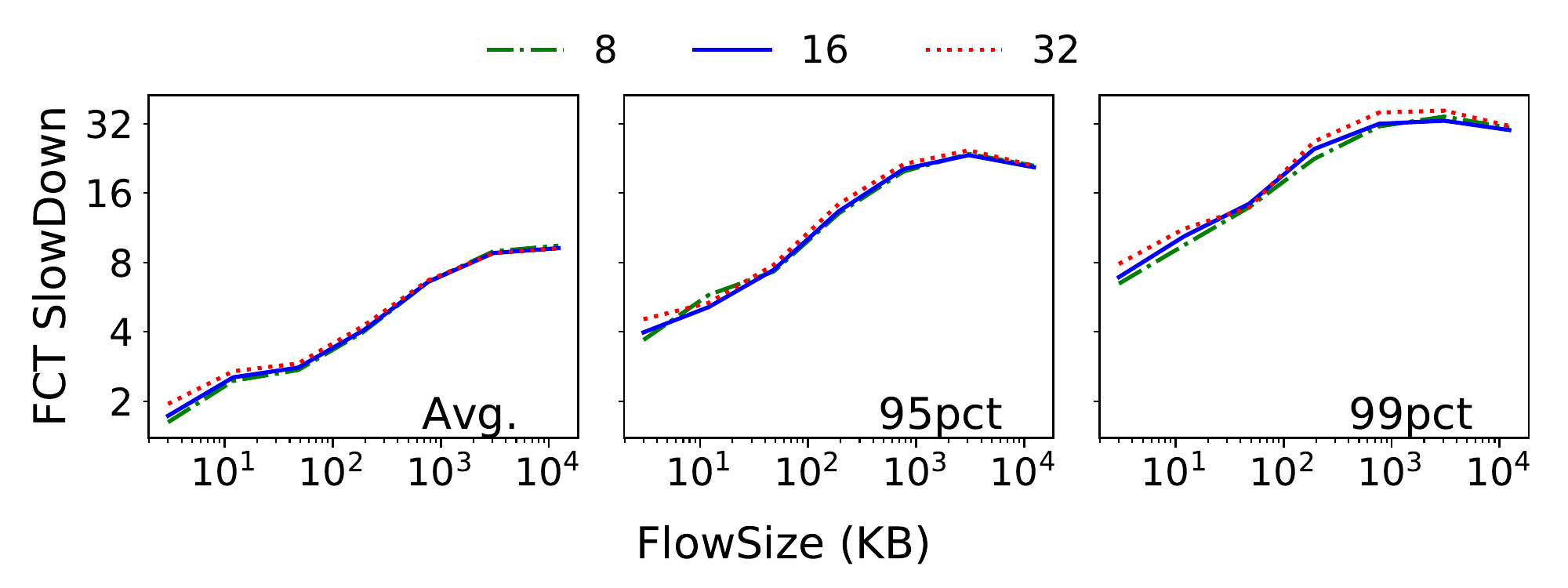}
         \vspace{-7mm}
        \caption{ExpressPass (Credit Buffer Size)}
        \label{fig:par_sense:expass_cbs}
    \end{subfigure}
    \begin{subfigure}[tbh]{\columnwidth}
        \includegraphics[width=\textwidth]{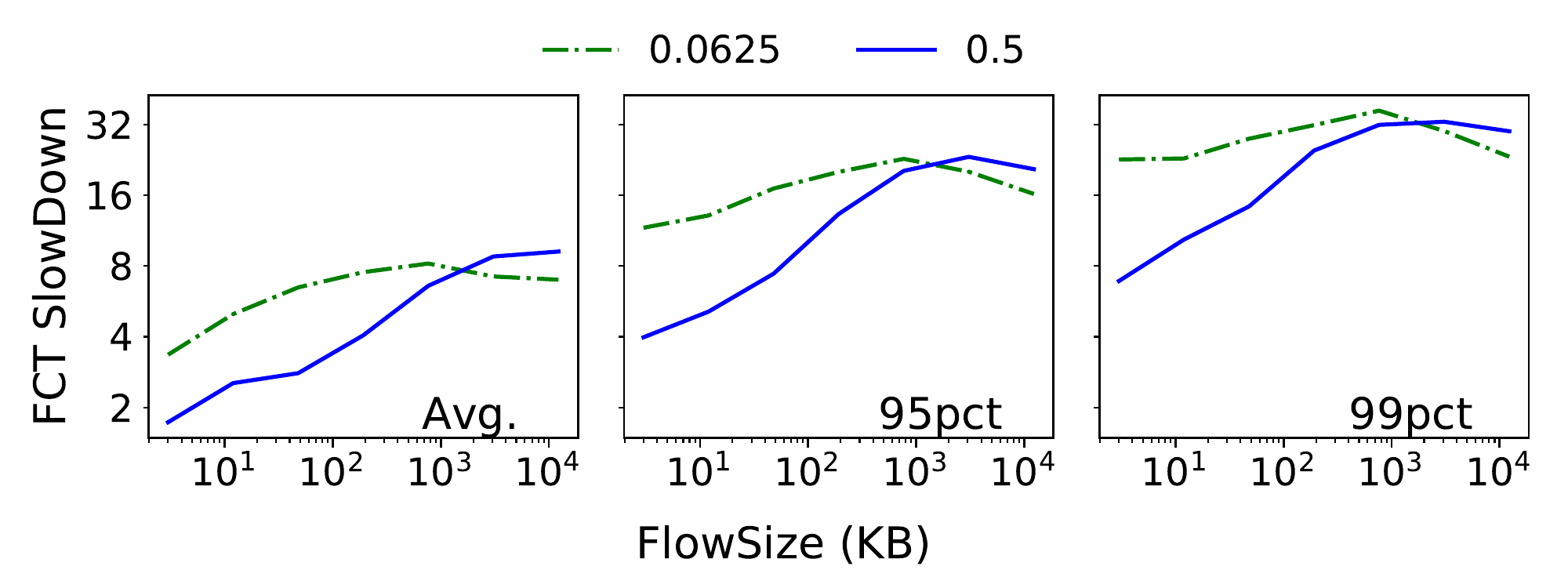}
         \vspace{-7mm}
        \caption{ExpressPass ($\alpha$)}
        \label{fig:par_sense:expass_a}
    \end{subfigure}
    \vspace{-3mm}
    \caption{\small \rev{99$^{th}$ percentile FCT slowdown for the Facebook workload, 60\% load without incast. Sensitivity to the choice of parameters in HPCC, DCTCP, and ExpressPass.}}
    \label{fig:par_sense}
    \vspace{-5mm}
\end{figure}

\rev{In this section, we perform sensitivity analysis to understand the impact of parameters on performance of HPCC, DCTCP and ExpressPass. We repeat the experiment in \Fig{fb:no_incast} (Facebook distribution with 60\% load). \Fig{par_sense} reports the average, 95$^{th}$ and 99$^{th}$ percentile flow completion times as we vary the parameters. In general, we observe that parameters present a trade-off between the latency of short flows (queuing) and the throughput of long flows (link utilization).}

\noindent
\rev{\textit{HPCC:} We vary the target utilization ($\eta$) from 90 to 98\%. As expected, increasing $\eta$ worsens the FCT of short flows but improves the FCT for long flows (marginally for both), see \Fig{par_sense:hpcc}.}

\noindent
\rev{\textit{DCTCP:} We vary the ECN marking threshold governed by parameters $K_{min}$ and $K_{max}$. Increasing the threshold increases the queuing at the switch, which increases FCT of short flows but improves link utilization (\Fig{par_sense:dctcp}). }

\noindent
\rev{\textit{ExpressPass:} Varying the credit buffer size has little impact on performance (\Fig{par_sense:expass_cbs}). We vary $\alpha$, which controls how the receiver credits are generated. Reducing $\alpha$ reduces ``credit waste'', improving the FCT of long flows. However, it also increases the FCT of short flows (\Fig{par_sense:expass_a}).}


\subsection{\rev{Impact of Spatial Locality}}
\label{app:locality}

  \begin{figure}[t]
    \centering
    \begin{subfigure}[tbh]{0.235\textwidth}
        \includegraphics[trim={0 0 0 4mm}, clip, width=\textwidth]{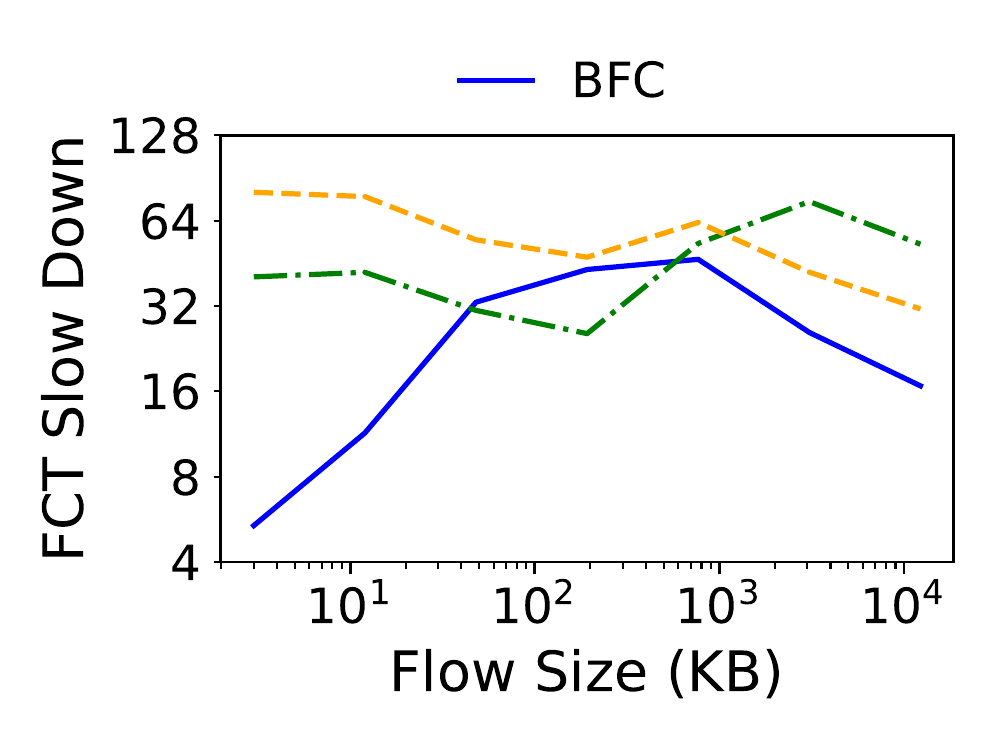}
        \vspace{-7mm}
        \caption{55\% + 5\% 100-1 incast}
        \label{fig:locality:incast}
    \end{subfigure}
    \begin{subfigure}[tbh]{0.235\textwidth}
        \includegraphics[trim={0 0 0 4mm}, clip,width=\textwidth]{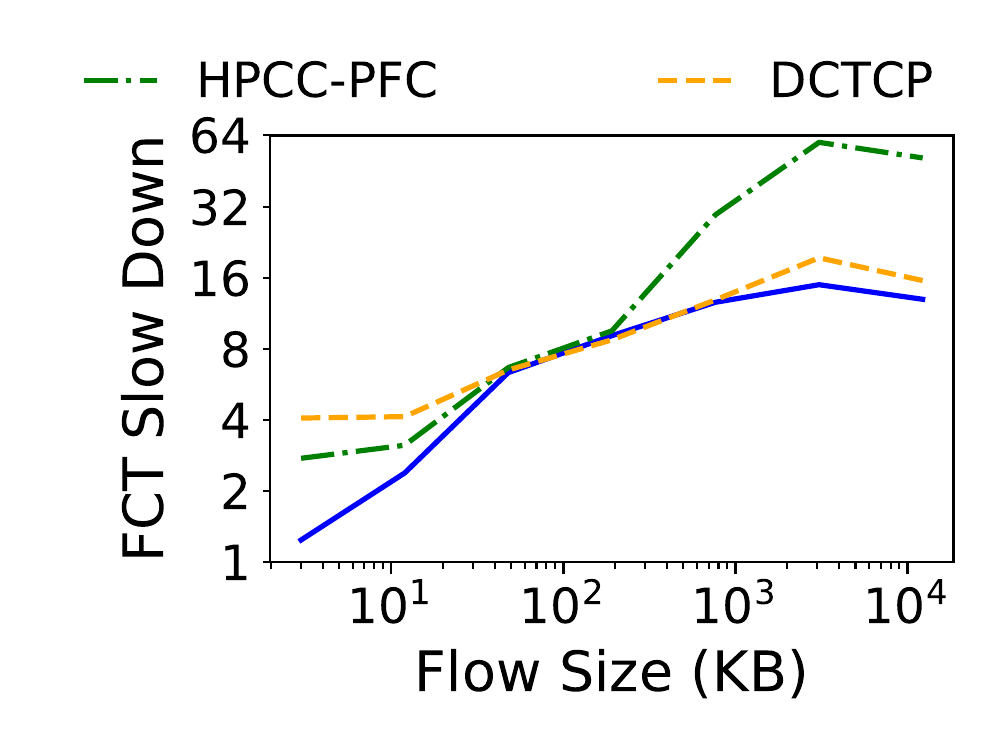}
        \vspace{-7mm}
        \caption{60\%}
        \label{fig:locality:no_incast}
    \end{subfigure}
    \vspace{-3mm}
    \caption{\small \rev{Impact of spatial locality. FCT slowdown (99$^{th}$ percentile) for Facebook distribution with and without incast.}}
    \label{fig:locality}
    \vspace{-5.5mm}
\end{figure} 

\rev{We repeated the experiment from \Fig{fb} with spatial locality in source-destination pairs such that the average load on all links across the network is same. \Fig{locality} shows the 99$^{th}$ percentile slowdowns. The trends are similar to \Fig{fb}.}

\subsection{Using TCP Slow-start}
\label{app:ss}

 \begin{figure}[t]
    \centering
    \begin{subfigure}[tbh]{0.235\textwidth}
        \includegraphics[width=\textwidth]{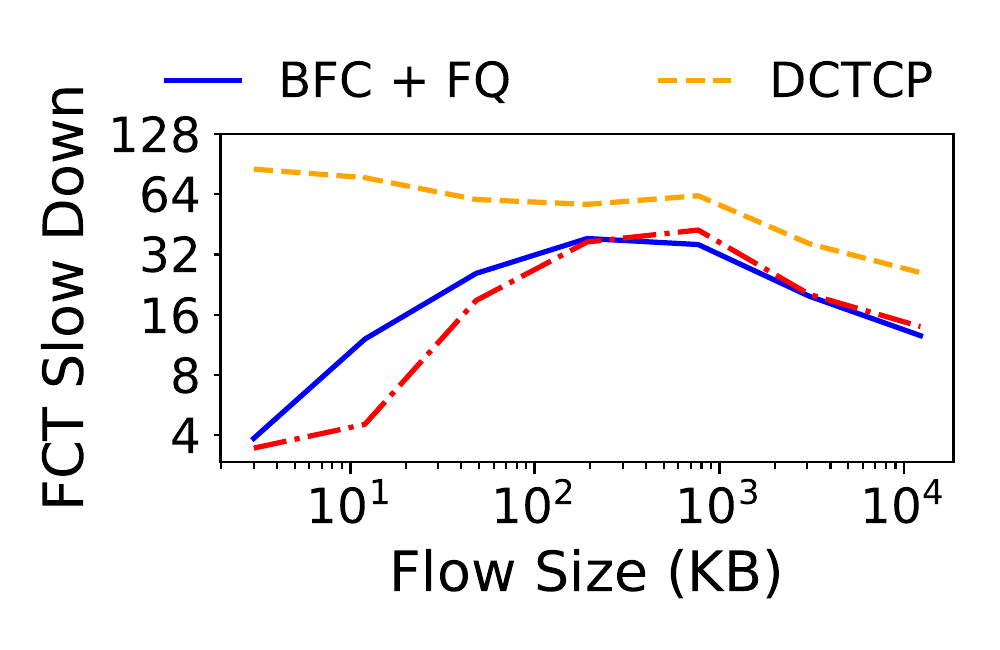}
        \vspace{-7mm}
        \caption{55\% + 5\% 100-1 incast\\(99$^{th}$ percentile FCT)}
        \label{fig:ss:incast_99}
    \end{subfigure}
    \begin{subfigure}[tbh]{0.235\textwidth}
        \includegraphics[width=\textwidth]{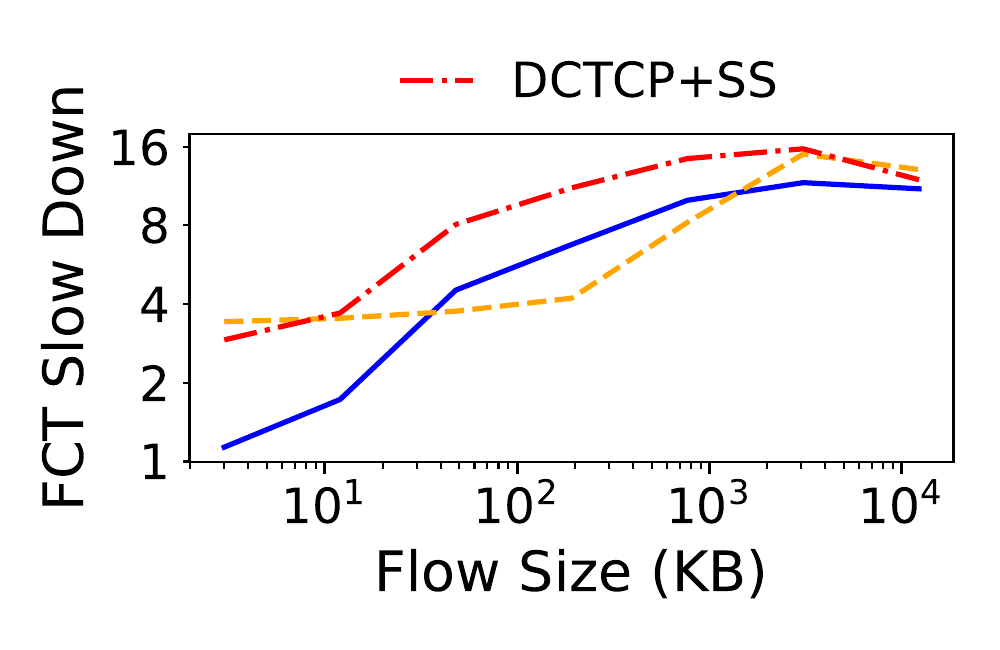}
        \vspace{-7mm}
        \caption{60\% \\(99$^{th}$ percentile FCT)}
        \label{fig:ss:no_incast_99}
    \end{subfigure}
        \begin{subfigure}[tbh]{0.235\textwidth}
        \includegraphics[width=\textwidth]{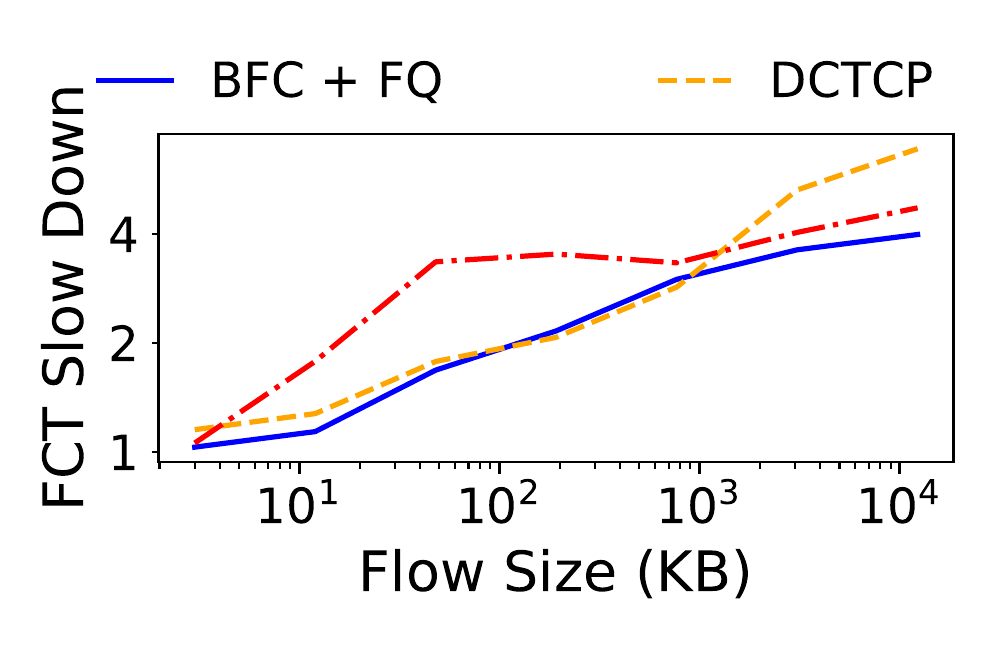}
        \vspace{-7mm}
        \caption{55\% + 5\% 100-1 incast\\(Median FCT)}
        \label{fig:ss:incast_median}
    \end{subfigure}
    \begin{subfigure}[tbh]{0.235\textwidth}
        \includegraphics[width=\textwidth]{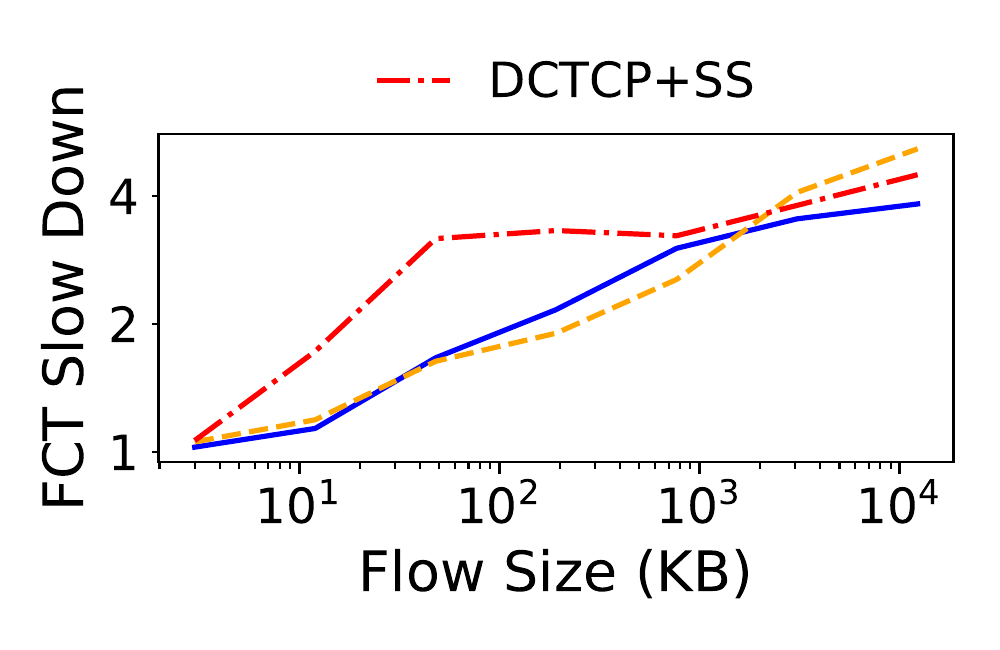}
        \vspace{-7mm}
        \caption{60\% \\(Median FCT)}
        \label{fig:ss:no_incast_median}
    \end{subfigure}
    \vspace{-3mm}
    \caption{\small Impact of using slow start on median and 99th percentile tail latency FCT slowdown, for the Facebook flow size distribution with and without incast (setup the same as \Fig{fb}). With incast, DCTCP + SS (slow start) reduces the tail FCT, but it increases median FCTs by up to 2 $\times$. In the absence of incast, DCTCP + SS increases both the tail and median FCT for short and medium flows.}
    \label{fig:ss}
    \vspace{-4mm}
\end{figure}

We also evaluate the impact of using TCP slow-start instead of starting flows at line rate in Figure~\ref{fig:ss}. We compare the original DCTCP with slow start (DCTCP + SS) with an initial window of 10 packets versus the modified DCTCP used so far (initial window of the BDP). The setup is same as \Fig{fb}. 

With incast, DCTCP + SS reduces buffer occupancy by reducing the intensity of incast flows, improving tail latency (\Fig{ss:incast_99}). However, slow start increases the median FCT substantially (\Fig{ss:incast_median}). Flows start at a lower rate, taking longer to ramp up to the desired rate. For applications with serially dependent flows, an increase in median FCTs can impact the performance substantially.

In the absence of incast, slow start increases both the tail (\Fig{ss:no_incast_99}) and median (\Fig{ss:no_incast_median}) FCT for the majority of flow sizes. In particular, short flows are still slower than with BFC, as slow start does not remove burstiness in buffer occupancy in the tail.

\subsection{Reducing contention for queues}
\label{app:isolate_incast}
To reduce contention for queues under incast, we tried a variant of BFC where the sender labels incast flows explicitly (similar to the potential optimization in ~\cite{homa}). 
BFC + IncastLabel assigns all the incast flows at an egress port to the same queue. This frees up queues for non-incast traffic, reducing collisions and allowing the scheduler
to share the link between incast and non-incast traffic more fairly.

 \begin{figure}[t]
    \centering
    \begin{subfigure}[tbh]{0.235\textwidth}
        \includegraphics[width=\textwidth]{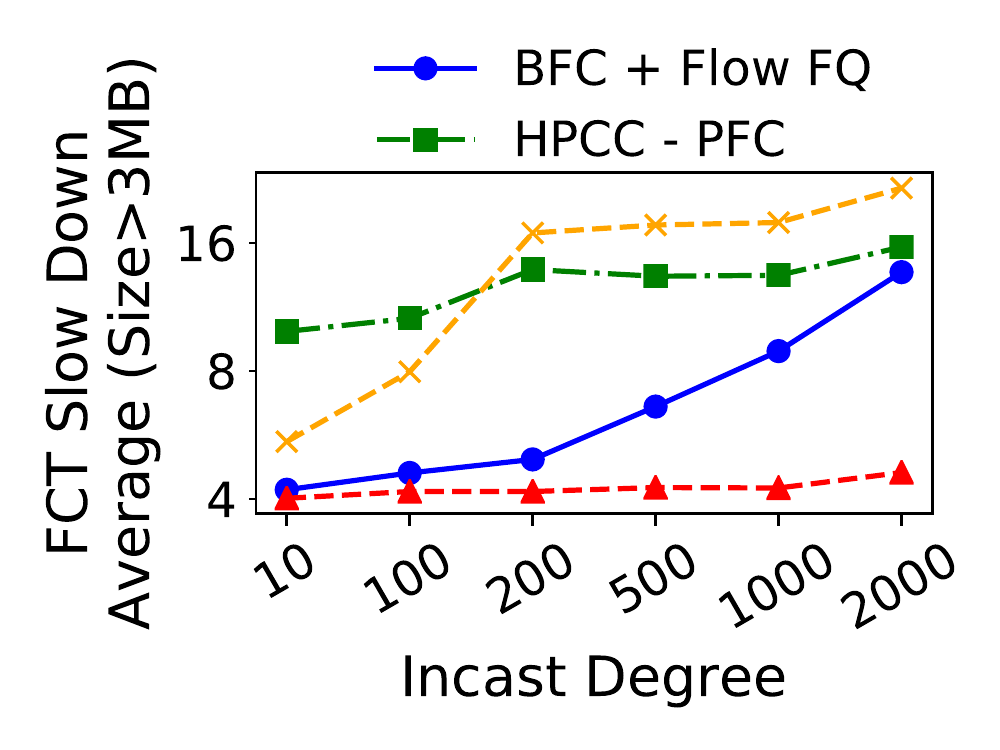}
         \vspace{-7mm}
        \caption{Average FCT for long flows}
        \label{fig:incast_isolate:long}
    \end{subfigure}
    \begin{subfigure}[tbh]{0.235\textwidth}
        \includegraphics[width=\textwidth]{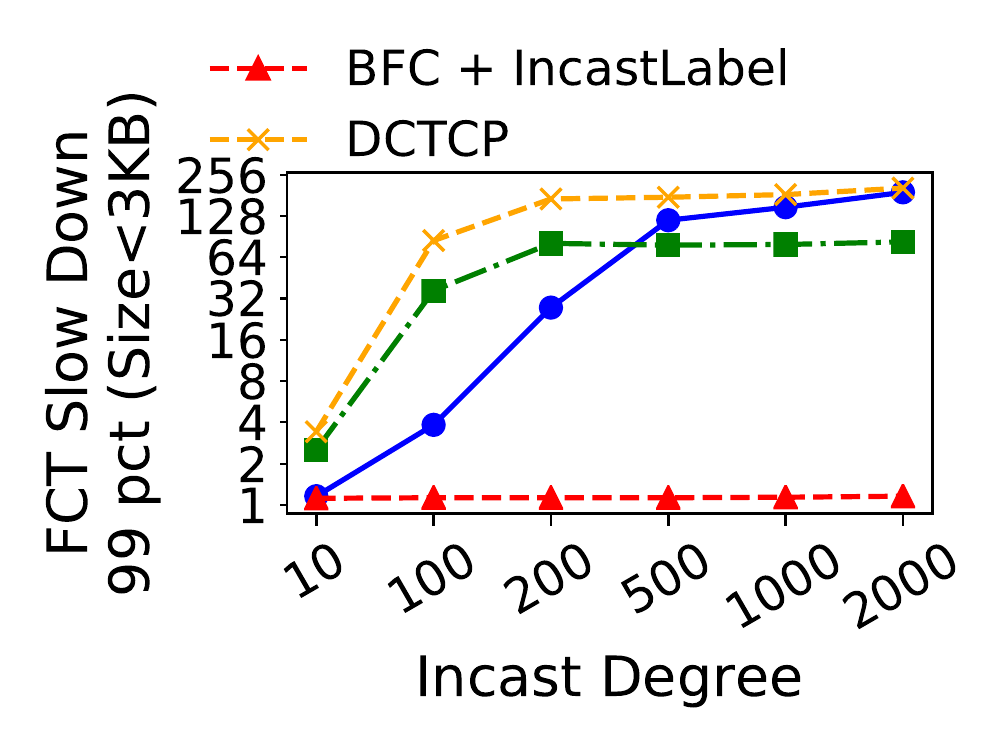}
        \vspace{-7mm}
        \caption{Tail FCT for short flows}
        \label{fig:incast_isolate:short}
    \end{subfigure}
    \vspace{-3mm}
    \caption{\small FCT slowdown for short and long flows as a function of incast degree. The x axis is not to scale. By isolating incast flows, BFC + IncastLabel reduces collisions and achieves the best performance.}
    \label{fig:incast_isolate}
    \vspace{-5mm}
\end{figure}

\Fig{incast_isolate} shows the performance of BFC + IncastLabel in the same setup as \Fig{incast_var}. The original BFC is shown as BFC + Flow FQ for per-flow fair queuing. BFC + IncastLabel achieves the best performance across all the scenarios. However, the FCTs for incast flows is higher compared to BFC + Flow FQ (numbers not shown here). When there are multiple incast flows at an ingress port, the incast flows are allocated less bandwidth in aggregate compared to per-flow fair queuing. 

While BFC + IncastLabel achieves great performance, it assumes the application is able to label incast flows, and so we use a more conservative design for the main body of our evaluation.

\subsection{Incremental Deployment}
\label{app:incr_deploy}
 \begin{figure}[t]
    \centering
    \begin{subfigure}[tbh]{0.235\textwidth}
        \includegraphics[width=\textwidth]{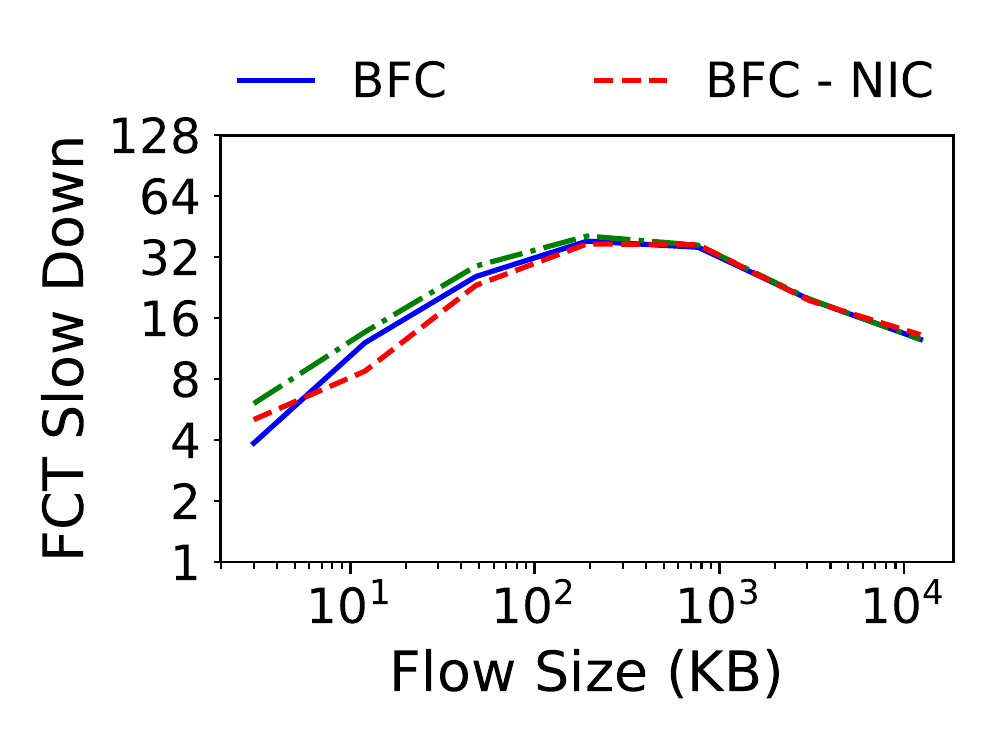}
         \vspace{-7mm}
        \caption{99$^{th}$ percentile FCT}
        \label{fig:deploy:fct}
    \end{subfigure}
    \begin{subfigure}[tbh]{0.235\textwidth}
        \includegraphics[width=\textwidth]{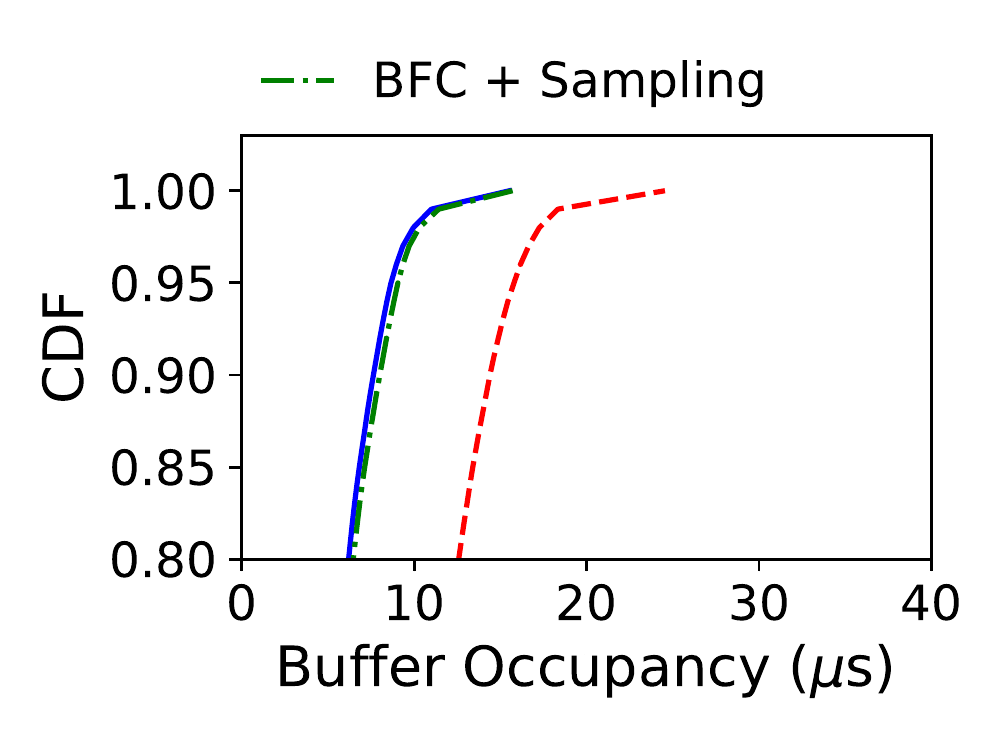}
        \vspace{-7mm}
        \caption{Buffer Occupancy}
        \label{fig:deploy:buffer}
    \end{subfigure}
    \vspace{-3mm}
    \caption{\small FCT slowdown (99$^{th}$ percentile) and buffer occupancy distribution for two BFC variants. When NICs don't respond to backpressure (BFC - NIC), BFC experiences moderate increased buffering. Using sampling to reduce recirculation (BFC + sampling) has marginal impact on performance.}
    \label{fig:deploy}
    \vspace{-5mm}
\end{figure}
We repeated the experiment in \Fig{fb:incast} in the scenario where i) BFC is deployed in part of the network; ii) The switch doesn't have enough capacity to handle all the recirculations. \Fig{deploy} reports the tail FCT and buffer occupancy for these settings.

\noindent\textbf{Partial deployment in the network:} We first evaluate the situation when BFC is only deployed at the switches and the sender NICs don't respond to backpressure signal (shown as BFC - NIC). To prevent sender NIC traffic from filling up the buffers at the ToR, we assume a simple end-to-end congestion control strategy where the sender NIC caps the in-flight packets for a flow to 1 end-to-end bandwidth delay product (BDP). As expected, BFC - NIC experiences increased buffering at the ToR (\Fig{deploy:buffer}). However, the tail buffer occupancy is still below the buffer size and there are no drops. Since all the switches are BFC enabled and following dynamic queue assignment, the frequency of collisions and hence the FCTs are similar to the orignal BFC.

\noindent\textbf{Sampling packets to reduce recirculations:} A BFC switch with an RMT architecture~\cite{bosshart13} recirculates packets to execute the dequeue operations at the ingress port. Depending on the packet size distribution of the workload, a switch might not have enough packet processing (pps) capacity or recirculation bandwidth to process these recirculated packets. In such scenarios, we can reduce recirculations by sampling packets. Sampling works as follows.

On a packet arrival (enqueue), sample to decide whether a packet should be recirculated or not. Only increment the pause counter and \texttt{\small size} in the flow table for packets that should be recirculated. The dequeue operations remain as is and are only executed on the recirculated packets. The \texttt{\small size} now counts the packets sampled for recirculation and residing in the switch. While sampling reduces recirculations, it can cause packet reordering. Recall, BFC uses \texttt{\small size} to decide when to reassign a queue. With sampling, \texttt{\small size} can be zero even when a flow has packets in the switch. This means a flow's queue assignment can change when it already has packets in the switch, causing reordering. However, sticky queue assignment should reduce the frequency of these events (\S\ref{s:backpressure}).

We now evaluate the impact of sampling on the performance of BFC (shown as BFC + Sampling). In the experiment, the sampling frequency is set to 50\%, \ie only 50\% of the packets are recirculated. BFC + Sampling achieves nearly identical tail latency FCT slowdowns and switch buffer occupancy as the orginal BFC. With sampling, fewer than 0.04\% of the packets were retransmitted due to packet reordering.

\subsection{Cross data center traffic}
\label{app:cross}
For fault tolerance, many data center applications replicate their data to nearby data centers (\eg to a nearby metro area). We evaluate
the impact of BFC on managing cross-data center congestion in such scenarios. 
We consider the ability of different systems to achieve good throughput for the inter-data-center traffic,
and we also consider the impact of the cross-data-center traffic on tail latency of local traffic, as
the larger bandwidth-delay product means more data is in-flight when it arrives at the bottleneck.

We created a Clos topology with 64 leaf servers, and 100\,Gbps links and 12\,MB switch buffers.
Two gateway switches connect the data centers using a 200\,Gbps link with 200\,$\mu$s of one-way delay (i.e. the base round
trip delay of the link is 400\,$\mu$s), or roughly equivalent to the two data centers being separated
by 50\,km assuming a direct connection.\cut{A longer hop will require more buffering in BFC to keep the link busy, 
and so we configured the buffer size of the gateway switch to be 64\,MB.}
The experiment consists of intra-data-center flows derived from the Facebook distribution (60\% load). Additionally, there are 20 long-lived inter-data-center flows in both the directions.

\begin{figure}[t]
    \centering
    \begin{subfigure}[tbh]{0.22\textwidth}
        \includegraphics[width=\textwidth]{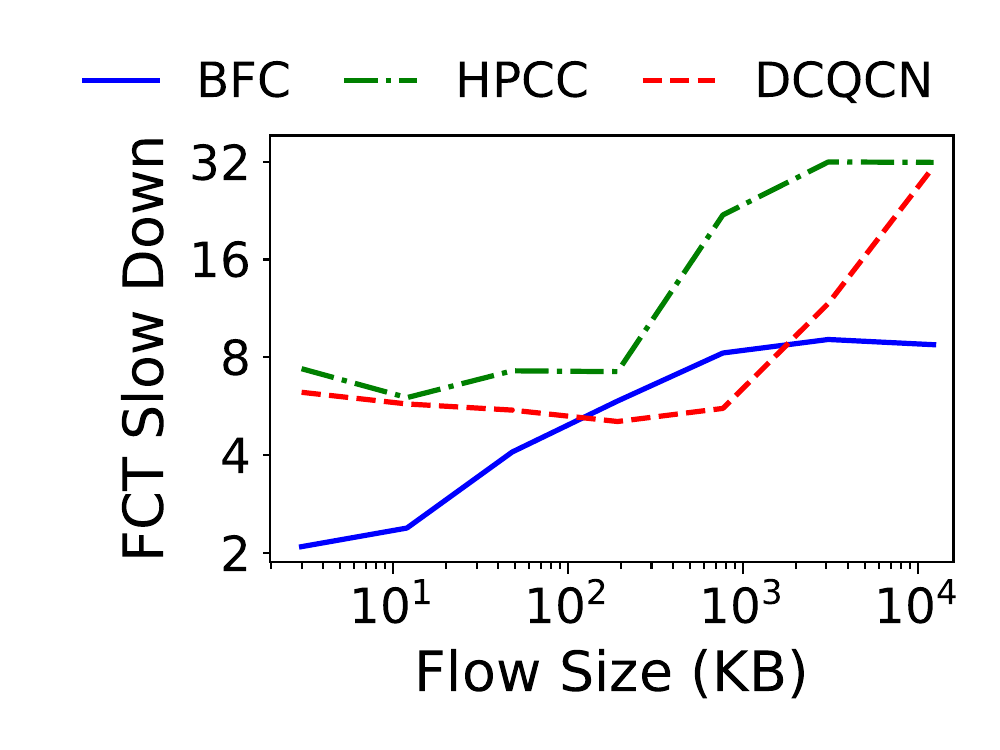}
         \vspace{-7mm}
        \caption{99$^{th}$ percentile FCT }
        \label{fig:cross:fct}
    \end{subfigure}
    \begin{subfigure}[tbh]{0.25\textwidth}
        \includegraphics[width=0.86\textwidth]{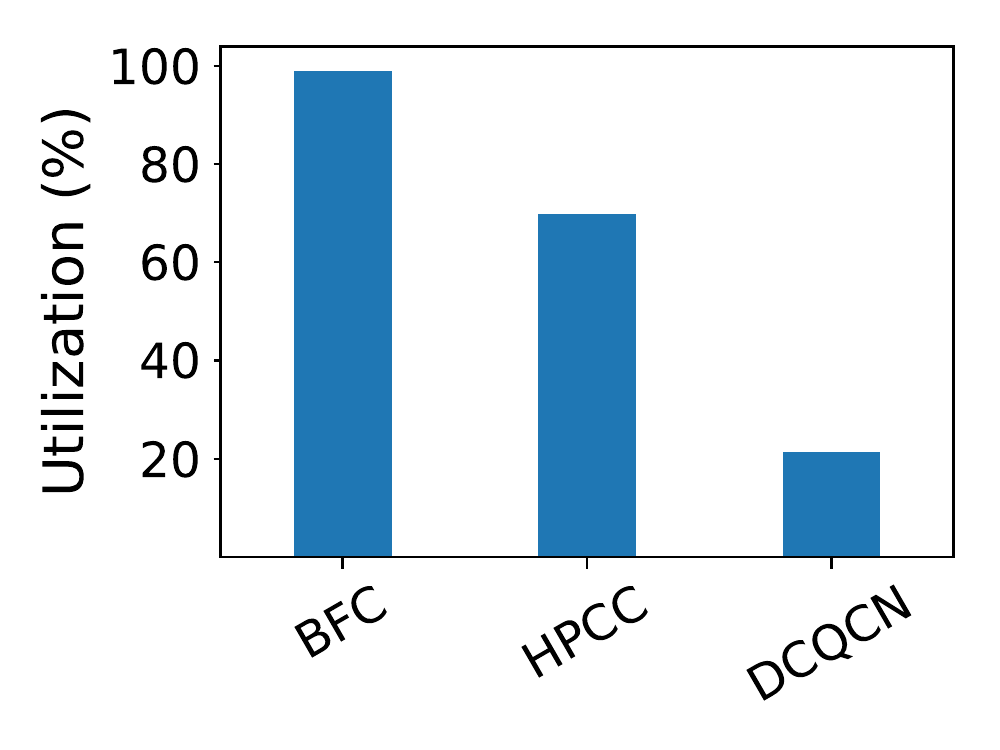}
        \vspace{-2mm}
        \caption{Utilization at the interconnect.}
        \label{fig:cross:utilization}
    \end{subfigure}
    \vspace{-3mm}
    \caption{\small Performance in cross data center environment where two data center are connected by a 200 $\mu$s link, for the Facebook workload (60\% load) with no incast traffic. The left figure shows the 99th percentile FCT slowdown for intra-data-center flows. The right figure shows the average utilization of the link connecting the two data centers.}
    \label{fig:cross}
    \vspace{-5mm}
\end{figure}

\Fig{cross:fct} shows the 99th percentile tail latency in FCT slowdown for intra-data-center flows for BFC, HPCC and DCQCN.\footnote{Data center operators have developed specialized protocols for better inter-data center link management~\cite{bbr}; 
comparing those to BFC is future work.} \Fig{cross:utilization} shows the average utilization of the link connecting the two data centers (interconnect), a proxy for the aggregate throughput of the long-lived inter-data-center flows.
BFC is better for both types of flows. 
With BFC, the link utilization of the wide area interconnect is close to 100\%, while neither HPCC nor DCQCN can maintain the link at full utilization, even with ample parallelism.  This is likely a consequence of slow end-to-end reaction of the inter-data-center flows~\cite{saeed2020annulus}. The congestion state on the links within a data center is changing rapidly because of the shorter intra-data-center flows. By the time an inter-data-center flow receives congestion feedback and adjusts its rate, the congestion state in the network might have already changed. When capacity becomes available, the inter-data-center flows can fail to ramp up quickly enough, hurting its throughput. 

Relative to the single data center case (cf. \Fig{fb:no_incast}), tail latency FCTs are worse for all three protocols, but the relative advantage of BFC is maintained. 
Where HPCC has better tail latency than DCQCN in the single data
center case for both short and medium-sized flows, once inter-data-center traffic is added, 
HPCC becomes worse than DCQCN. With bursty workloads, on the onset of congestion, the long-lived flow will take an end-to-end RTT to reduce its rate, and can build up to 1 BDP (or 500 KB) of buffering, hurting the tail latency of intra-data-center traffic. 
This has less of an impact on DCQCN because it utilizes less of the inter-data-center bandwidth in the first place.

In contrast, \bfc{} reacts at the scale of the hop-by-hop RTT.
Even though inter-data-center flows have higher end-to-end RTTs, on switches within the data center, 
\bfc{} will pause/resume flows on a hop-by-hop RTT timescale (2\,$\mu$s).
As a result, with BFC, tail latencies of intra-data-center flows are relatively unaffected by the presence of
inter-data-center flows, while the opposite is true of HPCC. 

\subsection{Physical queue assignment}
\label{app:qassign}
 \begin{figure}[t]
     \centering
     \begin{subfigure}[t]{0.22\textwidth}
    \includegraphics[width=1.0\textwidth]{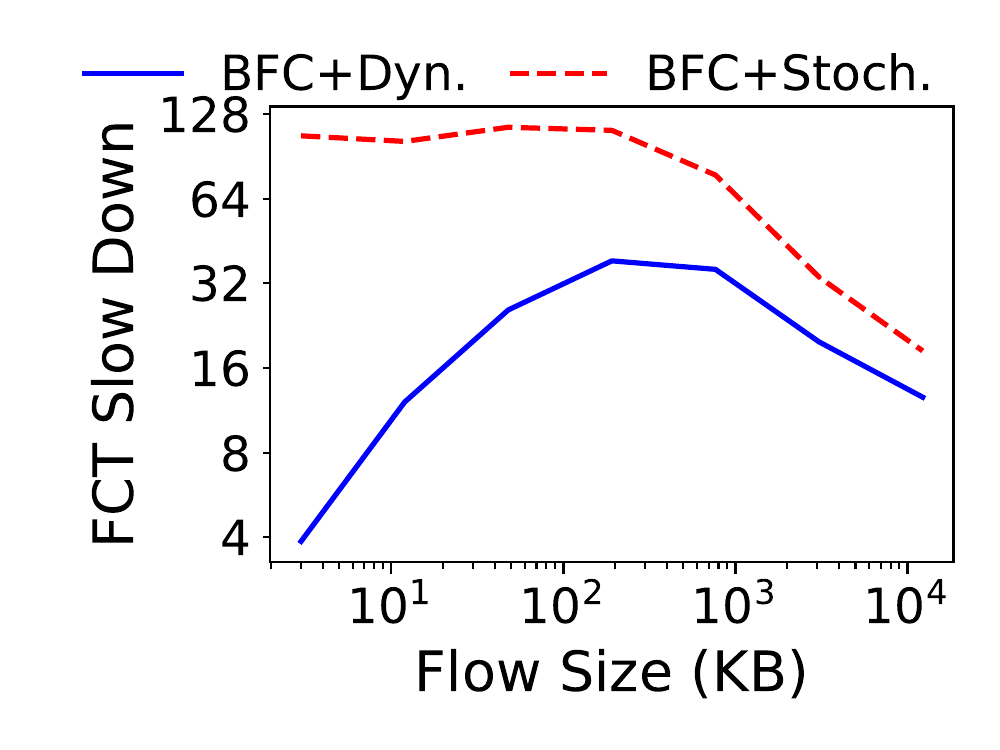}
    \vspace{-7mm}
    \subcaption{99$^{th}$ percentile FCT}
    \label{fig:indirection:fct}
    \end{subfigure}
     \begin{subfigure}[t]{0.22\textwidth}
    \includegraphics[width=1.0\textwidth]{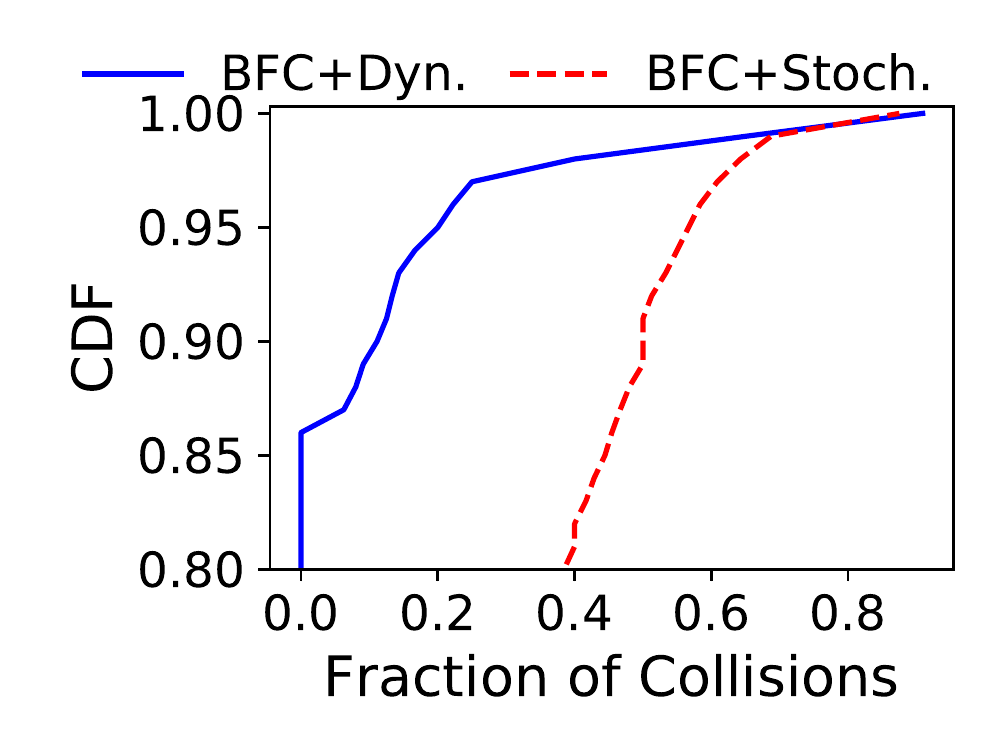}
    \vspace{-7mm}
    \subcaption{Collisions}
    \label{fig:indirection:collisions}
    \end{subfigure}
    \vspace{-3mm}
    \caption{\small Performance of BFC with stochastic queue assignment, for the workload
    in \Fig{fb:incast}. \bfc{} + Stochastic incurs more queue collisions leading to worse tail latency especially for small flows compared to \bfc{} + Dynamic.}
    \label{fig:indirection}
    \vspace{-5mm}
 \end{figure}
 To understand the importance of dynamically assigning flows to physical queues, we repeated the experiment in \Fig{fb:incast} with a variant of \bfc{}, \bfc{} + Stochastic, where we use stochastic hashing to statically assign flows to physical queues (as in SFQ). In \bfc{} (referred as \bfc{} + Dynamic here), the physical queue assignment is dynamic. To isolate the effect of changing the physical queue assignment, the pause thresholds are the same as \bfc{} + Dynamic.

 \Fig{indirection:fct} shows the tail latency. Compared to \bfc{}, tail latency for \bfc{} + Stochastic is much worse for all
 flow sizes. 
 Without the dynamic queue assignment, flows are often hashed to the same physical queue, triggering HoL blocking and hurting tail latency, even when there are unoccupied physical queues. \Fig{indirection:collisions} is the CDF of such collisions. \bfc{}+Stochastic experiences collisions in a high fraction of cases and flows end up being paused unnecessarily. Such flows finish later, further increasing the number of active flows and collisions.
 Even with incast, the number of active flows in BFC is smaller
 than the number of physical queues most of the time.

\subsection{Size of flow table}
\label{app:sensitivity}
\begin{figure}[t]
    \centering
    \begin{subfigure}[t]{0.45\textwidth}
        \includegraphics[width=\textwidth]{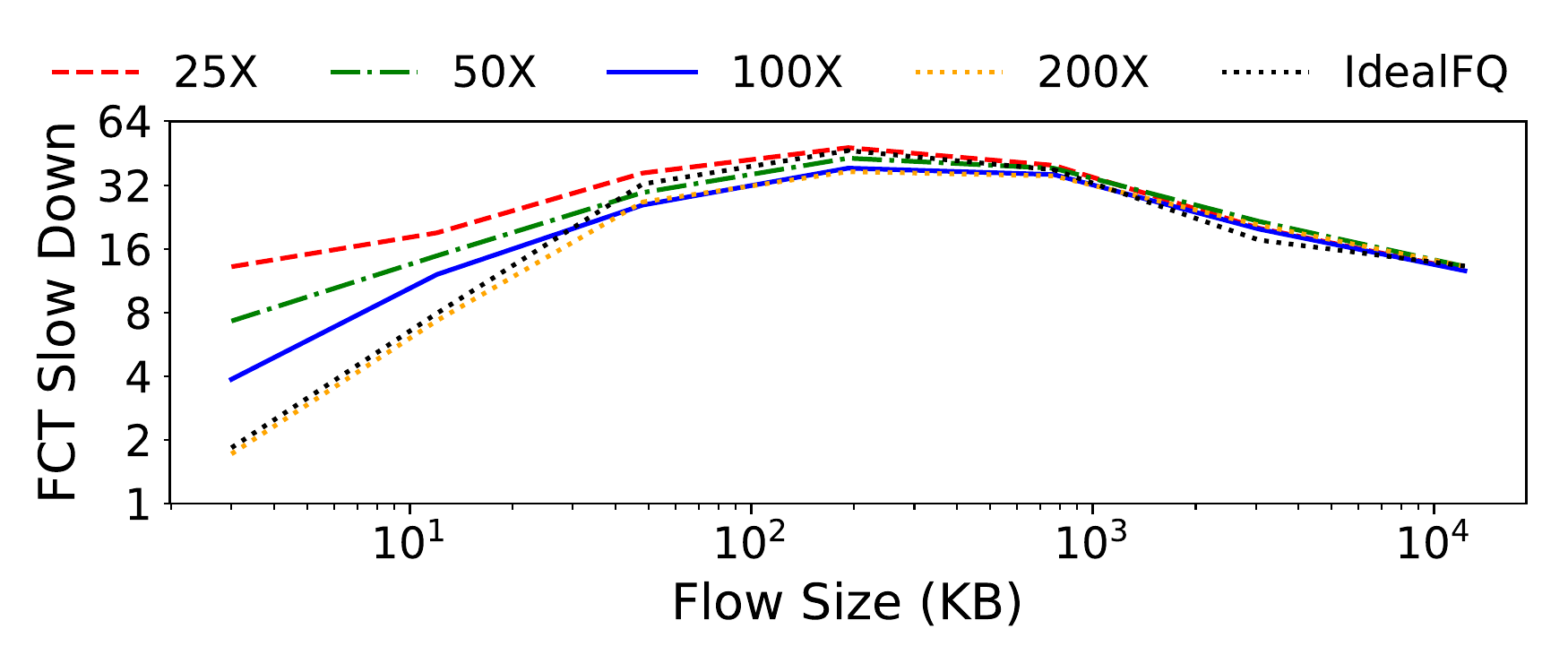}
        \vspace{-6mm}
        \caption{99$^{th}$ percentile FCT}
        \label{fig:impact_vq:fct}
    \end{subfigure}
    \vspace{-3mm}
    \caption{\small FCT slowdown (99th percentile) for BFC for different size flows as a function of the size of the flow table (as a multiple of the number of queues in the switch). The other experiments in the paper use a flow table of 100X. Further reducing the size of the flow table hurts small flow performance.}
    \label{fig:impact_vq}
    \vspace{-5mm}
\end{figure}
We repeated the experiment in \Fig{fb:incast}, but varied the size of the flow table (as a function of the number of queues in the switch). The default in the rest of the paper uses a flow table of 100X. \Fig{impact_vq} shows the tail latency as a function of flow size, for both smaller and larger flow tables. Reducing the size of the flow table increases the index collisions in the flow table. Each flow table collision means that those flows are necessarily assigned to the same physical queue. Tail latency FCTs degrade as a result, particularly for small flows and for smaller table sizes.
This experiment shows that increasing the size of the flow table would moderately improve short flow
tail latency for BFC.

\subsection{Incast flow performance}
\label{app:incast_fct}
\begin{figure}
    \centering
        \includegraphics[width=0.5\columnwidth]{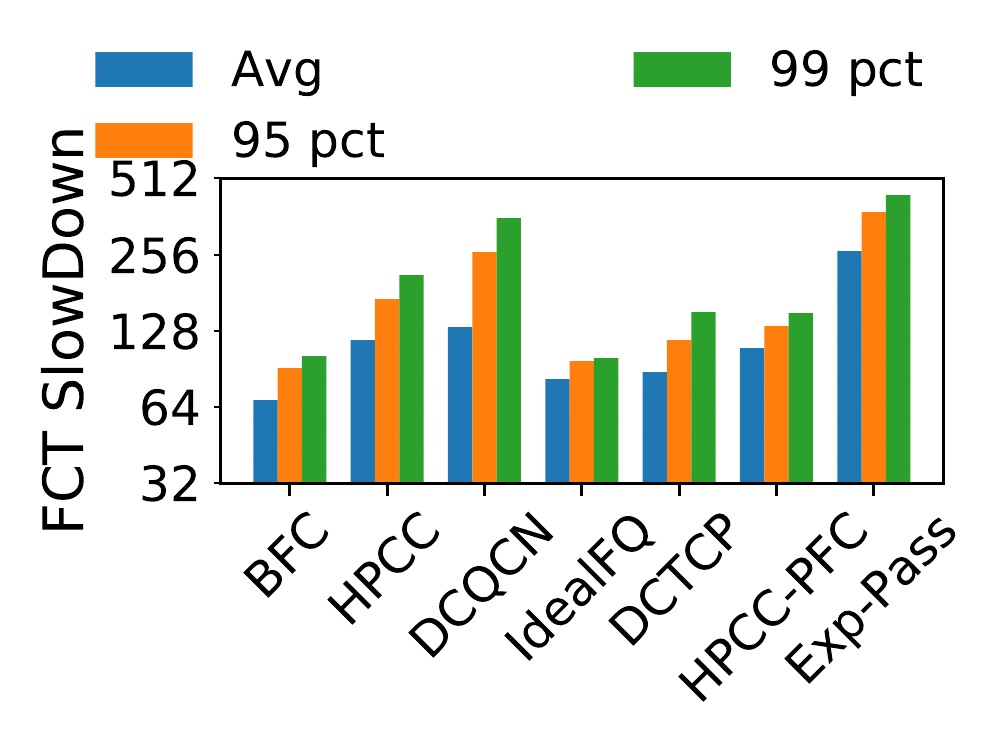}
        \vspace{-5mm}
        \caption{\small FCT slowdown for incast traffic. Slowdown is defined per flow.
        BFC reduces the FCT for incast flows compared to other feasible schemes. Setup from \Fig{google_incast}.}
        \label{fig:google_incast:fct_incast}
        \vspace{-7mm}
\end{figure}

\Fig{google_incast:fct_incast} shows the slowdown for incast flows for the Google workload used in \Fig{google_incast}. 
The benefits of BFC for non-incast traffic do not come at the expense of worse incast performance. Indeed, BFC improves the performance of incast flows relative to end-to-end congestion control, because it reacts faster when capacity becomes available at the bottleneck, reducing the percentage of time the bottleneck is unused while the incast is active.

\section{Deadlock prevention}
\label{app:deadlock}
We formally prove that BFC is deadlock-free in absence of cyclic buffer dependency. Inspired by Tagger~\cite{hu2017tagger}, we define a backpressure graph ($G(V, E)$) as follows:
\begin{enumerate}
    \item Node in the graph ($V$): A node is an egress port in a switch and can thus be represented by the pair <switchID, egressPort>. 
    \item Edge in the graph ($E$): There is a directed edge from $B \rightarrow A$, if a packet can go from $A$ to $B$ in a single hop (i.e., without traversing any other nodes) and trigger backpressure from $B \rightarrow A$. Edges represent how backpressure can propagate in the topology.
\end{enumerate}

We define deadlock as a situation when a node (egress port) contains a  queue that has been paused indefinitely. Cyclic buffer dependency is formally defined as the situation when $G$ contains a cycle.

\begin{theorem}
BFC is deadlock-free if $G(V, E)$ does not contain any cycles.
\end{theorem}

\noindent\textit{Proof:} We prove the theorem by using contradiction. 

Consider a node $A$ that is deadlocked. $A$ must contain a  queue ($A_q$) that has been indefinitely paused as a result of backpressure from the downstream switch. If all the packets sent by $A_q$ were drained from the downstream switch, then $A_q$ will get unpaused (\S\ref{s:backpressure}). There must be at least one node  ($B$) in the downstream switch that triggered backpressure to $A_q$ but hasn't been able to drain packets from $A_q$, i.e., $B$ is deadlocked. This implies, in $G$, there must be an edge from $B \rightarrow A$. Applying induction, for $B$ there must exist another node $C$ (at the downstream switch of $B$) that is also deadlocked (again there must be an edge from $C \rightarrow B$). Therefore, there will be an infinite chain of nodes which are paused indefinitely, the nodes of the chain must form a path in $G$. Since $G$ doesn't have any cycles, the paths in $G$ can only be of finite length, and therefore, the chain cannot be infinitely long. A contradiction, hence proved.

\smallskip
\noindent\textbf{Preventing deadlocks:}
To prevent deadlocks, given a topology, we calculate the backpressure graph, and pre-compute the edges that should be removed so that the backpressure graph doesn't contain any cycles. Removing these edges thus guarantees that there will be no deadlocks even under link failures or routing errors. To identify the set of edges that should be removed we can leverage existing work~\cite{hu2017tagger}.

To remove a backpressure edge $B \rightarrow A$, we use the simple strategy of skipping the backpressure operation for packets coming from $A$ going to $B$ at the switch corresponding to $B$.\footnote{To remove backpressure edges in PFC, Tagger uses a more complex approach that invloves creating new cycle free backpressure edges corresponding to the backpressure edges that should be removed. To ensure losslessness, Tagger generates backpressure using these new cycle free edges instead of the original backpressure edge. In our proposed solution, we forgo such requirement for simplicity.} Note that, a switch can identify such packets \emph{locally} using the ingress and egress port of the packet. This information can be stored as a match-action-table (indexed by the ingress and egress port) to check whether we should execute the backpressure operations for the packet.

For Clos topologies, this just includes backpressure edges corresponding to packets that are coming from a higher layer and going back to a higher layer (this can happen due to rerouting in case of link failures). Note that, usually the fraction of such packets is small (< 0.002\%~\cite{hu2017tagger}), so forgoing backpressure for a small fraction of such packets should hurt performance marginally (if at all).

\section{Impact of Pause Threshold}
\label{app:impact_th}
A consequence of the simplicity of BFC's backpressure mechanism is that a flow can temporarily run out of packets at a bottleneck switch while the flow still has packets to send. The pause threshold ($Th$) governs the frequency of such events. Using a simple model, we quantify the impact of $Th$.

Consider a long flow $f$ bottlenecked at a switch $S$. To isolate the impact of the delay in resuming, we assume that $f$ is not sharing a queue with other flows at $S$ or the upstream switch. Let $\mu_{f}$ be the dequeue rate of $f$ at $S$, \ie when $f$ has packets in $S$, the packets are drained at a steady rate of $\mu_f$. Similarly, let $\mu_f \cdot x$ be the enqueue rate of $f$ at the switch, \ie if $f$ is not paused at the upstream, $S$ receives packets from $f$ at a steady rate of $\mu_f \cdot x$. Here, $x$ denotes the ratio of enqueue to dequeue rate at $S$. Since $f$ is bottlenecked at $S$, $x > 1$. 

We now derive the fraction of time in steady state that $f$ will not have packets in $S$. We show that this fraction depends only on $x$ and $Th$, and is thereby referred as $E_f(x, Th)$.

The queue occupancy for $f$ will be cyclic with three phases. 
\begin{itemize}
\item Phase 1: $S$ is receiving packets from $f$ and the queue occupancy in increasing. 
\item Phase 2: $S$ is \emph{not} receiving packets from $f$ and the queue is draining.
\item Phase 3: $S$ is not receiving packets from $f$ while the queue is empty. 
\end{itemize}

The time period for phase 1 ($t_{p1}$) can be calculated as follows. The queue occupancy at start of the phase is 0 and $S$ is receiving packets from $f$. $f$ gets paused when the queue occupancy exceeds $Th$. The queue builds at the rate $\mu_f \cdot x - \mu_f$ (enqueue rate - dequeue rate). The pause is triggered after $\frac{Th}{\mu_f \cdot (x - 1)}$ time from the start of the phase. Since the pause takes an $HRTT$ to take effect, the queue grows for an additional $HRTT$. $t_{p1}$ is therefore given by:
\begin{align}
   t_{p1} &= \frac{Th}{\mu_f \cdot (x - 1)} + HRTT. 
\end{align}

The queue occupancy at the end of phase 1 is $Th + HRTT \cdot \mu_f \cdot (x - 1)$. The time period for phase 2 ($t_{p2}$) corresponds to the time to drain the queue. $t_{p2}$ is given by:
\begin{align}
    t_{p2} &= \frac{Th + HRTT \cdot \mu_f \cdot (x - 1)}{\mu_f}.
\end{align}

At the end of phase 2, there are no packets from $f$ in $S$. As a result, $S$ resumes $f$ at the upstream. Since the resume takes an $HRTT$ to take effect, the queue is empty for an $HRTT$. Time period for phase 3 ($t_{p3}$) is given by:
\begin{align}
    t_{p3} &= HRTT
\end{align}

Combining the equations, $E_f(x, Th)$ is given by:
\begin{align}
    E_f(x, Th) &= \frac{t_{p3}}{t_{p1} + t_{p2} + t_{p3}}\nonumber\\
           &= \frac{x -1}{\frac{Th}{HRTT \cdot \mu_f} \cdot x + (x^2 -1)}.
\end{align}

\begin{figure}[t]
     \centering
    \includegraphics[width=0.8\columnwidth]{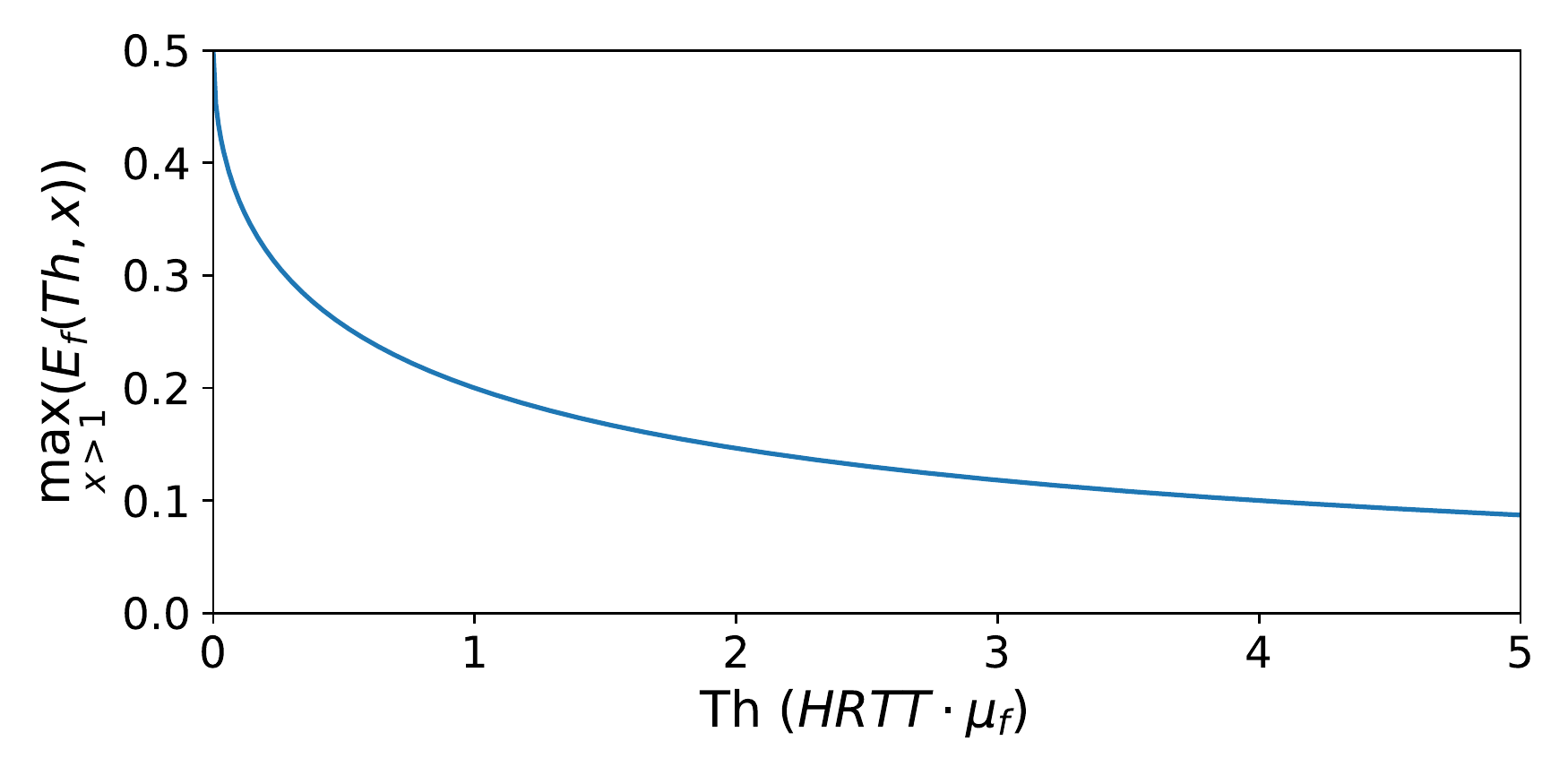}
    \vspace{-4mm}
    \caption{\small Impact of pause threshold ($Th$) on the metric of worst case inefficiency. Increasing $Th$ reduces the maximum value for the fraction of time $f$ can run out of packets at the bottleneck.}
    \label{fig:appendix_th}
    \vspace{-5.5mm}
 \end{figure}
 
Notice that for a given $x$, $E_f(x, Th)$ reduces as we increase $Th$. Increasing $Th$, increases the time period for phase 1 and phase 2, and the fraction of time $f$ runs out of packets reduces as a result. 

We now quantify the impact of pause threshold on the worst case (maximum) value of $E_f(x, Th)$. Given a $Th$, $E_f(x, Th)$ varies with $x$. When $x \to 1$, $(E_f(x, Th) \to 0$, and when $x \to \infty$, $(E_f(x, Th) \to 0$. The maxima occurs somewhere in between. More concretely, for a given value of $Th$, the maxima occurs at $x = \sqrt{\frac{Th}{HRTT \cdot \mu_f}} + 1$. The maximum value ($\max_{x > 1}\left(E_f(x, Th)\right)$) is given by:
\begin{equation}
    \max_{x > 1}\left(E_f(x, Th)\right) = \frac{1}{\left(\sqrt{\frac{Th}{HRTT \cdot \mu_f}} + 1\right)^2 + 1}.
\end{equation}

\Fig{appendix_th} shows how $\max_{x > 1}\left(E_f(x, Th)\right)$ changes as we increase the pause threshold. As expected, increasing the pause threshold reduces $\max_{x > 1}\left(E_f(x, Th)\right)$. However, increasing the pause threshold has diminishing returns. Additionally, increasing $Th$ increases the buffering for $f$ (linearly).

In BFC, we set $Th$ to 1-Hop BDP at the queue drain rate, \ie $Th = HRTT \cdot \mu_f$. Thereofore, the maximum value of $E_f(x, Th)$ is 0.2 (at $x = 2$). This implies, under our assumptions, that a flow runs out of packets at most 20\% of the time due to the delay in resuming a flow. 

Note that 20\% is the maximum value for $E_f(x, Th)$. When $x \neq 2$, $E_f(x, Th)$ is lower. For example, when $x = 1.1$ (\ie the enqueue rate is 10\% higher than the dequeue rate), $E_f(x, Th)$ is only 7.6\%. 

The above analysis suggests that the worst-case under-utilization caused by delay in resuming is 20\%. Note that in practice, when an egress port is congested, there are typically multiple flows concurrently active at that egress. In such scenarios, the under-utilization is much less than this worst-case bound, because it is unlikely that all flows run out of packets at the same time.
As our evaluation shows, with BFC, flows achieve close to ideal throughput in realistic traffic scenarios (\S\ref{s:eval}).

\end{document}